\documentclass[oneside,sn-mathphys,Numbered]{sn-jnl}

\usepackage{aas_macros}

\usepackage{graphicx}%
\usepackage{multirow}%
\usepackage{amsmath,amssymb,amsfonts}%
\usepackage{amsthm}%
\usepackage{mathrsfs}%
\usepackage[title]{appendix}%
\usepackage{xcolor}%
\usepackage{textcomp}%
\usepackage{manyfoot}%
\usepackage{booktabs}%
\usepackage{algorithm}%
\usepackage{algorithmicx}%
\usepackage{algpseudocode}%
\usepackage{listings}%
\usepackage{subcaption}
\usepackage{bigints}
\usepackage{hhline}

\usepackage{tabu}

\DeclareUnicodeCharacter{2212}{-}
\usepackage{geometry}
\newcommand{\dd}{\mathrm{d}}
\newcommand{\mpl}{M_\mathrm{Pl}}
\newcommand{\mplred}{M_\mathrm{Pl}}
\newcommand{\eqn}[1]{Eqn. (\ref{#1})}
\newcommand{\nn}{\nonumber \\}

\usepackage[normalem]{ulem}

\usepackage{table_macros}

\begin{document}

\title[Cosmology using numerical relativity]{Cosmology using numerical relativity}

\author*[1]{\fnm{Josu C.} \sur{Aurrekoetxea}}\email{josu.aurrekoetxea@physics.ox.ac.uk}

\author[2]{\fnm{Katy} \sur{Clough}}\email{k.clough@qmul.ac.uk}

\author[3]{\fnm{Eugene A.} \sur{Lim}}\email{eugene.lim@kcl.ac.uk}

\affil[1]{Astrophysics, Denys Wilkinson Building, University of Oxford, 
Keble Road, Oxford OX1 3RH, United Kingdom}

\affil[2]{School of Mathematical 
Sciences, Queen Mary University of London, Mile End Road, London E1 4NS, 
United Kingdom}

\affil[3]{Theoretical Particle Physics and Cosmology Group, Physics 
Department, King's College London, Strand, London WC2R 2LS, United Kingdom}


\abstract{This review is an up-to-date account of the use of numerical relativity to study dynamical, strong-gravity environments in a cosmological context. First, we provide a gentle introduction into the use of numerical relativity in solving cosmological spacetimes, aimed at both cosmologists and numerical relativists. Second, we survey the present body of work, focusing on general relativistic simulations, organised according to the cosmological history -- from cosmogenesis, through the early hot Big Bang, to the late-time evolution of universe. We discuss the present state-of-the-art, and suggest directions in which future work can be fruitfully pursued.  }

\maketitle

\tableofcontents

\section{Introduction}\label{sec:intro}

Einstein's theory of General Relativity (GR) revolutionised our understanding of cosmological evolution, and has provided the basis for modern precision cosmology. However, most studies rely heavily on the symmetry provided by large-scale spatial isotropy and homogeneity, as encapsulated by the spatially flat Robertson-Walker (RW) metric,
\begin{equation} \label{eq:RW_metric}
ds^2= -dt^2 + a^2(t) (dx^2 + dy^2 + dz^2)\,,
\end{equation}
assumming $c=1$. Where such a description holds, a preferred frame exists, and we can talk meaningfully about cosmological \emph{time} and \emph{space}, as defined by some observers aligned with the symmetry. The Hubble parameter $H(t) = \dot{a}/a$, describing the curvature of the metric in the time direction, sets the length scale of the problem at some given time $t$ (as $L \sim H^{-1}$). The symmetry in the metric relies on assumptions about both the underlying curvature and topology of the spacetime, and the distribution of matter. In particular, present-day observations \cite{Planck:2018vyg} provide direct evidence for isotropy, which combined with the Copernican principle imply that matter in the universe must be roughly spatially homogeneously distributed, with no $O(1)$ overdensities on Hubble scales\footnote{See however, \cite{Maartens:2011yx,Clarkson:2012bg} for caveats on inferring the homogeneity of the universe via observations along the lightcone.}.
Since, at the largest scales at least, this description does appear to correspond well to the current observable universe, most cosmological simulations can be accurately approximated by splitting the \emph{background} evolution of the scale factor $a(t)$ from the dynamics of the matter fields. 
For horizon-scale modes where the density is low and the fields are linear, a perturbative approach can be used, which is the usual approach for inflationary spacetimes (see e.g. \cite{Baumann:2022mni}).
For low density but highly non-linear field dynamics at the Hubble scale, one can numerically evolve the matter equations of motion on an expanding RW background, using the spatial average of the matter density to drive the evolution of the scale factor for consistency.
If the perturbations are large, but sub-horizon and non-relativistic (i.e. local backreaction matters), then the dynamics can be approximated using Newtonian gravity on an expanding RW background, a strategy that is the leading approach in large-scale N-body simulations of late-time cosmological evolution (see e.g. \cite{Springel:2005mi,Teyssier:2001cp}).

While numerical relativity (NR) is a standard tool in the computation of accurate gravitational waveforms in compact object astrophysics (i.e. black holes (BHs) and neutron stars (NSs)), its application to cosmology is not yet widespread. Nevertheless, numerical investigations in cosmological spacetimes have a long history, and some of the earliest works, mostly focussed on 1D planar cosmologies, deserve to be highlighted as pioneering work in the field \cite{Centrella1987,Matzner1982,CentrellaWilson1983,CentrellaWilson1984,Anninos1989,Kurki-Suonio:1987mrt,Shinkai1993,Shinkai1994}. In particular, works by Centrella, Matzner and their collaborators explored a wide range of topics including the collision of gravitational waves \cite{CentrellaMatzner1982}, planar shocks \cite{Centrella1980} and nucleosynthesis \cite{CentrellaMatzner1986}. More recent works that implement improvements in gauge choices, initial conditions and stability have built on these foundations to push forward our understanding of the universe. Cosmologists are now beginning to harness the considerable power of NR to explore strong gravity problems that are beyond the reach of standard perturbative treatments. 

\begin{figure}[t]
\centering
\includegraphics[width=\textwidth]{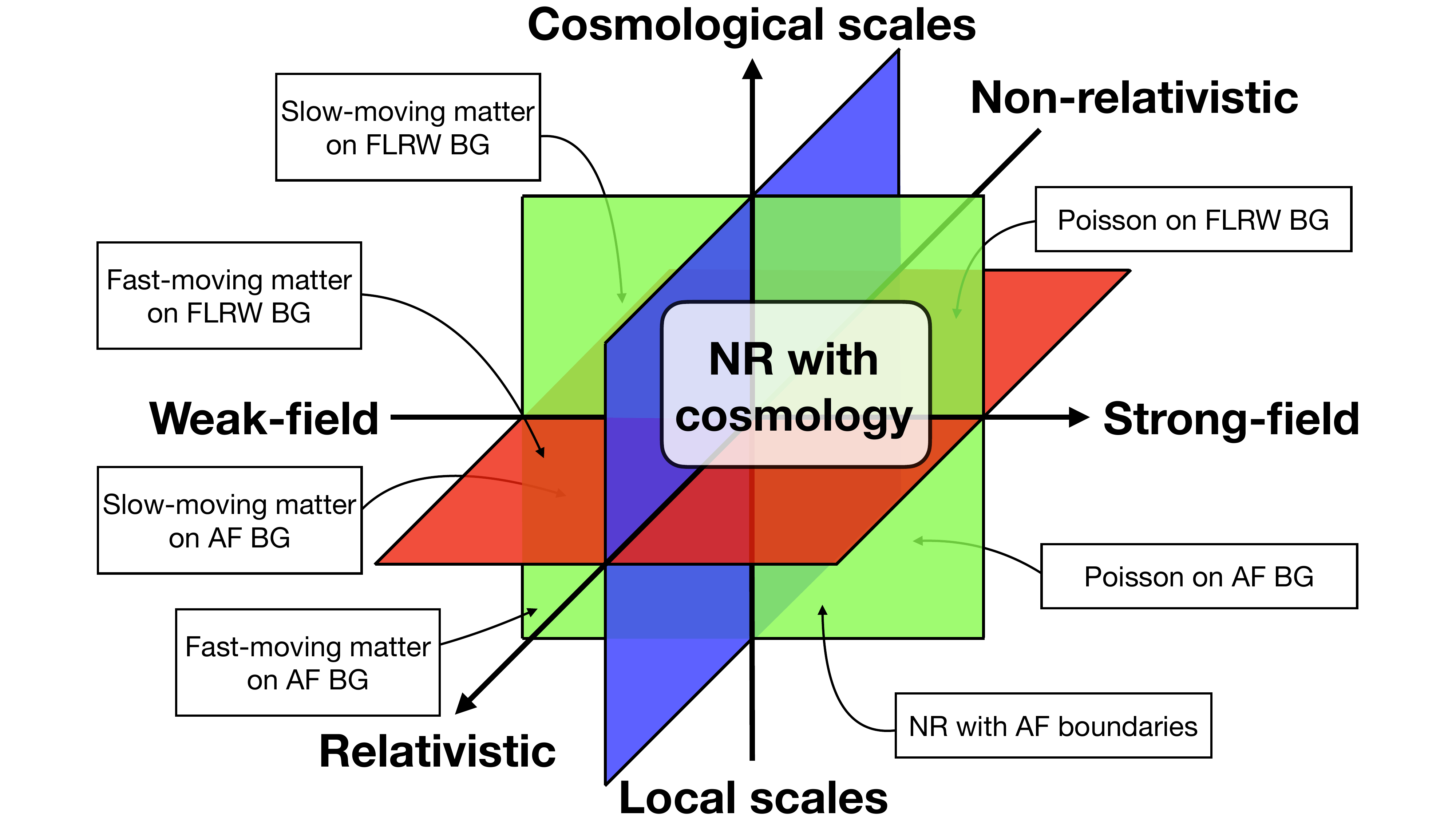}
\vspace{5pt}
\caption{In cases in which overdensities become large $\delta\rho/\rho_\mathrm{average} \sim 1$ on the length scale set by Hubble $L \sim H^{-1}$, and in the absence of any obvious symmetries, NR methods are required to make progress. BG and AF stand for \emph{background} and \emph{asymptotically flat}, respectively. }\label{fig:whenNR}
\end{figure}

Given the success of models relying on background RW symmetries, and the additional cost and complexity involved in NR, one should question the need to go beyond them to fully general relativistic simulations with no preferred frame. In general, three conditions should be met for an NR simulation to be justified:
\begin{enumerate}
    \item Local backreaction is large with relativistic matter exhibiting strong inhomogeneities in the density $\delta\rho/\rho_\mathrm{average} \sim 1$.
    \item The inhomogeneities are coherent on Hubble scales $L \sim H^{-1}$ or over Hubble times $T \sim H^{-1}$.
    \item Spacetime is dynamical and absent of symmetries that could simplify the description.
\end{enumerate}
The relation of cosmological NR simulations to other regimes where approximations can be made is illustrated in Fig \ref{fig:whenNR}.

One important but less obvious example is work that seeks to rule out the accumulation of secular or non-linear effects that may be lost when adopting a perturbative description. An example of this is work to address concerns about so-called \emph{backreaction} effects being unaccounted for in cosmological evolutions. As we will describe later in this review, these seem largely unfounded, with effects at percentage levels when studied systematically in full GR. Nevertheless, the simulations helped to advance discussions about how to correctly compare the non-linear and perturbative results, and the impact on next generation surveys and their systematics. 
Other examples of problems requiring NR that we will see in this review include the study of cosmological singularities, mechanisms like inflation and slow contraction that aim to act as smoothers of initially inhomogeneous spacetimes, and non-perturbative dynamics in reheating or primordial black hole (PBH) formation.

In this review, we focus on fully general relativistic simulations without approximations (i.e. not assuming a certain background, conformal flatness or perturbative or post-Friedmann description). This technically excludes some interesting and useful intermediate approaches between the non-linear and linear regimes that have been suggested in both NR and cosmology, e.g. \cite{Laguna:1999tk,Milillo:2015cva, Bonazzola:2003dm,Adamek:2015eda,Adamek:2020jmr,Barrera-Hinojosa:2019mzo, Barrera-Hinojosa:2020arz}, that we nevertheless briefly highlight where relevant. We provide useful background in Sec. \ref{sec:background}. Recognising that most cosmologists are not familiar with the methods of numerical relativists, and \emph{vice versa}, we start by providing two reviews of the topic from different angles -- first we explain cosmology to numerical relativists, and then we explain NR to cosmologists. We then review common technical and conceptual challenges of using NR in cosmological applications, which are themselves open topics of research. The resolution of some of these challenges may lead to new techniques that can be used in other applications of NR, and also to deeper insights about the nature of the cosmological spacetime in which we exist.

\begin{figure}[t]%
\centering
\includegraphics[width=\textwidth]{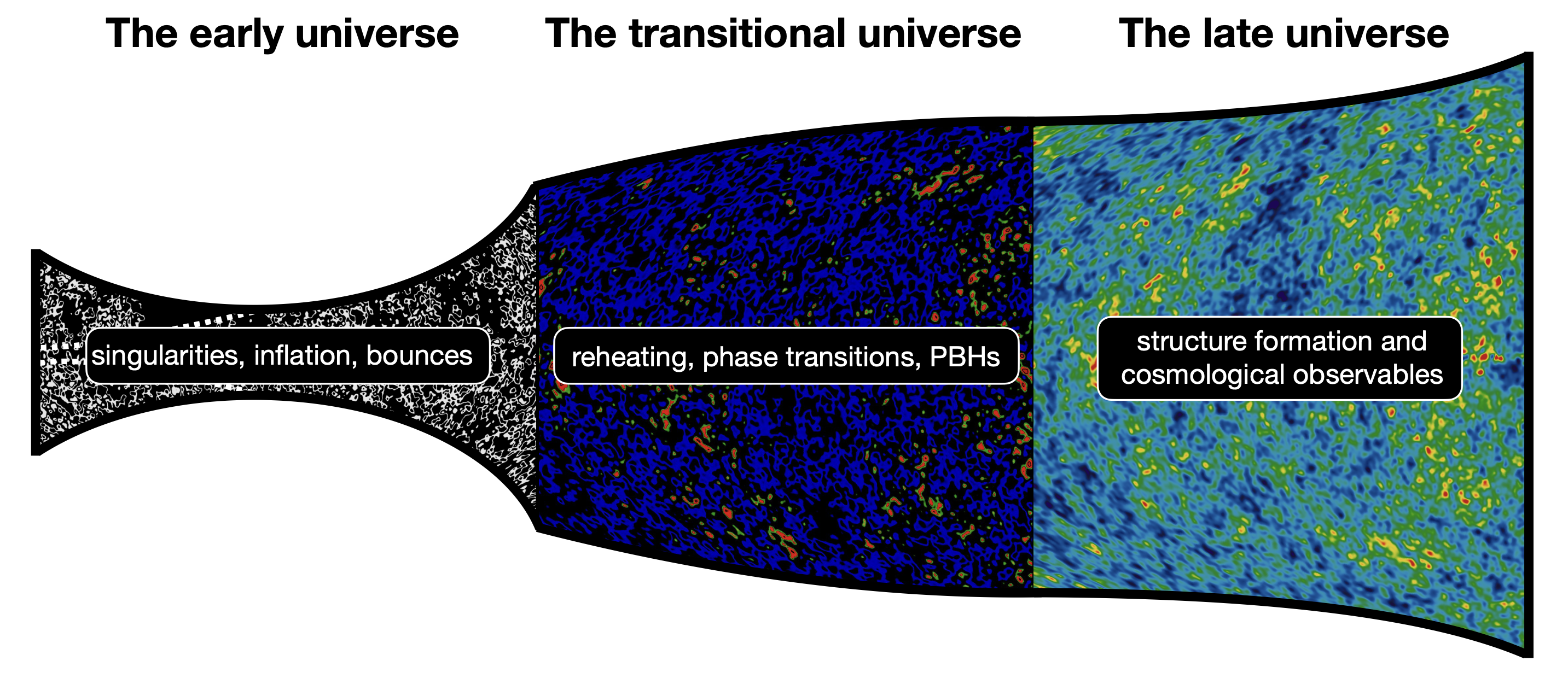}
\caption{We divide our review into work that focuses on the pre-Big Bang phase, which covers the period up to the end of inflation on this diagram. The post-Big Bang phase covers non-perturbative dynamics from the end of inflation to the emission of the CMB. The late-universe phase is the remainder of the diagram, which contains the standard cosmological history.}\label{fig:timeline}
\end{figure}

Next we provide a review of past work, divided into three broad sections in order of time, as illustrated in Fig. \ref{fig:timeline}. The first, Sec. \ref{sec:early_universe}, covers the most distant past of the universe, and questions about how it may have begun, including work on cosmological singularities, and attempts to reconcile our homogeneous universe with an inhomogeneous beginning. The second, Sec. \ref{sec:mid_universe}, describes strong gravity phenomena that may have occurred when our universe was already approximately homogeneous and isotropic on large scales, such as the non-perturbative dynamics in preheating and PBH formation. Finally Sec. \ref{sec:late_universe} covers the late universe, and in particular quantification of the potential effects of full GR on results from the approximate descriptions that are used in most large-scale cosmological simulations. We close with a summary table of the work done to date and a list of open directions.

We would highlight several excellent reviews on related topics, which complement the overview provided here, including \cite{Anninos:2001yc,Berger:2002st,Andersson:2006vk,Cardoso:2012qm,Lehner:2014asa,Choptuik:2015mma,Garfinkle:2016lcu,Coley:2017mna,Ijjas:2022qsv}.

\section{Background on numerical relativity and cosmology}\label{sec:background}

We start by providing sufficient background of both NR and cosmology to frame the later discussions on specific works. This is by no means a comprehensive description of either NR or cosmology, but aims to highlight the key features of each field that are relevant in the context of the other. With this background in place, we will discuss the key technical and conceptual issues associated with cosmological spacetimes in strong gravity regimes, namely boundary conditions, initial data, gauge choice, interpretation and units.

\subsection{Cosmology for numerical relativists} \label{sect:cosmo_for_NR}

We provide here only a brief description of cosmology in the context of NR. A good starting point for those new to cosmology are textbooks by Baumann \cite{Baumann:2022mni}, Weinberg \cite{Weinberg:2008zzc}, Mukhanov \cite{Mukhanov:2005sc}, Ryden \cite{Ryden_2016}, Peacock \cite{Peacock_1998}, Ellis, Maartens \& MacCallum \cite{Ellis_Maartens_MacCallum_2012} and Huterer \cite{Huterer_2023}.

Broadly defined, cosmology is the study of the evolution of the large-scale structure of the universe.  Before the 80s, this usually focussed on the study of the time evolution the scale factor $a(t)$ of the homogeneous Robertson-Walker (RW) metric in \eqn{eq:RW_metric}, with the key physical quantity
\begin{equation}
    H(t) \equiv \frac{\dot{a}}{a}~,\label{eqn:hubble_constant}
\end{equation}
which is clearly a variable that is a function of time but is often referred to as the \emph{Hubble parameter}, or when discussing its present day value the \emph{Hubble constant}.  Currently, there exists an observational \emph{tension} in measurements of its exact present day value $H_0$ (read as \emph{H-nought}) -- measurement of $H_0$ from observations of the Cosmic Microwave Background (CMB) and the clustering of galaxies suggest that $H_0 =67.97\pm 0.38~\mathrm{km}\mathrm{s}^{-1}~\mathrm{Mpc}^{-1}$  \cite{Planck:2018vyg,DESI:2024mwx}, while observations of astrophysical standard candles suggest $H_0 = 73.30 \pm 1.04~\mathrm{km}\mathrm{s}^{-1}~\mathrm{Mpc}^{-1}$ \cite{Riess:2021jrx,Freedman:2020dne}. Importantly, cosmological measurements are not direct measurements of the present day value but \emph{inferred} values from measurements at a different epoch, assuming a certain subsequent cosmological evolution. So the tension potentially provides evidence that the standard $\Lambda$CDM model (dark energy $\Lambda$ plus cold dark matter) is incomplete.

Inserting the RW metric into the Einstein equations provides the evolution of the single metric degree of freedom $a$ as a pair of equations known as the Friedmann-Lema\^{i}tre equations for a spatially flat universe
\begin{equation}
    \left(\frac{\dot{a}}{a}\right)^2 = \frac{8\pi G}{3}\rho~,\qquad \frac{\ddot{a}}{a} = -\frac{4\pi G}{3}(\rho +3p)~,\label{eqn:Friedman}
\end{equation}
where $\rho$ and $p$ are the energy density and pressure of the matter in the universe. Consistency with the metric symmetries constrains the matter to be a perfect fluid
\begin{equation}
    T_{\mu\nu} = (\rho+p)U_{\mu}U_{\nu} + Pg_{\mu\nu}~, \label{eqn:perfectfluid}
\end{equation}
with $U_{\mu}=(1,0,0,0)$ the local 4-velocity of a co-moving fluid parcel. The first equation of \eqn{eqn:Friedman} can be recognised as the Hamiltonian constraint, while the second equation describes the evolution of $H = \dot{a}/a$, which is the Raychaudhuri expansion equation in the FLRW limit. As we will see in the next section, this is equivalent to the evolution equation for the trace of the extrinsic curvature $K=-3H$ in the ADM formalism.

Conservation of \eqn{eqn:perfectfluid} leads to a continuity equation for the matter
\begin{equation}
    \dot{\rho} +3 \frac{\dot{a}}{a}(p+\rho)=0~.\label{eqn:continuity}
\end{equation}
Note that \eqn{eqn:continuity} does not close the system of equations; to do so we must impose an \emph{equation of state} relating $p$ and $\rho$
\begin{equation}
p=w\rho~,\label{eqn:eos}
\end{equation}
where $w$  is known as the \emph{equation of state parameter}, or simply \emph{double-u}\footnote{This parameter can in principle also depend on time, although in this case the equations do not close unless some other ansatz is made, for example assuming a dependence on the scale factor like that currently used to parameterise an evolving dark energy $w(a)=w_0+w_a(1-a)$, which was first introduced in \cite{Linder:2002et}.}. This parameterisation, whilst only an approximation for most actual fluids, allows cosmologists to study the impact of different components of the matter content on the cosmological evolution, i.e. $w=0$ for pressureless matter, $w=1/3$ for radiation and $w=-1$ for the cosmological constant. Where multiple matter types are present they are usually assumed independent and must be evolved separately, with their densities and pressures added to obtain the sources in \eqn{eqn:Friedman}.

On the largest observable scales, spacetime appears to be well-modelled by this \emph{FLRW} system of equations with a particular mix of matter types described by the $\Lambda$CDM model. Nevertheless, the universe possesses inhomogeneous structure such as clusters of galaxies, in which things like stars and physicists can emerge. These matter inhomogeneities are modelled by perturbations of the RW metric\footnote{Numerical relativists might notice that the perturbed RW metric \eqn{eqn:perturbed_RW} is the ADM-decomposition with different variable names. Historically this decomposition was the starting point for the development of the ADM form \cite{Bardeen:1980kt}. While in linear theory, the weakly hyperbolic nature of the ADM formalism is not a problem, this is not necessarily the case in fully non-linear GR \cite{Ijjas:2018cdm}.}
\begin{equation}
    ds^2  = -(1+2A)dt^2 + 2a(t)B_idx^idt + a^2(t)(\gamma_{ij}+E_{ij})dx^idx^j\,, \label{eqn:perturbed_RW}
\end{equation}
where $h_{00} = -2A$, $h_{0i}=aB_i$ and $h_{ij}= a^2E_{ij}$ are \emph{linear} perturbations $h_{\mu\nu}$ over the \emph{background} RW metric $g^{(0)}_{\mu\nu}$ given by \eqn{eq:RW_metric}, such that the true \emph{physical metric} is
\begin{equation}
    g_{\mu\nu} = g_{\mu\nu}^{(0)} + h_{\mu\nu}~, \label{eqn:perturbation_bg}
\end{equation}
and $\vert h_{\mu\nu}\vert \ll g_{\mu\nu}^{(0)}$.
Finally, the 3D spatial metric $\gamma_{ij}$ is maximally symmetric. Whilst one can include a spatially constant curvature, for simplicity we will consider only foliations that are spatially flat $\gamma_{ij} =\delta_{ij}$ for the remainder of this section. 

Cosmological perturbation theory (the study of the structure and evolution of cosmological perturbations) has yielded a remarkable amount of information about the universe, heralding the era of \emph{Precision Cosmology} (see Turner \cite{Turner:2022gvw} for a historical overview). The study is statistical in nature -- e.g. it is not important where individual galaxies are, but it is important how the separations of galaxies are statistically correlated with each other. The key observables of structure are correlation functions, $\langle X_1X_2\dots \rangle$ where $X_i$ are cosmological observables such as density perturbations or polarisation amplitudes, and the angle brackets denote an average over spatial hyperslicings. The most important quantity is the \emph{power spectrum} or two-point correlation function -- for example the matter power spectrum, which is the two-point correlation of the density perturbation $\langle \delta \delta \rangle$ where $\delta =\delta \rho/\rho$. As long as the perturbations remain small, in principle correlation functions at all orders  can be calculated  via perturbation theory (see for example \cite{Maldacena:2002vr,Weinberg:2005vy}). 

The evolution of the metric perturbations are given by linearising the Einstein equations as
\begin{equation}
G^{(0)}_{\mu\nu} + \delta G_{\mu\nu} =8\pi G (T^{(0)}_{\mu\nu} +  \delta T_{\mu\nu})~, \label{eqn:perturbed_EE}
\end{equation}
We first solve the background evolution $G^{(0)}_{\mu\nu} = 8\pi GT^{(0)}_{\mu\nu}$ to obtain the zeroth order terms for both the metric and the matter sector, and then plug those into the linearised Einstein equations $\delta G_{\mu\nu} = 8\pi G\delta T_{\mu\nu}$, which describes the inhomogeneous dynamics of the spacetime. We similarly linearise the matter equations of motion, which combined with \eqn{eqn:perturbed_EE}, will close the system of equations consistently. For the details of this (long but straightforward) calculation in the case of a cosmological background, see for example \cite{Baumann:2022mni}, and also the reviews of \cite{Tsagas:2007yx,Malik:2008im} for other approaches, and going beyond the linear regime.

We now clarify some of the nomenclature surrounding cosmological gauges, which can be confusing to non-cosmologists approaching it from an NR perspective. In GR, \emph{gauge freedom} refers to the freedom to choose coordinates, such that any physical measurement should remain invariant under a coordinate transform.  In general, this transformation can be non-linear, and applies to the whole spacetime metric and its slicing. In linear perturbation theory on a fixed background, this freedom is translated into the statement that any two perturbations $h_{\mu\nu}$ and $h'_{\mu\nu}$ are physically the same as long as they are related to each other by a \emph{gauge transformation} $\Delta h_{\mu\nu}$, which is a function of the background metric $g^{(0)}_{\mu\nu}$ and a gauge choice\footnote{In this discussion, we have chosen to follow the treatment of standard cosmology texts, which usually focuses on the more explicit coordinate-centric approach. There is an elegant geometrical picture to describe this equivalence, which is perhaps more familiar to the study of gravitational radiation as follows. One posits the existence of two spacetime manifolds which are related by a diffeomorphism $\phi : {\cal M}_b \rightarrow {\cal M}_p$, where ${\cal M}_b$  is the ``background spacetime'' and ${\cal M}_p$ is the ``physical spacetime'', with different tensor fields living on them. On ${\cal M}_b$ we define the ``background'' metric $g_{\mu\nu}^{(0)}$, and while on ${\cal M}_p$, we define the ``physical'' metric $g_{\mu\nu}$.  One can then treat linearized theory as the behaviour of tensor field perturbations $h_{\mu\nu}$ on ${\cal M}_b$, where $h_{\mu\nu}$ is the difference between the pull-back metric $\phi_*(g)_{\mu\nu}$ from ${\cal M}_p$ and the background spacetime metric, i.e. $h_{\mu\nu} =\phi_*(g)_{\mu\nu}-g_{\mu\nu}^{(0)}$ (notice the placement of the indices outside the bracket in the pullback of $g$).  In this picture, the gauge freedom at first order is generated by the choice of vector field $\epsilon^{\mu}$ on ${\cal M}_b$, and hence $h_{\mu\nu}$ in different gauges are physically equivalent. For gauge transformations at higher orders, please see \cite{Bruni:1996im,Sonego:1997np}. }. Specifically, suppose we have a \emph{linear} coordinate transform\footnote{Note that while the coordinate $x^{\mu}$ is not a spacetime vector, the infinitisimal transform $\epsilon^{\mu}$ is and hence its indices can be raised and lowered as usual.}
\begin{equation}
x^{\mu}\rightarrow x^{\mu} + \epsilon^{\mu}(x)~.\label{eqn:linear_transform}
\end{equation}
We can attribute this change only to the perturbation $h_{\mu\nu}(x)$, leaving the background $g^{(0)}_{\mu\nu}(x)$ unchanged. Here, the Greek indices refer to the coordinate basis of the $x$ coordinate system (see e.g. \cite{Weinberg:2008zzc}). One can think of $h_{\mu\nu}$ and $\delta T_{\mu\nu}$  as fields living on the background manifold with geometry $g^{(0)}_{\mu\nu}$.
Under the linear transform \eqn{eqn:linear_transform}, the metric perturbations transform as $h_{\mu\nu} \rightarrow h_{\mu\nu}(x) + \Delta h_{\mu\nu}(x)$ and the perturbations of the stress tensor as $\delta T_{\mu\nu} \rightarrow \delta T_{\mu\nu} + \Delta (\delta T_{\mu\nu})$ where
\begin{eqnarray}
\Delta h_{\mu\nu} &=& -g^{(0)}_{\sigma \mu}\partial_{\nu}\epsilon^{\sigma} -  g^{(0)}_{\sigma \nu}\partial_{\mu}\epsilon^{\sigma} - \partial_\sigma g^{(0)}_{\mu\nu}\epsilon^\sigma~, \label{eqn:gauge_transform_CPT} \\
    \Delta (\delta T_{\mu\nu}) &=& - T^{(0)}_{\sigma\mu}\partial_\nu\epsilon^{\sigma} - T^{(0)}_{\sigma\nu}\partial_\mu\epsilon^{\sigma} - \partial_{\sigma}T^{(0)}_{\mu\nu}\epsilon^{\sigma}~.
\end{eqnarray}
All perturbations that are related by this \emph{gauge transform} are physically equivalent. Thus there exists a \emph{redundancy} in the description of the perturbations that one can either remove by fixing a gauge (that is, choosing a specific $\epsilon^{\mu}$), or by choosing to work with \emph{gauge-invariant} variables \cite{Bardeen:1980kt,Mukhanov:1988jd,Mukhanov:1990me}.

To see how this is done, we note that since $h_{\mu\nu}$ are symmetric fields living on the maximally symmetric 3D spatial hyperslicings, we can decompose the $h_{i0} =aB_i$ and $h_{ij}= a^2E_{ij}$ components as follows
\begin{eqnarray}
    B_i &=& \partial_i B +\hat{B}_i~,\qquad \partial_i \hat{B}^i =0\nn
    E_{ij} &=&  2C\delta_{ij} + 2\partial_{(i}E_{j)} + \left(\partial_i\partial_j - \frac{1}{3}\delta_{ij}\nabla^2\right)E + \hat{E}_{ij}~, \nn
    \partial_i E^i&=&\partial_i \hat{E}^{ij}=\mathrm{Tr} \hat{E}_{ij}=0~, \qquad C=\mathrm{Tr} E_{ij}/3 ~. \label{eqn:SVTdecomp}
\end{eqnarray}
Here $A$, $B$, $C$ and $E$ in \eqn{eqn:SVTdecomp} are known as \emph{scalar-type} perturbations, $\hat{B}_i$ and $E_i$ are known as \emph{vector-type} perturbations and $\hat{E}_{ij}$ are known as \emph{tensor-type} perturbations.  There is a similar decomposition for $\delta T_{\mu\nu}$ (we refer the reader to \cite{Baumann:2022mni} for details).  This is called the \emph{Bardeen decomposition} or the \emph{S-V-T} decomposition \cite{Gerlach:1979rw,Bardeen:1980kt}. The power of this decomposition is that the scalar-type, vector-type and the tensor-type perturbations can be represented by spin-0, spin-1 and spin-2 eigenmodes of the Laplace-Beltrami operator respectively on the (background) spatial hyperslicing with $SO(3)$ symmetry, and hence are orthogonal to each other. Operationally, this means that they are decoupled from each other at the linear level -- one can consider the evolution of scalar, vector and tensor perturbations independently. Similarly, we can decompose $\epsilon_{\mu} = g^{(0)}_{\mu\nu}\epsilon^{\nu}$ into the time component $\epsilon_0$ and the space component $\epsilon_i = \partial_i \epsilon + \hat\epsilon_i$ where $\epsilon$ and $\epsilon_0$ are scalar-type variables, and the traverse $\hat\epsilon_i$ is a vector-type variable.  A choice of $\epsilon_0$, $\epsilon$ and $\hat\epsilon_i$ then fixes the gauge.  Using these variables, the transformations in \eqn{eqn:gauge_transform_CPT} are 
\begin{eqnarray}
    A \rightarrow  A + \partial_t \epsilon_0~, \quad
    C \rightarrow  C + 2H\epsilon_0~,\quad
    E \rightarrow  E -\epsilon ~,\quad
    aB \rightarrow  aB -\epsilon_0 - \partial_t\epsilon+2H\epsilon,
\end{eqnarray}
for the scalar-type variables,
\begin{eqnarray}
    a\hat{B}_i \rightarrow a\hat{B}_i - \partial_t\hat{\epsilon}_i+ 2H\hat{\epsilon}_i~,\quad
    a^2E_i \rightarrow a^2E_i - \hat{\epsilon}_i~,
\end{eqnarray}
for the vector-type variables and 
\begin{equation}
E_{ij} \rightarrow E_{ij}~,
\end{equation}
for the tensor-type variables. It is important to note that the labels \emph{scalar}, \emph{vector} and \emph{tensor} do not mean they are rank-0, 1 and 2 spacetime tensors. The scalar and vector-type perturbations \emph{are gauge-dependent} despite the lack of coordinate basis indices on the former. Meanwhile, the tensor-type perturbations are manifestly gauge-invariant -- the gauge transform variable $\epsilon^{\mu}$ does not contain a \emph{tensor-type} component. This nomenclature regarding gauge is standard in cosmology, but at odds with the NR perspective that we will discuss in Sec. \ref{sect:NR_for_cosmo}, where, for example, a scalar implies a truly coordinate independent quantity.

The two standard gauge choices commonly used in cosmological perturbation theory are the \emph{Newtonian gauge} and the \emph{synchronous gauge}. In the former, we can choose $\epsilon_0$ and $\epsilon$ to set two of the four scalar perturbations to zero
\begin{equation}
    B=0~,\qquad E=0~,
\end{equation}
while in the latter, we set 
\begin{equation}
    A=0~,\qquad B=0~.
\end{equation}
Similarly, we can choose $\hat{\epsilon}_i$ to set the gauge for the vectors -- usually $\hat{B}_i$ is set to zero. 
As noted above, the alternative is to work with \emph{gauge-invariant variables} -- these are linear combinations of the perturbations such that the gauge transformations cancel
\begin{equation}\label{eq:bardeen}
    \Psi = \frac{1}{a^2}\left(A+\partial_t(B-\dot{E})\right),\qquad \Phi = -C + \frac{1}{3}\gamma^{ij}\partial_i\partial_j E - \frac{H}{a^2}(B-\dot{E}),\qquad \Psi_i = \frac{1}{a}(\hat{B}_i-\dot{E}_i).
\end{equation}
Here $\Psi$, $\Phi$ and $\Psi_i$ are called \emph{Bardeen variables}, and are invariant under (linear cosmological) gauge transformations.

\begin{figure}
\centering
\includegraphics[width=\textwidth]{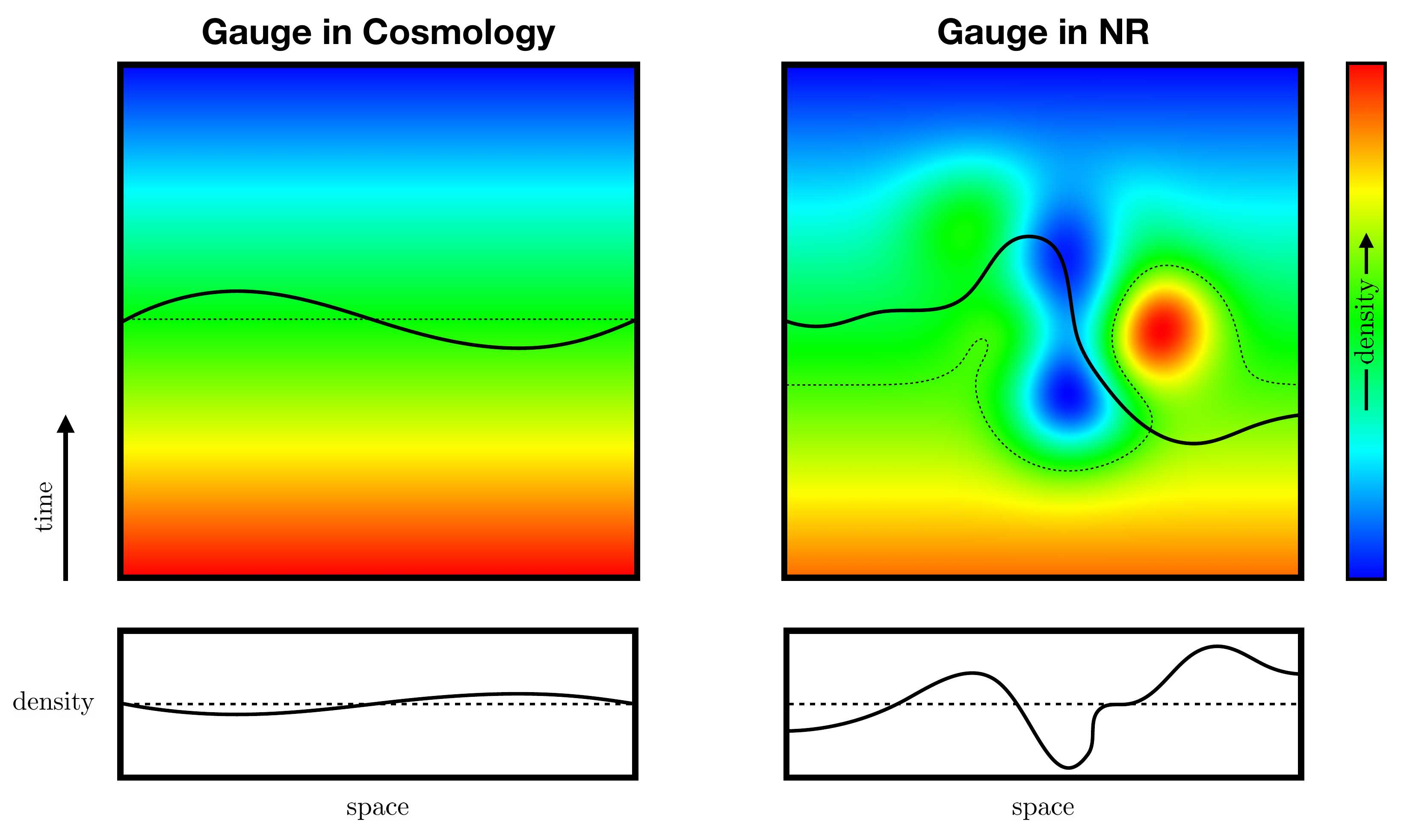}
\caption{Gauge in cosmology versus NR. (Left:) For cosmologists a change in \emph{gauge} implies a small perturbation to the slicing, but with the background FLRW slice about which the perturbation is defined always aligned with the symmetry of the spacetime. This can transform a constant density slice into one in which small perturbations are observed. (Right:) In the NR case, a change of gauge is a fundamentally different slicing of the spacetime (conventionally parameterised by a choice of the lapse and shift), which therefore potentially results in a (non-perturbative) change in the density observed on each constant-time slice. It is important to note that when there is no underlying symmetry, the notion of a preferred slicing is lost, and thus there is no choice of gauge in which we are able to recover FLRW-like observers.}
\label{fig-gaugecosmoNR}
\end{figure}

To summarise, in cosmological perturbation theory, the background metric $g^{(0)}_{\mu\nu}$ determines both the foliation and the coordinates on each hyperslice with normal $n^{\mu} = \partial_t$ -- the choice is already made here\footnote{One can of course foliate the RW spacetime with \emph{inhomogeneous} hyperslicings where the symmetry is not made manifest, but this would require a change of coordinates which mix up $t$ and $x^i$.}. Gauge freedom is restricted to linear transforms on the perturbation fields $h_{\mu\nu}$ and the corresponding matter stress tensor $\delta T_{\mu\nu}$, which manifests as a redundancy in the description of the perturbation variables.  Contrast this to the fully non-linear case in an Arnowitt-Deser-Misner \cite{Arnowitt:1962hi} (ADM) decomposition, as used in NR, where \emph{gauge-fixing} refers to the choice of the lapse $\alpha$ and shift $\beta^i$ functions, and fixes the temporal foliation and the spatial coordinates as described in Sec. \ref{sect:NR_for_cosmo} and \ref{sect:gauge_choice}. The difference is illustrated in Fig. \ref{fig-gaugecosmoNR}. 

In summary, when cosmologists describe scalar, vector and tensor perturbations, they refer to the invariant transformation properties of \emph{spatial} tensors and not spacetime tensors \cite{Bardeen:1980kt}. Thus as a consequence scalar perturbations are gauge dependent and tensors are gauge invariant under linear spacetime coordinate transformations.

This standard interpretation is problematic when we want to compare results between perturbation theory and NR, and in fact there is no obvious mapping between the two pictures once perturbations become large relative to the background. One must reformulate the question in an observer dependent way -- in terms of what particular geodesic observers moving through each spacetime would measure. This picture is non-trivial to reconstruct, and so less rigorous diagnostics like spatial averages are often used as a proxy, as will be discussed in Sec \ref{sec:interpretation}.

\subsection{Numerical relativity for cosmologists} \label{sect:NR_for_cosmo}

As for cosmology, we provide here only a brief description of NR in the context of the work that we will review. A good starting point for those new to NR is the student-friendly text by Baumgarte \& Shapiro \cite{Baumgarte:2021skc} and for those looking to understand the concepts here in more depth the textbooks by Alcubierre \cite{Alcubierre:2008co},  Baumgarte \& Shapiro \cite{Baumgarte:2010ndz}, Gourgoulhon \cite{Gourgoulhon:2007ue}, and Shibata \cite{Shibata_book}. 

The basic principle of NR is to solve the coupled matter-metric system of GR as a Cauchy or initial value problem, whereby we specify initial data for the matter and the metric on some space-like hypersurface, and then evolve it forward in \emph{time} (defined according to some chosen observers); that is, we specify the initial scenario, and then use the computer to find out what happens next. Such a formulation is naturally provided by the ADM decomposition of the Einstein equations, which was developed in the 1960s by Arnowitt, Deser, and Misner \cite{Arnowitt:1962hi} as a Hamiltonian formulation of GR.

A standard question when confronted with any formulation of PDEs is whether they are \emph{well-posed}, which is the demand that a unique solution exists that depends continuously on the initial data.  In the perturbative cosmology of the previous Sec. \ref{sect:cosmo_for_NR}, one separates the background evolution and the perturbation evolution equations, both of which are well-posed. In fully non-linear NR, the well-posedness of the formulation is less obvious and problems with well-posedness in NR have a long history. In particular, one of the early obstacles in NR was that the need for a well posed system of equations was not sufficiently appreciated, and since the ADM formulation \cite{Arnowitt:1962hi} was ill-posed, its direct numerical implementation was unstable. This problem can be surmounted by reformulating the evolution equations into one of several possible (locally) well-posed forms. For a detailed introduction to well-posedness in NR, see the review article by Hilditch \cite{Hilditch:2013sba}. One popular reformulation of the ADM system is the so-called Baumgarte-Shapiro-Shibata-Nakamura (BSSN) formalism \cite{Shibata:1995we,Baumgarte:1998te} based also on an earlier work with Nakamura, Oohara and Kojima \cite{Nakamura:1987zz}. An alternative formulation is Generalised Harmonic Coordinates (GHC), which was introduced by Friedrich \cite{Friedrich} in 1985 and based on the harmonic coordinates that had been shown to be well posed by Choquet-Bruhat already in 1952 \cite{Foures-Bruhat:1952grw}. GHC is well-known for its use in successfully evolving binary BH mergers for the first time by Pretorius \cite{Pretorius:2005gq}, but was applied to studies of the cosmological singularity before this by Garfinkle \cite{Garfinkle:2001ni}.  Another alternative is the tetrad formulation developed by Uggla \emph{et al.} \cite{Uggla:2003fp}, used by Garfinkle for the study of cosmological singularities \cite{Garfinkle:2007rv} and recent works on bouncing cosmologies (see Sec. \ref{sec:bounces}), which allows scaling out of the cosmological scale $H$. This helps the numerics to follow longer timescales during which $H$ is changing over a wide range of scales, as at the end of inflation or contraction towards a singularity, but necessitates the solution of an elliptic equation for the lapse at each step. There are some other early works that formulate hyperbolic equations with tetrads \cite{Estabrook:1996wa,Buchman:2003sq}, which were mostly abandoned when coordinate basis formulations achieved success with binary BH mergers, but may be worth revisiting to address some of the challenges of cosmology.

In this section we will mainly discuss NR in the context of the ADM formalism, focussing on the broad principles, which are similar for most of the more well posed formulations. However, anyone hoping to use NR codes in practise will need to familiarise themselves with one of the above formulations -- to keep this review focussed we leave a discussion of these to the NR textbooks mentioned above. 
The idea of slicing up spacetime into space-like hypersurfaces and then evolving them is familiar to most cosmologists, where the RW metric imposes a fixed foliation and coordinates upon each hyperslice as discussed in Sec. \ref{sect:cosmo_for_NR}. In NR, the choice of foliation and the coordinates imposed on it are usually \emph{dynamical} -- they respond to the local evolution of the matter and metric. This is a necessity for a well-posed evolution, but can also help with avoiding singularities that form and tracking changing scales over time on a finite numerical grid.

\begin{figure}
\centering
\includegraphics[width=0.8\textwidth]{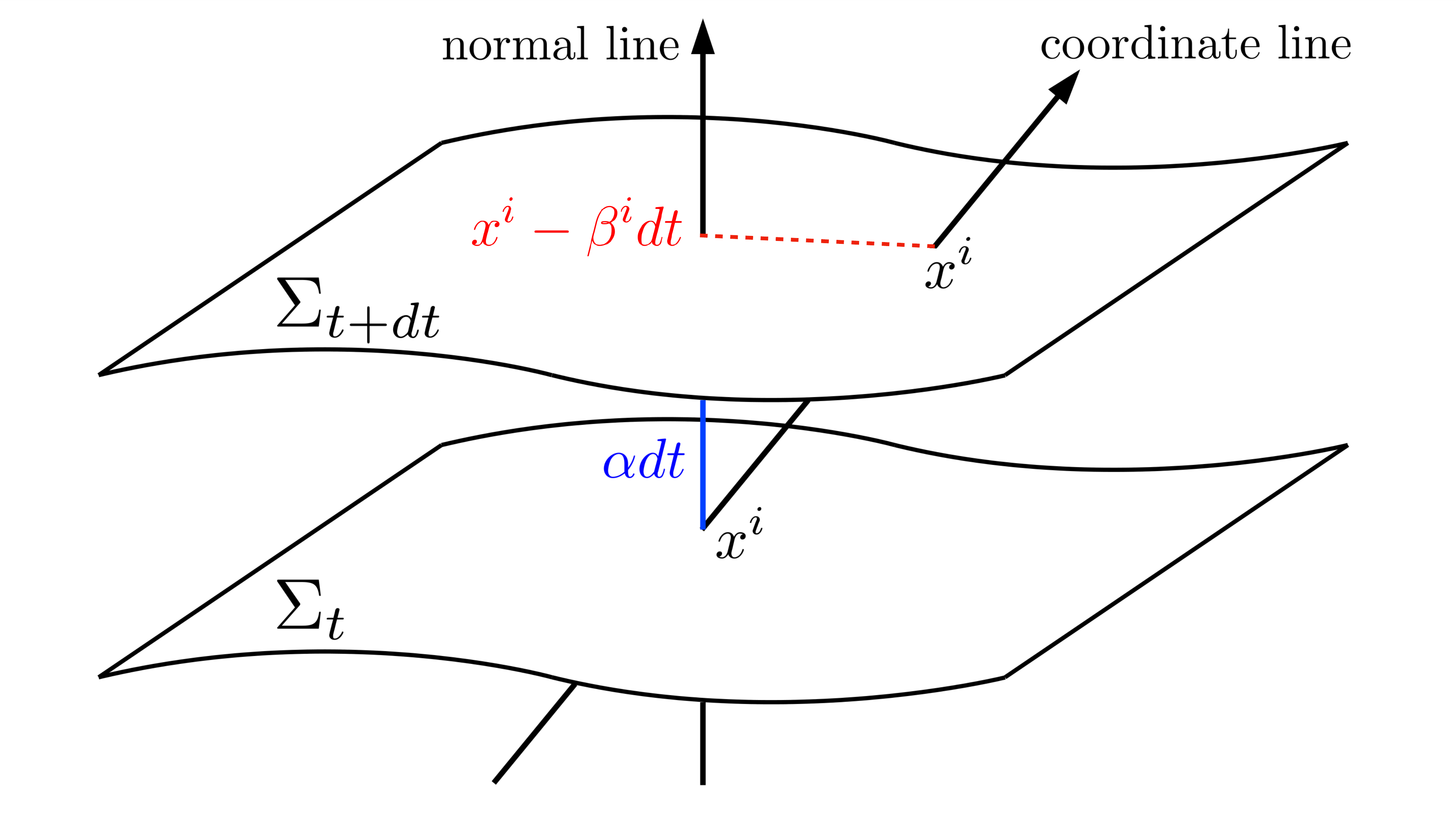}
\caption[$3+1$D Foliation of spacetime]{Foliation of a 4D spacetime into a 3D \emph{spatial} hyperslice, and a \emph{time-like} normal to that slice. The gauge variables -- the lapse $\alpha$ and shift $\beta^i$ -- are also illustrated. In this picture space is represented as a two dimensional surface, whereas in full GR each spatial slice is a 3D volume.}
\label{fig-Foliation}
\end{figure}

We now briefly review the ADM formalism \cite{Arnowitt:1962hi} and the way in which it is adapted for NR. The ADM formalism projects the 4D metric $g_{\mu\nu}$ into a foliation of spatial hypersurfaces with metric $\gamma_{ij}$ as in Fig. \ref{fig-Foliation}, with surfaces labelled by a constant time coordinate $t$. The normal vector to the slices is $n^\mu = (1/\alpha, \beta^i/\alpha)$. The coordinates $x^i$  runs over the three spatial directions, and the metric expressed in a coordinate basis adapted to the slicing is
\begin{equation} \label{eq:ADM_metric}
g_{\mu\nu} dx^\mu dx^\nu = -(\alpha^2 - \beta^i \beta_i) dt^2 + 2\beta_i dx^i dt + \gamma_{ij} dx^i dx^j ~.
\end{equation}
In NR, the coordinates $t$ and $x^i$ refer to the \emph{numerical} time and grid locations respectively, and, whilst they can be related to particular physical observers, these are not in any sense \emph{preferred} in the same way as cosmological observers, or the asymptotic observers in a Schwarzschild spacetime.  
The metric degrees of freedom describing proper lengths on the hypersurface are encoded in the 3-metric $\gamma_{ij}$, while $\alpha$ and $\beta^i$ are the \emph{lapse} and \emph{shift} variables respectively, which are generally thought of as gauge functions that define the choice of foliation and the coordinate labels imposed on it, although the clear separation of the metric degrees of freedom into \emph{physical} and \emph{gauge} is not easily defined. 
Note that we have implicitly chosen the coordinate basis defined by the Cartesian grid $x^{\mu}$ which is common in NR but not the only choice (for example the tetrad basis mentioned above or the dual foliation approach of Hilditch \cite{Hilditch:2015qea}).

Given that the equations of motion for the metric are second order in time, in NR we introduce the time derivative of the metric, in the form of the extrinsic curvature $K_{ij}$, to reorganise the equations as a coupled, first order in time system.  The extrinsic curvature is equivalently defined as the Lie derivative of the metric along the normal to the slice,
\begin{equation}
K_{\mu \nu} = -\frac{1}{2} \mathcal{L}_n g_{\mu \nu}  ~,
\end{equation}
or as the projection of the quantity $\nabla_\mu n_\nu$ into the spatial slice $K_{\mu\nu} = \gamma^\sigma_\mu \gamma^\rho_\nu \nabla_\sigma n_\rho$.  The spatial components of the extrinsic curvature in the adapted basis
\begin{equation}
K_{ij} = \frac{1}{2\alpha}(D_i \beta_j + D_j \beta_i -\partial_t\gamma_{ij}) ~, \label{eq:Kij}
\end{equation}
(where $D_i\equiv \gamma_{i}^\mu \nabla_\mu$) together with $\gamma_{ij}$, forms the phase space of metric degrees of freedom. 

\begin{figure}
\centering
\includegraphics[width=0.9\textwidth]{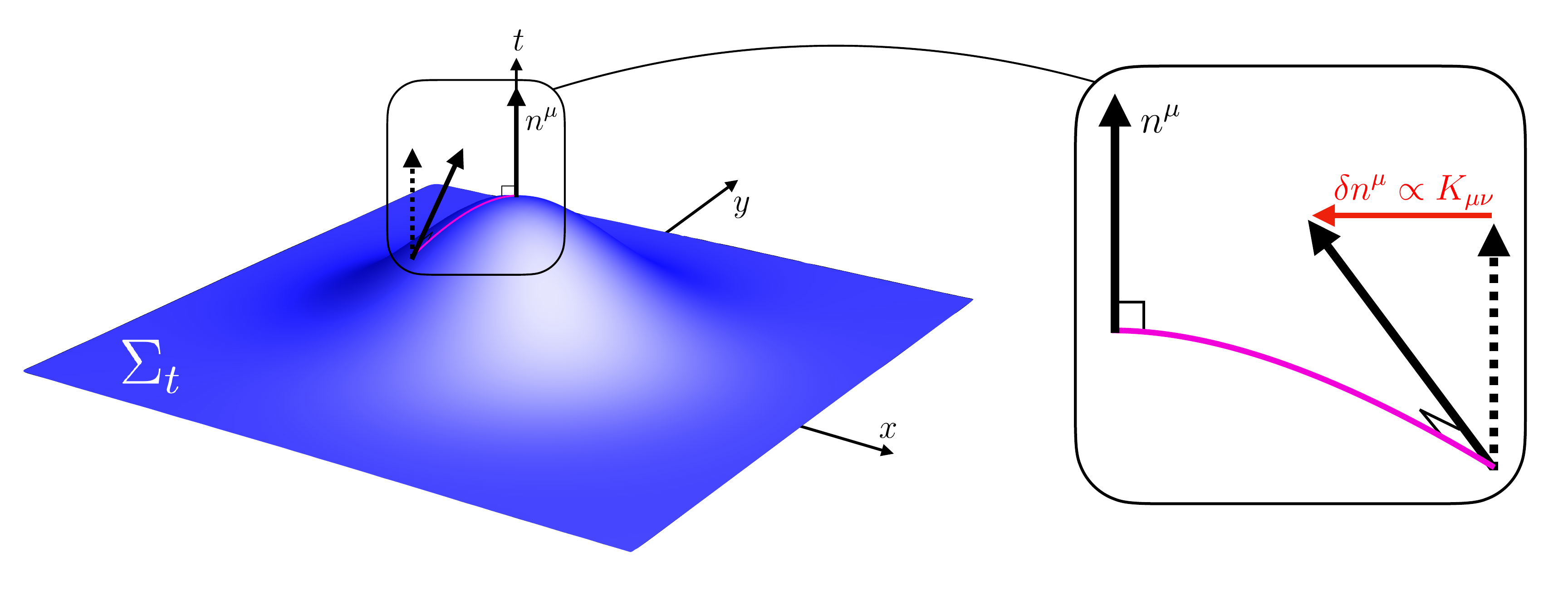}
\caption[Extrinsic curvature]{The extrinsic curvature $K_{\mu\nu}$ encodes the time derivative of the spatial metric $\gamma_{ij}$, which can be related to how the direction of the normal vector changes along the surface, projected into the surface $K_{\mu\nu} = \gamma^\sigma_\mu \gamma^\rho_\nu \nabla_\sigma n_\rho$. (Note that $n^\mu$ is orthogonal to the hypersurface in the Lorentzian sense since this is a spacetime diagram).
}
\label{fig-ExtrinsicCurvature}
\end{figure}

Physically, $K_{ij}$ defines how curved the 3D spatial slice is within the 4D spacetime in which it is embedded. A simple example of the difference between intrinsic curvature and extrinsic curvature is the surface of a cylinder -- the 2D surface is flat; were we to draw triangles on it, the angles would add up to 180 degrees. However, we consider it to be curved because the normal to the 2D surface changes direction in the 3D space in which it is embedded -- this latter concept of curvature, related to how the direction of the normal vector changes direction along the surface (recall that $K_{\mu\nu} \sim \nabla_\mu n_\nu$), is the extrinsic one, see Fig. \ref{fig-ExtrinsicCurvature}. 
Again cosmology provides a nice simple illustration of the concept, as for the RW metric \eqn{eq:RW_metric} we can identify the extrinsic curvature as being related to the time derivative of the spatial metric, and so to $a(t)$ and $H(t)$, as
\begin{equation}
\gamma_{ij} = a^2 \delta_{ij} ~, \quad K_{ij} \sim -H \gamma_{ij}~,
\end{equation}
and the trace $K = \gamma^{ij} K_{ij} =  -3H$. 
Note that some groups use the opposite sign convention (here we use that positive and negative $K$ corresponds to collapse and expansion, respectively). This reminds us that although our universe is roughly spatially flat, it is not at all flat in 4D spacetime, but actually curved in the time direction, by an amount quantified by $H$. We perceive the curvature of the metric in the time direction as an expansion of space in time, but it is really that geodesic observers are moving on a strongly curved spacetime and thus experience geodesic deviation along their trajectories.

The definitions so far are only geometric, and do not encode the physical dynamics of the Einstein equations. To do so the equations themselves must be decomposed into space-like and time-like parts. Doing this we obtain an evolution equation for the extrinsic curvature $K_{ij}$
\begin{multline}  \label{eq:dKdt}
\partial_t K_{ij} = \beta^k \partial_k K_{ij} + K_{ki} \partial_j \beta^k + K_{kj} \partial_i \beta^k - D_i D_j \alpha \\ 
+ \alpha \left({}^{(3)}R_{ij} + K K_{ij} - 2 K_{ik} K^k_j \right) + 4 \pi \alpha \left( \gamma_{ij}(S - \rho) - 2 S_{ij} \right)\,
\end{multline}
as well as the Hamiltonian and three momentum constraints
\begin{align}
\mathcal{H} &\equiv {}^{(3)}R_{ij} + K^2-K_{ij}K^{ij}-16\pi \rho = 0\,, \label{eq:Ham} \\
\mathcal{M}_i &\equiv D^j (\gamma_{ij} K - K_{ij}) - 8\pi S_i = 0\,, \label{eq:Mom}
\end{align}
which must be satisfied on every hypersurface. Here ${}^{(3)}R_{ij}$ is the Ricci tensor associated to the 3-metric, and the decomposed components of the matter stress-energy tensor (those measured by normal observers) are defined as
\begin{equation} \label{eq:matterdecomp}
\rho = n_\mu\,n_\nu\,T^{\mu\nu}\,,\quad S_i = -\gamma_{i\mu}\,n_\nu\,T^{\mu\nu}\,,\quad S_{ij} = \gamma_{i\mu}\,\gamma_{j\nu}\,T^{\mu\nu}\,,\quad S = \gamma^{ij}\,S_{ij} ~.
\end{equation}
The equations of motion for the matter fields (and for a fluid, their equation of state), close the system of equations. 

Again comparison of these ADM equations to the FLRW case is instructive. We see that the momentum constraints are trivially satisfied via isotropy, and the Hamiltonian constraint is simply the first of the Friedmann equations in \eqn{eqn:Friedman}. Whilst this is often thought of as an evolution equation for the scale factor, it is in fact a constraint, and the true dynamical equation for the metric quantity $a$ is the second Friedmann equation of \eqn{eqn:Friedman} as discussed in Sec. \ref{sect:cosmo_for_NR}\footnote{A good example of where treating the first Friedmann equation as the evolution equation will become problematic is going through a cosmological bounce where $H=0$ instantaneously. Then considering only the first Friedmann equation one would naively remain at $H=0$ forever, whereas in reality the spacetime will evolve according to the second Friedmann equation.}. The more complicated evolution equation for $K_{ij}$, \eqn{eq:dKdt}, is an extension of the 2nd Friedmann equation to the case where the quantities are spatially inhomogeneous and the gauge is non-trivial. For example, if $K_{xx}$ is greater than $K_{yy}$ this implies that (according to our chosen observers) the spacetime in the $x$ direction is collapsing faster than in the $y$ direction.

With the degrees of freedom and their equations of motion defined, the system can in principle be discretised and solved on a computer using standard numerical integration and differentiation techniques. However, as discussed above, due to the additional complication of ensuring the equations are well-posed, the ADM equations are not suitable for evolving (see Hilditch \cite{Hilditch:2013sba} for a review). These are modified, and auxiliary variables may be defined depending on the formulation (see e.g. the relevant chapters of \cite{Alcubierre:2008co, Baumgarte:2010ndz}).

The basic conceptual process of setting up and running a NR simulation is as follows, and we will provide further discussion of each step in the following sections:
\begin{enumerate}
\item Define the spatial boundary conditions, which may include assuming a certain topology of our universe e.g. $T^3 \times R$ -- see Sec. \ref{sec:boundary}.
\item Specify initial conditions that both satisfy the Hamiltonian and momentum constraints of GR, and are \emph{physically} the correct ones for the system to be studied -- see Sec. \ref{sec:initial_data}.
\item Define a gauge/coordinate evolution that is stable and well adapted to the problem -- see Sec. \ref{sect:gauge_choice}.
\item Extract and interpret meaningful diagnostic quantities and observables -- see Sec. \ref{sec:interpretation}.
\end{enumerate}

A key difficulty in any numerical simulation is the resolution of different length scales, since computers have a finite precision with which they can represent numbers. For example, if we want to study small length scale effects over long timescales (relative to the light crossing time of the physical length scale), it will be challenging. This is both a feature and a bug, as we have already seen that NR is most needed where the length scales of interest are similar and therefore perturbative approaches fail. However, there will still be many cases where the length scales in the problem change significantly in space or time, and therefore the numerics can break down. 
In particular, cosmological spacetimes with local overdensities introduce an additional  global scale (i.e. the Hubble parameter $H$) into the numerical system. This creates a challenge when our goal is to study local physics with a different scale.
In many cases we expect strong backreaction to be important on small scales, and while it is possible that the dynamics can decouple from the background cosmological evolution, there are intermediate cases where both scales need to be fully resolved. In cases where we need to follow a range of scales in different regions, one can use Adaptive Mesh Refinement (AMR) techniques to gain one or two orders of magnitude\footnote{See Radia \emph{et al.} \cite{Radia:2021smk} for a detailed description and application of AMR techniques in numerical relativity.}, but it is even better to use clever choices of the evolution variables that scale out expansion, as in the tetrad formulation of Uggla \emph{et al.} \cite{Uggla:2003fp} and BSSN-like formulations as used by Zilhao \emph{et al.} \cite{Zilhao:2012bb}. See also Daverio \emph{et al.} \cite{Daverio:2016hqi} for a proposed scheme for cosmological spacetimes that eliminates the fluid components through the constraint equations.

A crucial step in any numerical simulation is convergence testing. These tests are often performed using three simulations with low, medium and high resolutions $\Delta_\mathrm{LR}>\Delta_\mathrm{MR}>\Delta_\mathrm{HR}$. The relative error of a quantity $u$ between simulations should then decrease when increasing the resolution as
\begin{equation}
    \lim_{\Delta \rightarrow 0} \frac{\vert u_\mathrm{HR} - u_\mathrm{MR}\vert}{\vert u_\mathrm{MR} - u_\mathrm{LR}\vert} \approx \frac{\Delta_\mathrm{HR}^n-\Delta_\mathrm{MR}^n}{\Delta_\mathrm{MR}^n-\Delta_\mathrm{LR}^n}\,,
\end{equation}
where $n$ is the order of convergence, which should match the order of the finite difference stencils used, see \cite{Alcubierre:2008co} for more details. When the exact solution of a quantity is known, two resolutions $\Delta_\mathrm{LR} > \Delta_\mathrm{HR}$ are sufficient. In NR, this allows convergence tests of the Hamiltonian and momentum constraints to be carried out using two simulations as
\begin{equation}
    \lim_{\Delta \rightarrow 0} \frac{\mathcal{H}_\mathrm{HR}}{\mathcal{H}_\mathrm{LR}} = \lim_{\Delta \rightarrow 0} \frac{\mathcal{M}_\mathrm{HR}}{\mathcal{M}_\mathrm{LR}}  \approx \left(\frac{\Delta_\mathrm{HR}}{\Delta_\mathrm{LR}}\right)^n\,,
\end{equation}
where $\Delta\rightarrow 0$ indicates that this limit is expected to hold only when the resolution is sufficiently high.
We emphasise here that the \emph{smallness} of the constraint violation is not particularly meaningful in isolation, although one can compare it to the sum of the absolute magnitudes of the terms that appear in the constraints, as an approximate measure of the typical energy or momentum scales in the simulation (which may change over time in an evolving cosmology). The most important test is that it remains bounded in the simulation (i.e. with slower than exponential growth) and converges at the appropriate order when resolution is increased. It is common to excise regions within apparent horizons and at the boundaries (where constraint violation is inevitable but should not affect the region of interest) when assessing this convergence.

\subsection{Boundary conditions}
\label{sec:boundary}

The most important difference between NR applications to astrophysical systems and to that of cosmological systems is that, at the largest scales, the universe is not asymptotically Minkowski/flat. For most of its evolution, it is well described by an FLRW universe with a characteristic scale at time $t$ set by the Hubble parameter $H(t)$. The connection between the astrophysical use of NR and the reality of our cosmology is that the length and timescales for most astrophysical phenomena are much shorter than the Hubble length and time $\sim H^{-1}$, and so they do not feel the cosmological evolution. One can then safely assume that spacetime is asymptotically flat, for example, in a \emph{local} phenomenon such as a BBH merger.  
On the other hand, in an NR simulation of cosmological spacetimes, we are often interested in the \emph{global} structure over cosmological timescales, e.g. how are the matter fields distributed? Is the spacetime inflating? etc. Therefore, the fact that the universe is not Minkowski on average will play a role and the choice of spatial boundary may impact on the physical result of the simulation.

\begin{figure}
\centering
\includegraphics[width=0.5\textwidth]{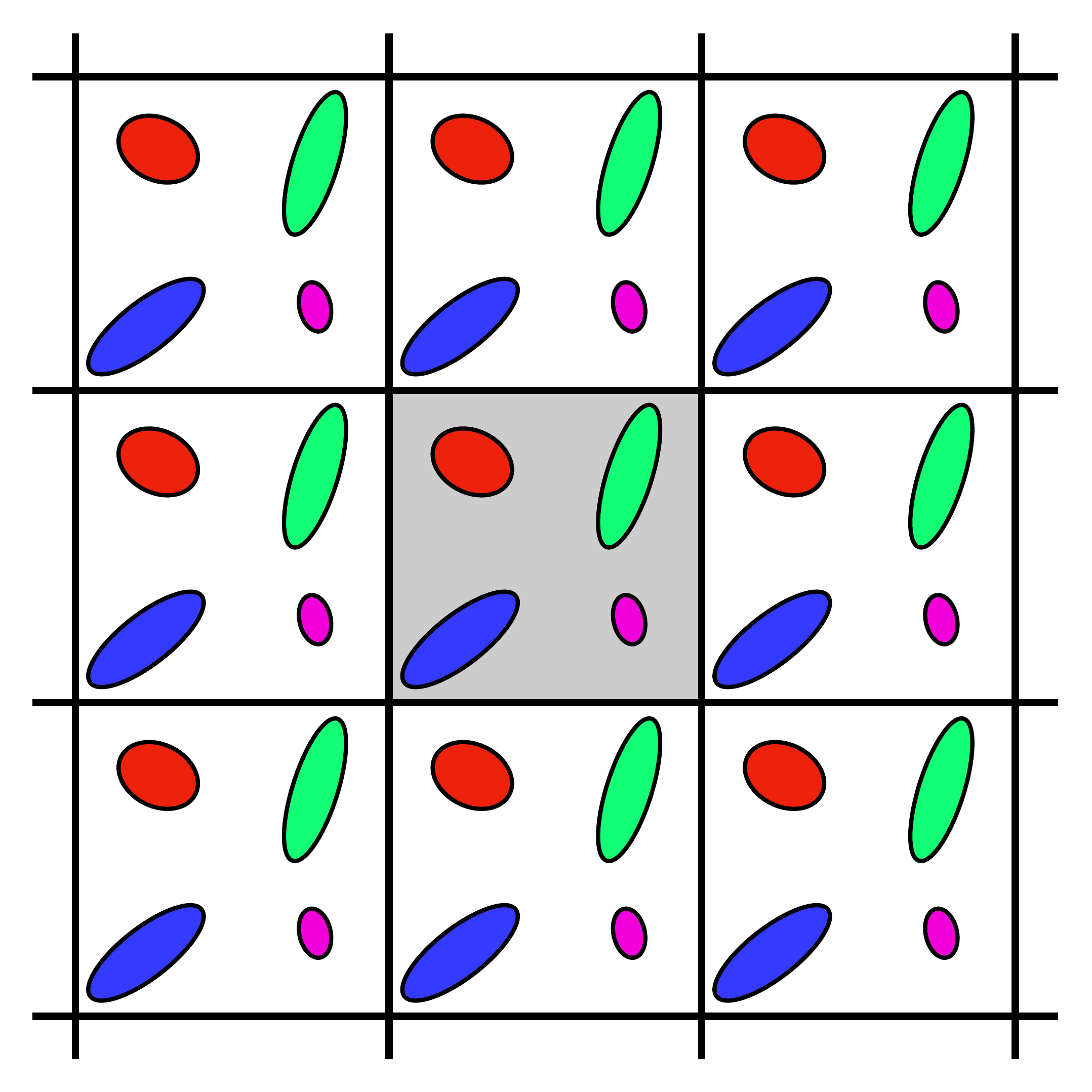}
\caption[Periodic boundary conditions]{Periodic boundary conditions are often used in cosmological simulations, with the idea of treating the computational domain as being a \emph{typical} patch of the entire larger universe, rather than being truly topologically a torus on the relevant scales.}
\label{fig-Periodic}
\end{figure}

The most common and straight-forward solution is to apply periodic boundary conditions, usually with the idea of treating the computational domain as being a \emph{typical} patch of the entire larger universe, rather than being truly topologically a torus on the relevant scales, see Fig. \ref{fig-Periodic}.  In late-time post-Big Bang Nucleosynthesis cosmological evolution this is a well-justified choice as the spacetime is observed to be \emph{statistically} homogeneous and isotropic\footnote{Specifically, correlation functions $\langle XY\dots\rangle$ are invariant under translation and rotation.}. In the early universe the choice is less easy to justify -- we simply do not know what the topology of our universe is on scales beyond the cosmological horizon. Imposing periodic boundary conditions in Cartesian coordinates restricts the topology of space to be a 3-Torus, and hence the Ricci scalar of the spatial hypersurface $^{(3)}R$ has to be negative somewhere or $^{(3)}R=0$ everywhere if there are no BHs \cite{10.4310/jdg/1214432678}.  
Operationally, periodic boundary conditions also impose restrictions on the choice of initial conditions \cite{Aurrekoetxea:2022mpw} -- we will discuss this in Sec. \ref{sec:initial_data} later. 
Finally, periodic boundary conditions impose a discrete symmetry in the size of the domain with a physical scale that is potentially evolving since it depends on the gauge. Thus one must ensure that the scale of the physics being investigated is smaller than the scale of this discrete symmetry, or accept that the results may depend on the choice of boundaries. This issue has recently been raised in the context of studies of inflation in \cite{Garfinkle:2023vzf}, and more generally in N-body studies of cosmological evolution in \cite{Coley:2023iol,Racz:2020nsv}.

Alternative approaches have been taken, in which one instead matches to a particular FLRW spacetime asymptotically, see for example \cite{Niemeyer:1999ak,Shibata:1999zs,Corman:2022rqo,East:2016anr}. If compactified coordinates are used as in \cite{Corman:2022rqo,East:2016anr} it is possible to consistently impose the boundary conditions at asymptotic infinity, but since this imposes the homogeneous solution at sufficiently large distances, it removes any ``global'' backreaction from the inhomogeneities. In other codes boundary conditions are imposed at large but finite distances from the system of interest. In such a case, there is no guarantee that such a choice will be \emph{consistent} -- imagine we choose to match to an asymptotically de Sitter region during a simulation of inflation \cite{Shibata:1993fx}, and some dynamics of the field at the interior of the domain propagate waves out to the boundary. This should disturb the boundary in some way and so rigid de Sitter becomes inconsistent and creates unphysical reflections of the waves. One may be able to fix such a case by applying de Sitter plus outgoing waves as is done for Sommerfeld boundary conditions in an asymptotically flat spacetime. However, the problem is more general -- the choice of boundaries has a physical impact if the timescales of interest allow them to communicate with the physical region of interest, and therefore their choice is part of the specification of the physical situation under study.

This is an area in which further work is needed, and new ideas would be welcome. Some options might be to revisit alternative boundary choices that have previously been explored in NR, such as absorbing or constraint preserving boundary conditions \cite{Calabrese:2001kj,Bona:2004ky,Sarbach:2004rv,Buchman:2006xf,Rinne:2007ui,Rinne:2008vn,Vano-Vinuales:2017qij,Vano-Vinuales:2023yzs,Vano-Vinuales:2024tat}, to try to compactify the domain as in Corman \emph{et al.} \cite{Corman:2022rqo} and recent work for gravitational wave extraction \cite{Peterson:2023bha,Gautam:2021ilg,Vano-Vinuales:2014koa,Gasperin:2019rjg}, or to impose that the boundary tracks a cosmological horizon or restricts the domain to that bounded by the null lightcones emanating from the boundaries. 

\subsection{Initial data for cosmological simulations}
\label{sec:initial_data}

One of the key difficulties in NR is how to set initial conditions that both satisfy the constraint equations and represent the physical spacetime of interest. In this section, we will focus on the issues that arise in cosmological NR simulations. We refer the reader to the standard NR texts \cite{Alcubierre:2008co, Baumgarte:2021skc,Baumgarte:2010ndz,Shibata_book} for more details on the methodologies for finding initial data in NR.

Initial data for the gravitational degrees of freedom consists of the 6 components of the spatial metric $\gamma_{ij}$ and the 6 components of the extrinsic curvature $K_{ij}$ at each point on the initial hypersurface. For a given initial matter configuration, $\{\gamma_{ij},K_{ij}\}$ must satisfy the Hamiltonian and momentum constraints Eqns. (\ref{eq:Ham}-\ref{eq:Mom}), which, in the most general case, represent a set of four coupled, elliptic PDEs. Configurations that satisfy these equations are said to be \emph{on the constraint surface}. The four constraints remove 4 degrees of freedom from the initial data -- the remaining 8 (4 of which are physical, and 4 gauge) together with the 4 gauge variables $\{\alpha, \beta^i\}$ must be chosen according to physical principles or knowledge about the system. A key challenge is that, even when we have an initial solution which is on the constraint surface, \emph{it might not reflect the physical problem that we are trying to investigate}. This is because, for a given matter configuration, described by the stress-tensor $T_{\mu\nu}$, there may be multiple (and potentially an infinite number of) metric configurations that satisfy the Einstein equations that are not physically equivalent. For example, one can usually add some additional transverse traceless component to the extrinsic curvature $K_{ij}$ (roughly additional gravitational wave content). This will change the physical spacetime, but the constraints can still be satisfied. 

There is no \emph{one-size-fits-all} method to find initial conditions in NR. Nevertheless, there are two standard methods for finding initial conditions -- the \emph{conformal traverse-traceless} (CTT) and the \emph{conformal thin sandwich} (CTS) approaches \cite{York:1998hy,Pfeiffer:2002iy}, which as their names imply, perform a conformal transformation of the 3-metric
\begin{equation}
\gamma_{ij} = \psi^4 \bar{\gamma}_{ij}\,.
\end{equation}
The simplifying assumption of conformal flatness $\bar{\gamma}_{ij}=\delta_{ij}$ immediately removes 5 degrees of freedom from the 6 of the metric $\gamma_{ij}$, leaving the the \emph{conformal factor} $\psi$ to be specified\footnote{Note that in making this choice we have already restricted the physical space of the solutions, as we have used up our 4 coordinate degrees of freedom in the initial data that are independent of the lapse and shift. A good example of this is that there is no slicing of the Kerr spacetime that is conformally flat, so this choice would prevent us from finding such a solution, even if other quantities were chosen appropriately.}. The 6 degrees of freedom in $K_{ij}$ are decomposed into
\begin{equation}
    K_{ij} = A_{ij} + \frac{1}{3}\gamma_{ij}K\,, \label{eqn:Kij_decompose}
\end{equation}
where $A_{ij}$ can be further decomposed into a traceless-traverse component $\bar{A}^{ij}_{\mathrm{TT}}$ and a longitudinal component $\bar{A}^{ij}_{\mathrm{L}}$,
\begin{equation}\label{eqn:Aij_conf}
    \bar{A}_{ij} = \psi^2A_{ij}\,,\qquad \bar{A}^{ij} = \bar{A}^{ij}_{\mathrm{TT}} + \bar{A}^{ij}_{\mathrm{L}}\,.
\end{equation}
The 6 components  -- 1 in $K$, 2 in $\bar{A}^{ij}_{\mathrm{TT}}$ and 3 in $\bar{A}^{ij}_{\mathrm{L}}$ -- combined with $\psi$, represent the remaining 7 components that can be set. Given the 4 constraints in \eqn{eq:Mom}, this means that we still need to choose values for 3 components. The choice of which components to specify, and which are resolved using the constraints is dependent on both the physical problem at hand, and the simplicity (or not) of finding solutions. 

In the special case when the stress-tensor has zero momentum density $S_i=0$, the momentum constraint \eqn{eq:Mom} becomes
\begin{equation}
    {\cal M}_i = D^{j}(\gamma_{ij}K - K_{ij})=0~,  \label{eqn:Si=0}
\end{equation}
and hence $D_i K=D^{j}K_{ij}$. A simple and often used \emph{ansatz} is to set $K=\mathrm{constant}$ and $\bar{A}_{ij}=0$ so that the momentum constraint is trivially satisfied. Such a choice leaves $\psi$ to be determined, and this is obtained by solving the Hamiltonian constraint, an elliptic equation in $\psi$ that can be solved with standard multigrid techniques,
\begin{equation}
    8\bar{D}^2\psi -\psi \bar{R}- \frac{2}{3}\psi^5K^2 + \psi^{-7}\bar{A}_{ij}\bar{A}^{ij}=-16\pi\psi^5 \rho~, \label{eqn:Ham_constraint}
\end{equation}
(where we have restored $\bar{A}_{ij}$ and a general conformal metric for completeness). Here $\bar{D}$ is the covariant derivative associated with the conformal 3-metric $\bar{\gamma}_{ij}$ and $\bar{R}$ is its Ricci scalar.

In more general cases, one in principle has to solve the coupled set of constraints \eqn{eq:Mom}. Both the CTT and the CTS methods \cite{York:1998hy} (or now more commonly the extended CTS method (XCTS) \cite{Pfeiffer:2002iy,Pfeiffer:2004nc}) that were initially developed for astrophysical simulations such as BH and NS mergers have been successfully used in cosmological NR simulations.  In the former, the freely specified variables are $\{\bar{\gamma}_{ij},K, \bar{A}^{ij}_{\mathrm{TT}}\}$ together with the gauge variables $\alpha$ and $\beta^i$, while in the XCTS case the freely specified variables are 
$\{\bar{\gamma}_{ij}, \partial_t\gamma_{ij},K,\partial_t K\}$ and $\alpha$ and $\beta^i$ are the \emph{determined} variables. 

In both of these methods, the conformal factor $\psi$ is determined by solving the Hamiltonian constraint, an elliptic equation \eqn{eqn:Ham_constraint}. However, solutions to elliptic equations do not always exist, and when they do, they are not necessarily unique. Indeed, if periodic boundary conditions are imposed as is common in cosmological simulations (see Sec. \ref{sec:boundary}), elliptic equations of the form of \eqn{eqn:Ham_constraint} do not possess unique solutions or are necessarily trivial because the Laplacian operator $\bar{D}^2$ is not invertible on a conformally flat spatial slice with periodic boundary conditions. For periodic boundary conditions, a set of \emph{integrability conditions} must be obeyed such that not only are the constraints in \eqn{eq:Mom} satisfied, but their integrals over the periodic domain are zero (more specifically, one needs to ensure that the source of the elliptic equation is in the adjoint of the Laplacian operator, see \cite{Garfinkle:2020iup,Bentivegna:2013xna}). For example, for the case where $\bar{A}_{ij}=0$ and $\bar\gamma_{ij}=\delta_{ij}$, the integrability conditions
\begin{equation}
\int dV\left(\frac{2}{3}\psi^5K^2-16\psi^5\rho\right)=0\,,\qquad \int dV\left(\frac{2}{3}\psi^6\partial_i K +8\pi\psi^6S_i\right)=0\,,
\end{equation}
must be satisfied. This can lead to difficulties in constructing initial conditions for cosmological spacetimes, and limit the choice of initial conditions that can be explored. 

The presence of fundamental matter fields plays a crucial role in many cosmological simulations. In such cases, matter is represented by one or more fundamental fields such as the scalar inflaton, which presents a special challenge in setting initial conditions. This is because unlike fluids, which can be represented by macroscopic variables such as energy density $\rho$ and pressure $p$, these variables themselves are functions of the fields \emph{and}  their gradients. For example, for a single scalar field $\phi$ with potential $V(\phi)$, the energy density and momentum are
\begin{equation}
    \rho = \frac{1}{2}\psi^{-4}\gamma^{ij}\partial_i \phi\partial_j \phi + \frac{1}{2}\Pi^2 + V(\phi)\,,\qquad S_i = -\Pi\partial_i\phi~,~
\end{equation}
where the conjugate momentum $\Pi \equiv \alpha^{-1}(\partial_t\phi + \beta^i\partial_i\phi)$. This is a complication because standard methods to ensure the elliptic equation \eqn{eqn:Ham_constraint} has unique solutions, that involve scaling the energy and momentum densities \cite{Baumgarte:2006ug,Walsh:2006au}, do not easily extend to fundamental fields, especially in the case of a non-trivial potential.

The problem arises due to the maximal principle \cite{York:1978gql}. 
Setting $\bar{A}_{ij}=0$ and $\bar{\gamma}_{ij}=\delta_{ij}$ for simplicity and linearizing \eqn{eqn:Ham_constraint} via $\psi\rightarrow \psi_0 + u$, we get the following inhomogeneous linear PDE
\begin{equation}
\left(\partial^j\partial_j - q(x)\right)u = \sigma(\psi_0)~,\qquad q(x) = \frac{5}{12}\psi_0^4K^2-10\pi\psi_0^4\rho~,
\end{equation}
where the source $\sigma(\psi_0) = -2\psi_0^5\rho - \partial^j\partial_j\psi_0 + (1/12)\psi_0^5K^2$.  If the boundary is Dirichlet and $q(x)>0$ everywhere in the domain and its boundary, then the maximal principle implies that the only solution is $u=0$ and hence a unique solution for $\psi$ in \eqn{eqn:Ham_constraint} exists. Conversely, if $q(x)<0$ somewhere, then this cannot be guaranteed and a unique solution may not exist. A standard trick to render the solution unique is to rescale $\rho \rightarrow \tilde{\rho}\psi^{-5}$, which switches the sign of the $\rho$ term in $\sigma$ and ensures that $q(x)>0$. If $\rho$ itself depends on $\psi$ and $\phi$, such a rescaling may not always allow $\phi$ to be simply reconstructed, although in Corman \& East \cite{Corman:2022alv}, which developed the CTS method of \cite{East:2012zn}, a method of rescaling the momentum that preserved the ratio of the gradient and kinetic energy in the solution was successfully employed for a massive potential, along with bespoke analytic solutions that ensure the momentum constraint is always satisfied as proposed by Garfinkle \& Mead \cite{Garfinkle:2020iup}. Corman \& East \cite{Corman:2022alv} show that for the specific choices they made unique solutions exist, although in more general cases the uniqueness of the Hamiltonian constraint is potentially problematic.

If rescaling is not possible, one can still attempt to numerically solve the constraints to find \emph{a} solution (if it exists) -- if a trial solution is close to a solution and the perturbations are weak, this method can still work. Alternatively, in Aurrekoetxea et. al. \cite{Aurrekoetxea:2022mpw}, it was proposed that instead of specifying $K$ and solving the Hamiltonian constraint \eqn{eqn:Ham_constraint} for $\psi$ as an elliptic equation, one instead specifies $\psi$ and then solves for $K$ as an \emph{algebraic} equation (up to a sign). This so-called \emph{CTTK} method has been successfully employed in cosmological NR problems \cite{Aurrekoetxea:2023jwd,deJong:2023gsx,Elley:2024alx,Brady:2023dgu} and was inspired by the Friedmann equations where $H \sim K$ varies with $\rho$. It also works for BBH scenarios \cite{Bamber:2022pbs,Aurrekoetxea:2023jwk,Aurrekoetxea:2024cqd}, but note that as a result of the spatial variation in $K$, solving the momentum constraint cannot be avoided.

To date most works use a conformally flat metric $\bar{\gamma}_{ij}=\delta_{ij}$ as an initial condition. Whilst in the context of simulating cosmological spacetimes this may seem a natural choice, we should be mindful of several things. As pointed out by Garfinkle \& Mead \cite{Garfinkle:2020iup}, this imposes a constraint on the choice of $A_{ij}$ and $S_i$ -- there exist matter configurations that do not source a conformally flat spatial metric hyperslice (in particular, those with significant anisotropy). One good example is a uniform momentum density in a single direction, which does not obviously conflict with the periodicity imposed but is prevented by the integrability conditions in the conformally flat case.  This is a strong simplification in terms of the possible initial conditions and may be important in simulations to test the robustness of inflation and bouncing spacetimes (see Sec. \ref{sec:inflation} and Sec. \ref{sec:bounces}). Going beyond conformal flatness requires a generalisation of the methods described above, but should not be conceptually problematic -- this would be an interesting avenue for future work and permit studies of more general cases in cosmological spacetimes.

Other tricks have been employed to solve the Hamiltonian and momentum constraints by making judicious choices of the matter fields, such as solving for the momentum of the scalar field or its the energy density given an inhomogeneous choice for the conformal factor $\psi$ (e.g. \cite{Giblin:2015vwq,Joana:2020rxm}), or introducing extra fields to cancel out or simplify the overall matter sources in the constraints (e.g. \cite{Goldwirth:1991rj,Easther:1999ws,Bastero-Gil:2010tpb}).  Such methods are a good choice where we are not too concerned about the exact form of any perturbations we are introducing, and just want constraint satisfying data in which perturbations are present in the metric. 

More work on connecting the physical properties of the initial spacetime to the choices of free variables would be valuable for cases where we would like more control over the data. We also note that at present there seems to be no fully general, publicly available initial condition solver for cosmological spacetimes, unlike codes for asymptotically flat spacetimes of which there are several\footnote{Although we note that a version of the CTTK solver of Aurrekoetxea \emph{et al.} \cite{Aurrekoetxea:2022mpw} is currently being prepared for public release as part of the GRTL repository \url{https://github.com/GRTLCollaboration}.}. One exception is \texttt{FLRWSolver}, developed by Macpherson \emph{et al.}, but this is specialised to initialise data for cosmological perturbations arising from inflation for studies of late-time cosmology. More open-sourcing of such tools would lower the barrier for new researchers undertaking NR simulations, which would ultimately benefit the field.

\subsection{Gauge choices} 
\label{sect:gauge_choice}

There are two key considerations that set the requirements for a choice of gauge in numerical simulations -- firstly, and most importantly, is the stability of the evolution. The second consideration is that we would like to efficiently sample the spacetime, and want to distribute the coordinates across the region of interest in a way that captures the relevant physical scales, but does not oversample them. As discussed above, in an ADM-based formalism, the gauge freedom is broadly encapsulated in the lapse and shift functions, $\alpha$ and $\beta^i$.  

In the case of zero shift $\beta^i$, the lapse function $\alpha$ determines the ratio between the proper time experienced by a normal/Eulerian observer and the coordinate time that passes for the observers at fixed spatial coordinates. 
A comparison of the ADM metric \eqn{eq:ADM_metric} and RW metric \eqn{eq:RW_metric} reminds us that in the spatially homogeneous case, the lapse is related to the reparameterisation of the time coordinate, for example, moving from conformal time $\eta$ to proper time $t$ as $dt = a d\eta$, or simply choosing the unit of time. In an inhomogeneous spacetime, the effect is more complex and in particular a spatially varying lapse results in an acceleration of the coordinate observers relative to the normal ones. 

The shift vector $\beta^i$ determines how the coordinate points relating to each observer on the slice are relabelled as we move forward in time as illustrated in Fig. \ref{fig-Foliation}. Again in a homogeneous case this is easy to visualise: imagine $\beta^x = -1$ and so we relabel our $x$ coordinate as $x+dt$ after a time $dt$ elapses. This is just a Galilean transformation in the $x$ coordinate, and would make an object that was originally at a fixed coordinate appear to translate with a coordinate speed of $dx/dt = 1$. However, the spatially varying shift case is less easy to interpret and introduces a physical distortion of the spacetime -- we see from \eqn{eq:Kij} that its gradients contribute to the extrinsic curvature $K_{ij}$, for example.
In most NR simulations both $\alpha$ and $\beta^i$ are functions of both $x^i$ and $t$.  

Setting $\alpha=1$ and $\beta^i=0$ is a possible choice of gauge  --  this gives a \emph{geodesic slicing} in which the numerical grid corresponds directly to the normal or geodesic observers. In the context of NR simulations of cosmology, if deviations from an FLRW spacetime are small, then this gauge is equivalent to setting the \emph{synchronous gauge} in perturbation theory (see Sec. \ref{sect:cosmo_for_NR}). Unfortunately, despite its simplicity, geodesic slicing is a poor choice of gauge when strong localised overdensities and/or BHs form, since gravitating systems tend to focus geodesic congruences, causing coordinate singularities. Therefore, both the gauge is usually allowed to vary spatially and in time during a simulation. In particular, in the moving puncture approach, evolution equations are imposed on $\partial_t \alpha$ and $\partial_t \beta^i$ as functions of the metric variables, or equivalently in the approach of GHC, some appropriate gauge source functions $H^\mu$ are specified. More details on these approaches are given below. 

The choice of gauge evolution is usually tied to the physical problem at hand, and we will discuss particular NR gauge choices used in cosmological works in the remainder of this section. In Sec. \ref{sec:interpretation} we will discuss the difficulties in interpretation that the use of a dynamical gauge introduces. However, let us emphasise already here that where the spacetime is highly inhomogeneous, the notion of a preferred time-like observer has already been lost. Thus even in the case where we chose the lapse and shift to be trivial, there is no \emph{right gauge} and we would not be able to recover FLRW observers, as we discussed in Sec. \ref{sect:cosmo_for_NR}, see also Fig. \ref{fig-gaugecosmoNR}.

There is no magic bullet in the choice of gauge in NR, but there are now some well-established principles for what works. A good choice of gauge should typically aim to do some or all of the following (a) avoid hitting singularities in a finite time  (b) maintain well-posedness of the system of PDEs (c) settle into an approximately stationary state when the spacetime possesses an approximate time-like Killing vector (d) avoid stretching the hyperslice by driving the spatial metric towards conformal flatness (e) ensure the discrete timesteps respect the CFL condition for numerical stability -- i.e. ensure that the ratio of proper time and proper distance is less than the $\mathcal{O}(1)$ number determined by the time integration method used $d\tau/dl < \mathcal{O}(1)$. Choosing a gauge in NR is often a matter of trial and error (and some good luck). 

Cosmological spacetimes that are uniformly expanding are often more forgiving than BH ones when it comes to the choice of gauge -- as long as overdensities remain small, geodesic slicing, with the lapse  $\alpha=1$ and the shift $\beta^i=0$ can be stable, and this choice has been employed successfully in several works (e.g. \cite{Giblin:2015vwq, Bentivegna:2015flc}). As discussed above, in the limit of a homogeneous spacetime, this choice matches the standard FLRW synchronous gauge.  The stability arises because the expansion of the normal observers counteracts the tendency for them to focus as a result of overdensities. However, once significant overdensities form, focusing of geodesics will lead to coordinate singularities forming in finite time and one of the standard NR gauges or extensions thereof must be used. 

The standard approach to evade these singularities is the \emph{moving puncture} approach, where the lapse and shift variables are given dynamical evolution equations. For the lapse evolution, most simulations use a particular numerically robust class of foliation defined by the so-called \emph{Bona-Masso slicing} condition \cite{Bona:1994dr}
\begin{equation}
    \partial_t \alpha = -\alpha^2f(\alpha)K~,\label{eqn:bona_masso}
\end{equation}
where $K = \mathrm{Tr}(K_{ij})$ is the trace of the extrinsic curvature and $f(\alpha)$ is some (normally positive definite) function of $\alpha$. For $K>0$ (contracting regions that normally occur around overdensities), $\alpha\rightarrow 0$ is driven to small values to prevent the coordinate observes from focussing and creating a coordinate singularity (or reaching a physical one in the case of BHs). Often this \emph{collapse of the lapse} is thought of as slowing the passage of time so one never reaches the singularity, but its effect is mainly to create a gradient in the lapse that generates an outward acceleration of the normal observers relative to the overdensity, thus stabilising them relative to freely-falling geodesic observers. One can also think of it as an attempt to dynamically drive $K$ to zero.

Similarly, imposing a specific evolution for the shift $\beta^i$ can help to avoid singularities and stabilise the evolution -- one effective condition is the \emph{Gamma-driver} \cite{Alcubierre:2002kk,Alcubierre:2000yz}
\begin{equation}
\partial_t \beta^{i}=\frac{3}{4} B^i~, \qquad \partial_t B^i = \partial_t \tilde\Gamma^{i} - \eta B^i~,\label{eqn:shift_gauge}
\end{equation}
where $\eta>0$ is a dimensionful constant (which in BH spacetimes is of order $\sim 1/M_{\mathrm{ADM}}$ but in cosmological spacetimes would be roughly set by the total mass of any matter collapsing into overdensities) that damps the shift evolution, and $B^i$ (not to be confused with the $B_i$ variable in \eqn{eqn:perturbed_RW}) is an auxiliary variable. The Gamma-driver avoids slice stretching by seeking to minimise $\tilde\Gamma^i \sim -\partial_j\tilde{\gamma}^{ij}$, and hence preventing the formation of coordinate singularities. Using \eqn{eqn:bona_masso} with $f(\alpha)=1/\alpha$ (\emph{$1+\log$} slicing) and \eqn{eqn:shift_gauge} combined is known as the \emph{moving puncture gauge}, which allows for the stable evolution of moving BHs.  This gauge permitted the successful simulations of BH mergers \cite{Baker:2005vv,Campanelli:2005dd}, without the explicit excision of singularities that is required in the GHC formalism\footnote{A common gauge used in analytic solutions of GR is the \emph{harmonic gauge}, which imposes a condition on the coordinates $x^{\mu}$ such that they satisfy the equation
\begin{equation}
    \Box x^{\mu}=0~,\label{eqn:harmonic_gauge}
    \end{equation}
In the GHC formalism, the idea is to drive this wave equation with a set of \emph{source functions} $H^{\mu}$ \cite{Garfinkle:2001ni,Pretorius:2004jg} 
\begin{equation}
\Box x^{\mu} = H^{\mu}~, \label{eqn:GHC_gauge}
\end{equation}
where $H^{\mu}$ is evolved dynamically with a set of differential operators ${\cal L_{(\mu)}}$ that can be freely specified as a function of the spacetime coordinates, the metric and its derivatives, and the source functions and their derivatives. Their exact form depends on the physics being investigated and can be chosen in such a way as to achieve slicings similar to those discussed for the moving puncture gauge. \eqn{eqn:GHC_gauge} and its derivatives then form a set of constraint equations, which can be substituted into the Einstein equations to form a set of well-posed equations of motion for the metric. 
}
\cite{Pretorius:2004jg,Pretorius:2005gq}. Both the standard GHC and moving-puncture gauges have been successfully applied to NR evolutions of cosmological spacetimes, including cases in which BHs form \cite{East:2015ggf,Clough:2016ymm,Clough:2017efm, Aurrekoetxea:2019fhr,Corman:2022rqo,Corman:2022alv}\footnote{Recently, an alternative \emph{shock-avoiding} choice of $f(\alpha) = 1 + \kappa/\alpha^2$, first proposed by Alcubierre \cite{Alcubierre:1996su,Alcubierre:2002iq}, has been found to be efficient in evolving spacetimes that contain collapsing matter \cite{Baumgarte:2022ecu}, but this has yet to be applied to cosmological evolutions.}.

One can encounter problems when applying the BH gauges to cosmological spacetimes. For example, in the moving puncture gauge, because $\alpha$ is driven by $\mathrm{sgn}(K)$ so as to \emph{slow down} evolution in regions of collapsing spacetime when $K>0$, in cosmological expansion $K<0$ can lead to the rapid growth of the lapse $\alpha \gg 1$. Then, each numerical time-step leads to a large leap in cosmic time, which can cause instability and accuracy issues, especially if the lapse varies significantly across the domain and other timescales of interest exist in the problem. For example, in the context of reheating, the oscillation of the field around the minimum can become under-resolved when the lapse grows too large. 

To circumvent this issue, one can implement a more cosmological spacetime-friendly gauge. In Giblin \& Tishue \cite{Giblin:2019nuv}, a modification of the 1+log slicing condition was used in which
\begin{equation}\label{eq:gauge_Kbar}
    \partial_t\alpha = -2\alpha(K - \langle K \rangle )~,
\end{equation}
where $\langle K\rangle$ is the spatial average of $K$ over the hyperslice -- the idea is to set the rate of change of $\alpha$ such $\partial_t{\alpha}=0$ when space is homogeneous, since $K = \langle K \rangle$. This gauge choice approximately ensures that simulation time is identified with the cosmic time coordinate rather than conformal time. In GHC, something similar can be achieved by setting $H^t = K_0$, as done by East \emph{et al.} \cite{East:2016anr} and Corman \emph{et al.} \cite{Corman:2021osa}.
A similar gauge was used by Kou \emph{et al.} \cite{Kou:2021bij}, given by
\begin{equation}\label{eq:gauge_rad}
    \partial_t \alpha = -\frac{2\tau \alpha^2}{1+\alpha^2}\left(K - \langle K \rangle\right)
\end{equation}
with $\beta^i=0$. This is a version of the radiation gauge $g^{ij}\Gamma_{ij}^\rho = 0$, and has the late-time asymptotic solution $K\rightarrow \langle K\rangle$ and $\partial_t\alpha=0$ for an approximately homogeneous spacetime. Elley \emph{et al.} \cite{Elley:2024alx} introduced the following gauge obtained by setting $f(\alpha) = e^{-\alpha}/\alpha$ in the Bona-Masso condition
\begin{equation}
    \partial_t \alpha = -\alpha e^{-\alpha}K+\beta^i\partial_i \alpha~.
\end{equation}
Again, the idea is to prevent a runaway growth of $\alpha$ in homogeneous (and hence expanding) regions, while still avoiding the focusing of observers via the Bona-Masso condition in collapsing regions. Another approach to avoid violating the CFL condition due to the growing lapse, used in East \emph{et al.} and Corman \emph{et al.} \cite{East:2015ggf,Corman:2022rqo,Corman:2022alv}, is adapting the timestepping by dividing it by the maximum of the lapse over the numerical domain.

An alternative approach in the tetrad formulation proposed by Uggla \emph{et al.} \cite{Uggla:2003fp}, requires choosing \emph{constant mean curvature} slices with homogeneous (but time-dependent) $K(t)$ to foliate the spacetime. Such a choice typically requires solving an elliptic equation for the lapse $\alpha$ at every numerical time step, since the aim is to achieve
\begin{eqnarray}
    \partial_t K \sim -D^2 \alpha = 0~,
\end{eqnarray}
which is potentially more computationally expensive and may lead to issues of uniqueness and existence of solutions, or gauge instabilities when evolving black hole spacetimes. Nevertheless, it has the value of possessing a clear physical interpretation -- these are observers for which local expansion is homogeneous even if the matter distribution is not (though clearly these will not generically be geodesic observers). Other interesting recent propositions are the use of mean curvature flow, as in \cite{Doniere:2023ebv}, or a reference metric approach as in \cite{Giblin:2017juu}. The former requires implicit timestepping methods, which are less commonly used than explicit ones. However, the additional technical complexity may be worthwhile for achieving better gauge behavior in certain cases.

In general, we view the question of gauges in cosmology as having been insufficiently studied. There is a tendency (including by the authors of this review!) to reuse existing methods for BH spacetimes rather than develop and explore new and potentially more well-adapted paradigms. We hope to see new developments in this area in the future.

\subsection{Interpretation of quantities in an inhomogeneous spacetime}
\label{sec:interpretation}

Given the freedom and dynamical construction of the foliation of spacetime into spatial hypersurfaces described in the previous section, it is clear that any interpretation of the numerical results must take into account the potential dependence on the gauge. 
 
One issue is that for any higher-ranked tensor beyond the spacetime scalar, the component values depend on the basis choice -- for example for $X_{\mu}$ the physical meaning of $X_0$, $X_1$ etc depends on what basis the index $\mu$ is associated with. Even when accounting for that (or using true spacetime scalars \cite{Ijjas:2023bhh, Garfinkle:2023vzf,Munoz:2022duf}), what you see on the numerical grid at some coordinate time $t$ still depends on the choice of foliation.  This is particularly important when comparing the numerical results from different simulations. 
In principle, once a simulation is complete, one can ``refoliate the spacetime'' via coordinate transforms into any foliation one might desire, but in practice this is extremely difficult. It may also be the case, given the finite extent of the domain in time and space, that simulations with different foliations capture substantially different regions of the spacetime.

An important distinction is between what we will refer to as \emph{gauge dependence} and \emph{slicing dependence}. Consider the specific example of the energy density of the ADM decomposition $\rho = n^\mu n^\nu T_{\mu\nu}$ where $n^{\mu}$ is the unit vector orthonormal to the hypersurfaces of the foliation. This is a spacetime scalar, and therefore it must be gauge independent. However, it is the energy density measured by a particular observer, specifically the one following the normal congruence $n^\mu$, and therefore if we change our slicing, and hence the physical trajectory that we label $n^\mu$, we change the physical scalar that we calculate. Therefore the value of $\rho$ that we measure depends on the slicing -- it is \emph{slicing dependent}. If we were able to track the original normal observer in the new gauge, and calculate the energy density associated with them, then the value would remain the same, but this is not usually done in practise. Mapping between specific observers in inhomogeneous spacetimes is not straightforward (especially if one changes the initial data), and a simulation in a different gauge will normally be associated with a different set of observers, and their associated observables. A further complication is that what we are most interested in physically is $\rho_R = T_{\mu\nu} u^\mu u^\nu$, the density of the fluid in its rest frame, with matter 4-velocity $u^\mu$, since this corresponds to the gauge-invariant density perturbations on large scales \cite{Yoo:2014sfa,Desjacques:2016bnm,Bertacca:2012tp,Bruni:2011ta}. However, since $u^\mu$ in general may have vorticity, it then may not be hypersurface orthogonal. Therefore a slicing corresponding to observers in the rest frame of the fluid (i.e. $n^\mu = u^\mu$) may not exist.

A related problem is the tendency to use \emph{average} quantities across the spatial slices as some measure of the evolution. We must be careful in interpreting these since there is no possible time-like observer who can measure such quantities -- only a god-like observer who can freeze time and move around the spatial slice would ever be able to measure a quantity at space-like separated points. 
Even if the quantity is a spacetime scalar, its integral over the surface is slicing dependent, and so the time evolution may say more about the gauge than the physics. In particular, seeing some quantity blow up can often be a sign of a coordinate singularity forming, rather than a physical instability.
This observation is particularly important in the context of quantifying cosmological observations, which are made on a lightcone and not spatial hypersurfaces.  ``Global'' measures such as the spacetime's Petrov classification can also be useful to understand the broad structure of the solutions, and this has been investigated by Munoz \& Bruni \cite{Munoz:2022duf,Munoz:2023rwh}, although it is again hard to map them to observables.

The most natural and well-motivated physical quantities to track in an evolution are the ones measured by geodesic observers. This still involves some ambiguity -- e.g. should they be comoving with the expansion on the initial slice or on the final slice (which would require them to be tracked backwards in time)? Such methods have been employed in late-time cosmological simulations, as we will detail in Sec. \ref{sec:late_universe}, where precision is important, but in early universe cases the use of quantities associated with coordinate observers and tracking of spatial averages are the common diagnostics. This is a reasonable approach where the question is fairly binary -- e.g. does inflation occur somewhere in the spacetime, or not? Does a BH form (as measured by an apparent horizon forming) or not? However, in cases where the question is more precise, e.g. \emph{in which regions that are causally connected to some section of the initial data does inflation occur?}, then the diagnostics need to be carefully formulated. The development and use of geodesic-based diagnostic tools in order to facillitate comparison in different gauges is an area where further effort could be usefully expended.

A common observable for which accurate predictions are crucial is the gravitational-wave spectrum. At the linear level, a non-local gauge-invariant GW density $\rho_\mathrm{GW}$ can be constructed by averaging over a region larger than the wavelengths. In the FLRW framework, the conventional method to compute it involves extracting the transverse-traceless (TT) parts of the tensor perturbations that are sourced by the matter fields
\begin{equation}
    \rho_\mathrm{GW}=\frac{1}{32\pi G}\langle\dot{h}_{ij}^\mathrm{TT}\dot{h}^{ij\mathrm{TT}}\rangle\,,
    \qquad 
    \ddot{h}_{ij} + 3H\dot{h}_{ij} - \frac{1}{a^2}\nabla^2 h_{ij} = 16\pi G S_{ij}\,,
\end{equation}
and can be projected as
\begin{equation}
    h_{ij}^\mathrm{TT}(t,x)=\int\frac{\dd^3k}{(2\pi)^3} e^{i \mathbf{k\cdot x}} \Lambda_{ij,kl}(\mathbf{\hat{k}})
    h_{lm}(\mathbf{t,\hat{k}})\,,
\end{equation}
where the projector operator is defined as $\Lambda_{ij,kl}(\mathbf{\hat{k}}) \equiv P_{il}(\mathbf{\hat{k}}) P_{jm}(\mathbf{\hat{k}}) - P_{ij}(\mathbf{\hat{k}}) P_{lm}(\mathbf{\hat{k}})/2$, with $\mathbf{\hat{k}}=\mathbf{k}/k$ and $P_{ij}(\mathbf{\hat{k}}) = \delta_{ij}-\hat{k}_i \hat{k}_j$. 

\begin{figure}[t]
\centering
\includegraphics[width=\textwidth]{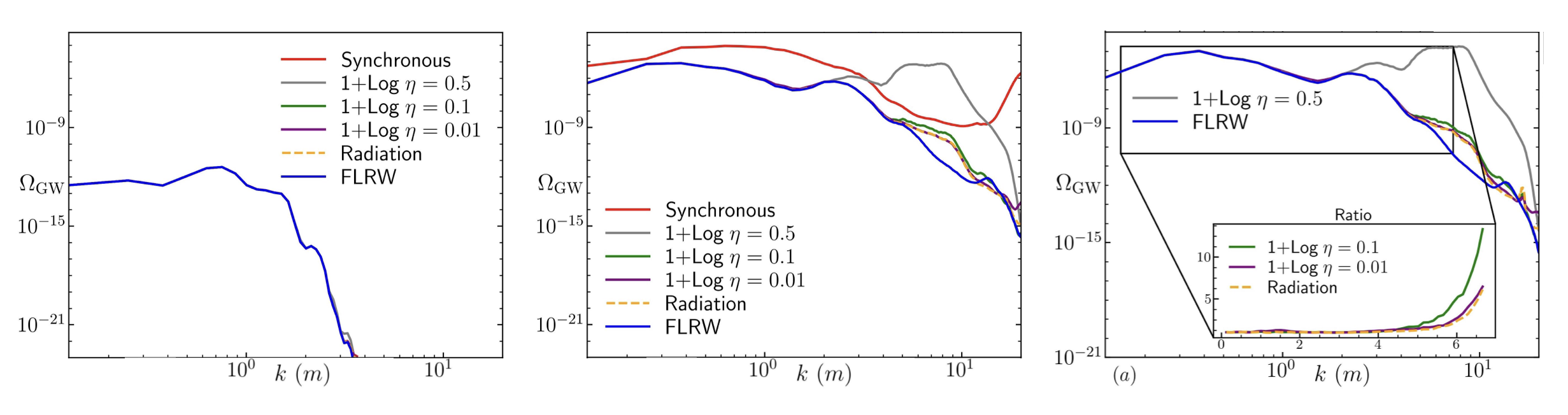}
\caption{Time evolution (from left to right) of GW spectra for a preheating model with different NR gauge choices. The figure shows the breakdown of the synchronous gauge in reliably extracting the GW spectra. The insets in the bottom panels depict the ratio between the GW spectrum in the corresponding gauge condition and that of the FLRW simulation. Figure from Kou \emph{et al.} \cite{Kou:2021bij}.} \label{fig:gw_comparison}
\end{figure}

In NR simulations, GWs are often extracted by projecting $\tilde{A}_{ij}$ (defined in Eqns. (\ref{eqn:Kij_decompose}--\ref{eqn:Aij_conf})) into the TT gauge
\begin{equation}
    \rho_\mathrm{GW}=\frac{1}{8\pi G}\langle\tilde{A}_{ij}^\mathrm{TT}\tilde{A}^{ij\mathrm{TT}}\rangle\,,
\end{equation}
Note, however, that although the GW spectra can be directly extracted by projecting out the transverse-traceless part of the extrinsic curvature, this is only approximate as it assumes $\dot{h}_{ij}\approx -2 \tilde{A}_{ij}$, which may not be true when perturbations are large. The GW spectrum $\Omega_\mathrm{GW}\equiv \rho_c^{-1}\dd \rho_\mathrm{GW}/\dd\log k$ can be computed from lattice simulations of length $L$ as \cite{Adshead:2023mvt}
\begin{align}
   \Omega_\mathrm{GW}(k)\approx  \frac{3}{2\pi^2 L^3}\frac{k^3}{\langle K\rangle^2}\sum_{i,j}\vert \tilde{A}_{ij}(k,\tau)\vert^2\,.
\end{align}
Kou \emph{et al.} \cite{Kou:2021bij} followed a similar approach defining an expanding background metric to be subtracted from the full metric accounting for the conformal factor $\psi$, with $\delta\gamma_{ij}=\gamma_{ij}-\langle\psi^{4}\rangle \delta_{ij}$. 
The energy density in GWs was then computed as
\begin{equation}
\rho_\mathrm{GW}=\frac{1}{32\pi G}\langle\delta{\gamma}_{ij;0}^\mathrm{TT}\delta{\gamma}^{ij\mathrm{TT}}_{;0}\rangle\,,
\end{equation}
where \emph{${;0}$} is the covariant derivative compatible with the background metric. In Fig. \ref{fig:gw_comparison} we show their results, where they compared the extracted $\Omega_\mathrm{GW}$ from oscillon preheating using a rigid FLRW scheme and NR simulations with different gauges: (i) synchronous gauge (geodesic slicing): $\alpha=1$ and $\beta^i=0$; (ii) A combination of the 1+log slicing and Gamma driver shift conditions, Eqn. \eqref{eq:gauge_Kbar}; (iii) Radiation gauge in Eqn. \eqref{eq:gauge_rad}. This comparative approach aimed to understand the impact of NR gauge choices on the extraction of gravitational waves in a fully relativistic setting, and they concluded that the extracted GWs are highly sensitive to the choice of gauge. Synchronous gauge (geodesic slicing) was found to be an unreliable gauge condition for extracting GW spectra, as it contained redundant gauge modes that artificially enhanced the GW power spectrum by 1-2 orders of magnitude.
This example highlights the importance of carefully choosing and interpreting diagnostics from NR simulations. In particular, one needs to think carefully about measuring physical observables for GWs. Ideally, for example, the simulation would evolve to an approximately FLRW state at late times, so that we could then extract the locally averaged gravitational wave energy density measured by late time comoving observers. Measurement of GWs in the non-linear part of the evolution is inherently affected by gauge issues, since there is no natural ``background plus perturbation'' split.

\subsection{Units}

Cosmologists often use units in which $\hbar=c=1$ for studies of the early universe, where quantities are described in terms of the reduced Planck mass $\mplred = \sqrt{\hbar c/8\pi G}$. In this system of units masses $\sim \mplred$ and lengths $\sim 1/\mplred$ , unlike the geometric units used in most simulations in NR where $G=c=1$ and both lengths and masses have units of $[M]$ where $M$ is some physical mass scale in the problem (e.g. $M = M_\odot$). So, for example, energy density has units of $[\mplred]^4$ in Planck units and $[M]^2$ in geometric units. Another important case to highlight is that the scalar field is dimensionless in geometric units, as opposed to having a mass dimension in Planck units. A value of 1.0 for the scalar field is therefore always very large and corresponds to a Planckian energy scale, regardless of what physical scale [M] is taken to represent. To convert a quantity X with units of $[M]^n$ in geometric units to its value Y in units of $[\mplred]^m$, one uses
\begin{equation}
    Y  = X (M/\mplred)^n ~. \label{eqn:units}
\end{equation}
In most simulations we are dealing with length scales well above the Planck length\footnote{Note also the presence of the $8\pi$ factor in the Planck mass, which corresponds to expressing the Einstein equations as $G_{\mu\nu} = T_{\mu\nu} / M_\mathrm{Pl}^2$. $8\pi$ is not an order 1 number (despite cosmologists often setting it to 1!), so if in NR one explicitly includes the $8\pi$ factor in the ADM equations (as most NR codes do), one needs to take care when converting units that it is correctly accounted for.} (where we expect the classical description to hold), and so $(M/M_\mathrm{Pl}) \gg 1$.

In NR, there remains a free mass/length scale $M$ in the problem that once set determines the physical values of the results. This is usually set such that the code values are order 1 numbers, which is advantageous for numerical reasons. For example, in an equal-mass binary BH simulation, the BH bare masses are chosen to be $M\sim 0.5$, with exact values set such that the total ADM mass is exactly 1. This simulation describes equally well a merger of supermassive BHs and solar mass ones, it simply requires the results to be scaled appropriately between the two. So in NR papers one sees quantities like lengths and times described in units of $M$, which in this example is the total ADM mass. The fact that the BH mass and not its Schwarzschild radius was chosen as the NR reference scale is an unfortunate historical convention that causes a lot of unnecessary confusion for those coming from Planck units. It is much better to think of the fundamental scale as being set not by the BH mass, but rather by the Schwarzschild radius relating to the curvature created by such a mass. Consider for example the \emph{mass parameter} $\mu$ that appears in the Klein-Gordon equation
\begin{eqnarray}
    g^{\mu\nu} \nabla_\mu \nabla_\nu \phi = \mu^2 \phi ~.
\end{eqnarray}
In an NR simulation of a scalar field and a BH, the physical scale of interest in this system could be set by either $\mu$, or the mass $M_\mathrm{BH}$ of the BH; once one is chosen, the other is fixed. Say that we choose to set the code values $M_\mathrm{BH}=\mu=1$. This does \emph{not} mean that if we choose the BH to represent a mass of $M_\odot$ then the scalar particle mass is $M_\odot$, which is clearly nonsense. It means that the Compton wavelength of the particle is of the same order as the Schwarzschild radius of the BH, which is about 1km. The correct particle mass $m_p$ can be obtained by recognising that $\mu = m_\mathrm{p} c/\hbar$ in geometric units does not have units of $[M]$ but rather $[M]^{-1}$ (or more sensibly, inverse length), so applying the formula in \eqn{eqn:units} (with $m=1$ and $n=-1$)
\begin{eqnarray}
    \mu_\mathrm{Planck} &=& \mu_\mathrm{Geo} \left(\frac{M_\odot}{\mplred} \right)^{-1} \mplred \nn
                &=& 1 \left(\frac{2 \times 10^{30}\, {\rm kg}}{2 \times 10^{-8}\, {\rm kg}} \right)^{-1} \times (10^{28}\, {\rm eV}) \nn
                &=& 10^{-10}\, {\rm eV} ~.
\end{eqnarray}
which one can verify corresponds to particle with a Compton wavelength of order $\sim$ km. The key takeaway here should be that in NR simulations $\hbar \neq 1$ and this not only changes the \emph{values} but also the \emph{dimension} of the quantities. It is important to go back to the real physical unit (in terms of mass, length and time) and check the dimension in each unit system to get the right transformation, with $[M]\sim [L] \sim [T]$ in geometric units and $[M]\sim 1/[L] \sim 1/[T]$ in Planck units.

\section{The early universe}\label{sec:early_universe}

We start our review of cosmological NR works with those studying the beginning of our universe, or (depending on one's preferred paradigm) those describing previous epochs before the standard hot Big Bang cosmology commences. The broad goal of these works is to understand the nature of the cosmological singularity, to connect it to our currently observed homogeneous and isotropic universe, or to remove it entirely by postulating a \emph{bounce} in the cosmological evolution that would see the universe transition between collapsing and expanding phases.

Most of these works have a common theme in that they seek to understand how \emph{generic} and \emph{robust} certain processes are within the context of classical GR in cosmological spacetimes. We can only observe one Universe, and so in this context NR provides the potential for numerical experiments to understand what other possibile universes may have been generated if we assume a certain paradigm. Such studies are open to philosophical arguments about anthropics or how to precisely define terms like \emph{generic} or \emph{robust} (in particular, the so-called \emph{measure problem}, see for example \cite{Schiffrin:2012zf}). Nevertheless, they can at the very least provide material with which to inform the debate, and at most they provide strong statements about consistency in models invoked to explain the observed features of our universe. 
In the final subsection we review work to identify potential signatures from inflationary bubble collisions, which provides a good example of attempts to connect such early universe paradigms with concrete observables.

\subsection{Cosmological singularities and mixmaster universes}
\label{sec:singularities}

Some of the earliest work in NR was applied to understanding the nature of singularities (see \cite{Berger:2002st} for a review). Here we focus on aspects related to the cosmological singularity case, which has also been discussed in the reviews \cite{Berger:2002st,Garfinkle:2016lcu,Anninos:2001yc,Coley:2017mna,Andersson:2006vk,Cardoso:2012qm,Choptuik:2015mma}. Note that here when we discuss time evolution, usually the evolution is running backwards \emph{towards} the singularity from some initial state that represents our universe at a later time. As a result the spacetimes are collapsing and terms that normally dilute with expansion may become dominant, such as curvature and anisotropy.

The fact that rewinding our universe unavoidably results in a singularity in the absence of exotic matter types is a consequence of Penrose's singularity theorems \cite{Penrose:1964wq}. However, these theorems simply tell us that null geodesics cannot be extended beyond some future event, and not about the nature of the singularity that is reached. Belinskii, Khalatnikov, and Lifshitz (BKL) conjectured that the singularity in generic gravitational collapse is space-like, local, and oscillatory \cite{Belinsky:1970ew}; more specifically, that at each point they approach a state that is well-described by the vacuum, homogeneous Bianchi type IX (Misner's \emph{Mixmaster} \cite{Misner:1969hg}) cosmology. 

The statement that the singularity is \emph{space-like} means that it cannot be observed before it is reached. The \emph{local} aspect of the spacetime is crucial in ensuring that the dynamics of neighbouring points become decoupled (i.e. spatial gradients become negligible in terms of their physical effects on the local evolution). In the resulting Mixmaster cosmology, the spacetime transitions through a series of Kasner spacetimes for which the metric is
\begin{eqnarray}
    g_{\mu\nu} dx^\mu dx^\nu = -dt^2 + t^{2p_1} dx^2  + t^{2p_2} dy^2  + t^{2p_3} dz^2
\end{eqnarray}
where the exponents $p_i$ are constants (a spatially flat FLRW cosmology is a special case where the values of $p_i$ are equal). In the vacuum case the exponents must satisfy $\sum_i p^i = \sum_i p_i^2 =1$, which implies that one direction must expand whilst the other two contract. The transitions between epochs with different Kasner exponents are discrete and chaotic, but have a well-defined map, see Barrow \cite{Barrow81}. They occur when the local spatial curvature $^{(3)} R$ becomes dominant (as it may in a closed FLRW cosmology or where the anisotropy of the spatial metric grows during collapse). In cosmologies where the $^{(3)} R$ never dominates the spacetime will remain stuck in a particular Kasner case. This is then referred to as an Asymptotically Velocity Term Dominated (AVTD) singularity since only the kinetic term in the Hamiltonian drives the evolution at late times. The BKL conjecture is that the Mixmaster dynamics are the most generic case, with AVTD spacetimes only occurring where particular symmetries are imposed. Therefore the nature of the generic singularity is of an \emph{oscillatory} Mixmaster type.

Evidence in support of the BKL conjecture has been obtained from several NR works. The first studies by Berger \& Moncrief \cite{Berger:1993ff} for $T^3 \times R$ vacuum Gowdy spacetimes that have two spatial symmetry directions found only AVTD behaviour, as did later extensions to Gowdy spacetimes with one space-like U(1) symmetry or other topologies \cite{Berger:1995wg,Hern:1997uf,Garfinkle:1999ev,Berger:1998wr}. However, works with more generality, and including matter, were found to be consistent with a BKL picture \cite{Berger:1998vxa,Garfinkle:2003bb,Weaver:1997bv,Berger:1998xx,Berger:2000hb,Berger:2001jy}. The general vacuum case with $T^3 \times R$ topology but no spatial symmetries was first studied by Garfinkle in \cite{Garfinkle:2003bb} using a scaled tetrad formalism of Uggla \emph{et al.} \cite{Uggla:2003fp}. The connection of BKL singularities to the null singularities expected in generic BHs or other asymptotically flat spacetimes is still under investigation \cite{Saotome:2010tc,Garfinkle:2012zh}.

\begin{figure}[t]
\begin{minipage}{0.59\textwidth}
\includegraphics[width=\textwidth]{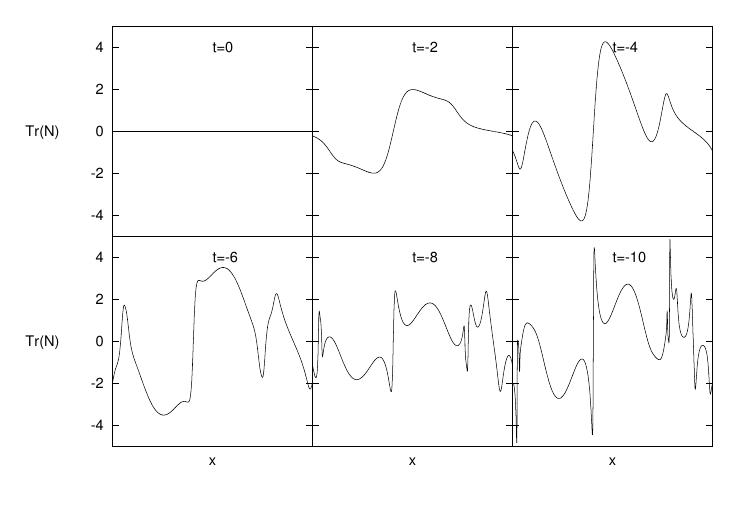}
\end{minipage}
\hfill
\begin{minipage}{0.38\textwidth}
\vspace{-14pt}
\includegraphics[width=\textwidth]{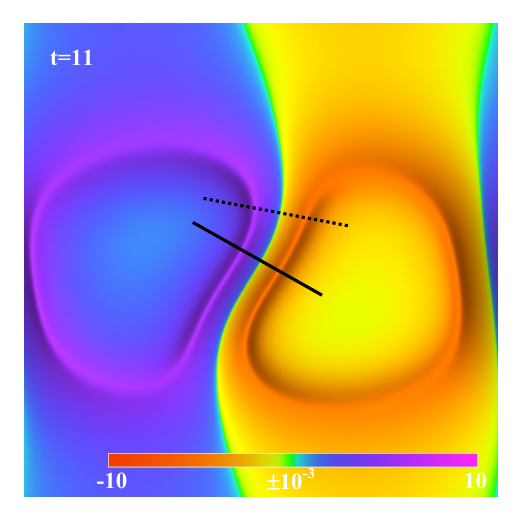}
\end{minipage}
\caption{(Left:) Spike development in the metric quantity $\mathcal{N}^\alpha_\alpha$ (the quantity that vanishes on a spike surface) in a 1+1D simulation. (Right:) Spikes in $\mathcal{N}^\alpha_\alpha$ in a 2+1D simulation manifest as curves. From Garfinkle \& Pretorius \cite{Garfinkle:2020lhb}.} \label{fig:spikes}
\end{figure}

Whilst numerical evidence has mostly supported the BKL conjecture, the initial work by Berger \& Moncrief \cite{Berger:1993ff} was also responsible for the discovery of the existence of a new feature of small-scale structure, now called \emph{spikes}. These occur where points are stuck in an old Kasner epoch while points around them transition to new ones. The term spikes is a bit of a misnomer because these are generically 2D surfaces in a 3+1D spacetime, but since they are codimension one surfaces in a spacetime with symmetry, and were mostly studied in 3+1D spacetimes with 2 symmetry directions, they then appear to be point-like in plots like the left panel of Fig. \ref{fig:spikes}. Studies of spikes have been performed in $G_2$ cosmologies \cite{Lim:2009dg} and Gowdy spacetimes \cite{Garfinkle:2004tu,Garfinkle:2003yq,Berger:1997nn}. Recently adaptive mesh refinement (AMR) has been used by Garfinkle \& Pretorius to permit the study of cases with only one symmetry direction (so that the \emph{spikes} manifest as closed curves \cite{Garfinkle:2020lhb}, see Fig. \ref{fig:spikes}. The numerical simulations are found to be consistent with an analytical formula for the shape of the spikes that suggests that whilst the gradients at spikes may grow arbitrarily large, their impact on the dynamics nevertheless decreases as the singularity is approached. This then implies that the spikes may be only transient features, and that the BKL picture holds even at these somewhat special points.

The BKL conjecture is also sometimes extended with the statement that \emph{matter does not matter}, implying that the vacuum dynamics are sufficient to describe the approach to the singularity. This is evident for most regular matter types, for which the the growth in the energy density during collapse is subdominant to the growth in anisotropy (which grows as $\propto 1/a^6$). However, one exception is a scalar field in which the energy density may grow at the same rate, or greater depending on its potential, which can suppress the BKL behaviour \cite{Garfinkle:2001ni,Andersson:2000cv,Berger:1999tp,Curtis:2005va}. This fact is used as a smoothing mechanism in bouncing scenarios, as will be discussed in Sec. \ref{sec:bounces}. 

The simulations used to study singularities inspired a number of novel numerical techniques, for example the tetrad formulation that permits the overall scaling of the expansion to be factored out \cite{Uggla:2003fp,Garfinkle:2007rv}, tracking swarms of test particles to measure gravitational waves \cite{Berger:1994qc} and symplectic integration schemes for accuracy and stability \cite{Richter:2008pr,Berger:1996gk}. The use of GHC by Garfinkle \cite{Garfinkle:2001ni} inspired the successful solution to the binary BH problem by Pretorius \cite{Pretorius:2004jg,Pretorius:2005gq}, nicely demonstrating how results from different areas of NR can find applications in other physical scenarios.

\subsection{Inflation and the initial condition problem}
\label{sec:inflation}

Cosmic inflation \cite{Guth:1980zm, Starobinsky:1980te, Linde:1981mu,Albrecht:1982wi, Linde:1983gd} is a theory of the early universe in which a period of nearly exponential expansion occurs, providing an explanation of why the universe on large scales is spatially flat and homogeneous, despite the expectation that the cosmological singularity will be chaotic and, due to its \emph{local} nature, uncorrelated at separated points. Inflation also provides a natural mechanism to generate an approximately scale-invariant spectrum of primordial perturbations that seed structure formation, as observed in the CMB \cite{Planck:2018jri}. The simplest mechanism for inflation is a single scalar field, the inflaton, that slowly rolls down a roughly flat region of its potential. To be consistent with this picture, sufficiently \emph{generic} (presumably inhomogeneous) initial data should still permit the scalar field potential to dominate and at least some of the universe to undergo a period of inflation. Single-field models of inflation are classified as high-scale or low-scale models, depending on the energy scale associated with inflation, or as large-field or small-field models depending on the field range over which the minimum required 60 e-folds of inflation occurs (some care is needed in interpreting the different terms \cite{Linde:2016hbb}). In cases where the inhomogeneities are strongly backreacting at the Hubble scale (which is predominantly the case for large-field models), NR simulations may be needed to determine whether this happens\footnote{Early work on using NR techniques to investigate stochastic inflation has been done by Launay, Rigopoulos \& Shellard \cite{Launay:2024qsm} and Florio \& Shellard \cite{Florio:2024pgm}.}.

The problem of initial conditions for inflation has been studied using many analytic and semi-analytic methods (see Brandenberger \cite{Brandenberger:2016uzh} for a review). Whilst anisotropy was shown analytically to remain subdominant in work by Turner \& Widrow \cite{Turner:1986tc}, the general case of strong inhomogeneities required a numerical treatment.
Using NR, early work by Goldwirth \& Piran \cite{Goldwirth:1989pr,Goldwirth:1989vz,Goldwirth:1991rj}, focused on spherically symmetric inhomogeneities in an inflationary theory, assuming a closed universe to avoid the issue of imposing boundary conditions. The initial work included a massless scalar or fluid \emph{radiation} component along with the inflaton scalar field, the latter of which was subject to a range of potentials. 
These simulations suggested that inflation would begin as long as there is initially a homogeneous region of size $R \sim H_\mathrm{inf}^{-1}$ at the inflationary part of the potential, where $H_\mathrm{inf}$ is the inflationary scale. 

\begin{figure}[t]
\begin{minipage}{0.46\textwidth}
\includegraphics[width=\textwidth]{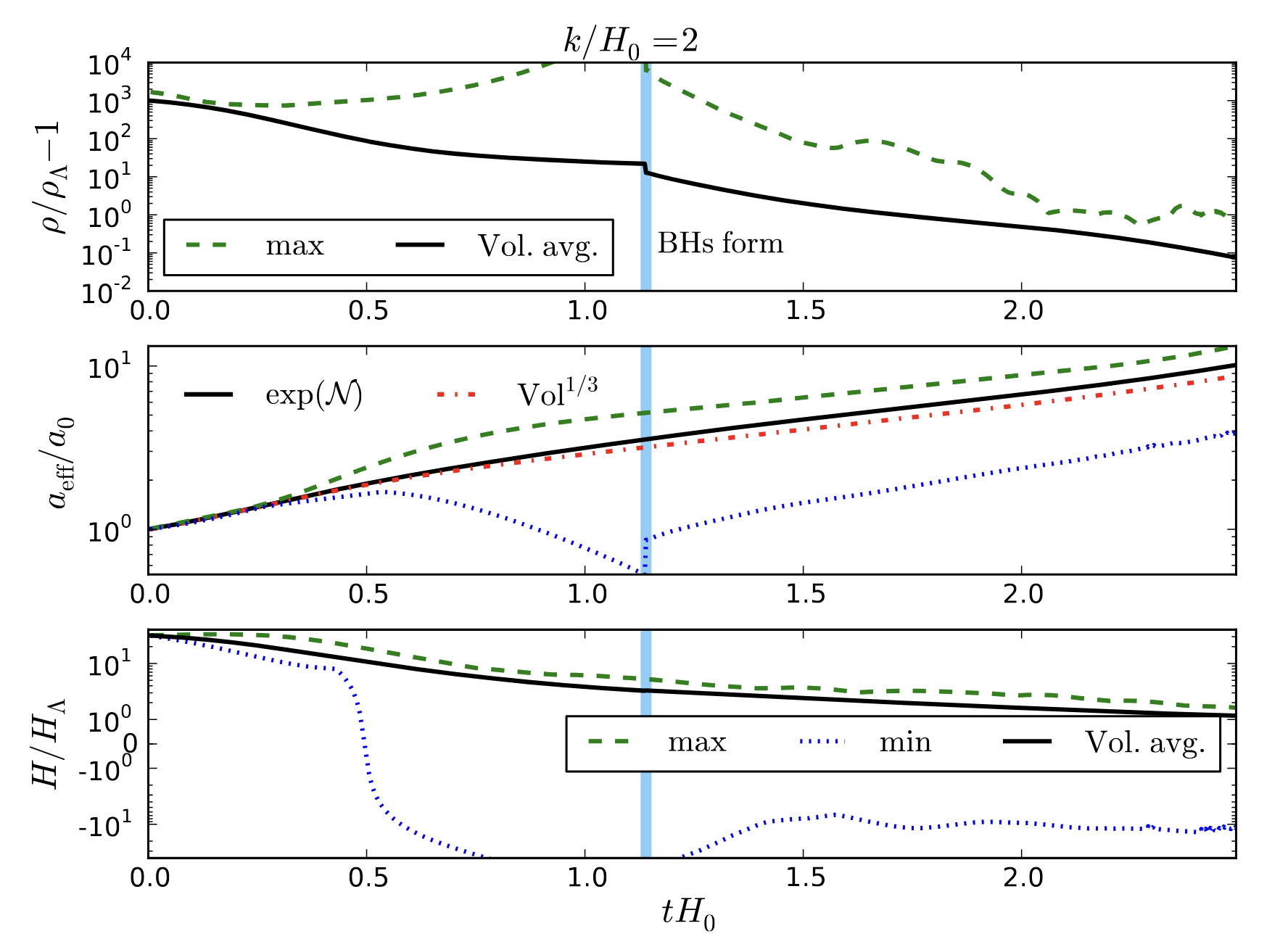}
\end{minipage}%
\hfill
\begin{minipage}{0.52\textwidth}
\vspace{8pt}
\includegraphics[width=\textwidth]{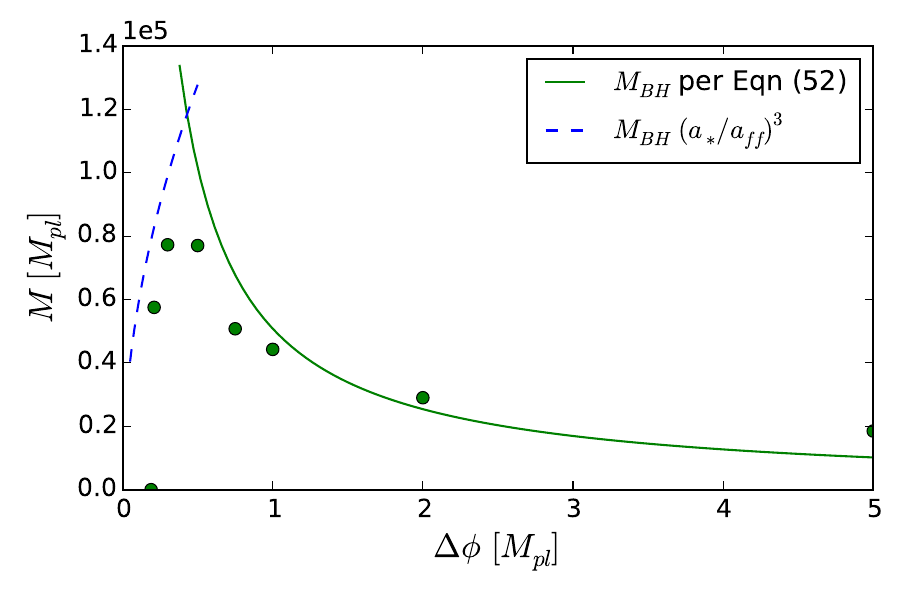}
\end{minipage}%
\caption{(Left:) Figure from East \emph{et al.} \cite{East:2015ggf} showing the first successful simulations of high scale inflation in the 3+1D case in which BHs form. (Right:) Figure from Clough \emph{et al.} \cite{Clough:2016ymm} showing the mass of the BHs formed versus the size of the initial perturbations in a large field model, showing that the maximum mass of the BH that can be formed by such perturbations is bounded.} \label{fig:inflation}
\end{figure}

For high-scale inflation, a number of simulations have supported the claim that inflation persists with generic initial conditions. The first 3+1D simulations with periodic boundary conditions by Laguna, Kurki-Suonio \& Matzner \cite{Laguna:1991zs,Kurki-Suonio:1993lzy}, building on earlier 1D work with Centrella and Wilson \cite{Kurki-Suonio:1987mrt}, showed that inflation could occur even if kinetic and gradient energies were initially comparable to the potential energy density. However, these simulations were limited to initial conditions that did not lead to BH formation.  Simulations leading to BH formation were carried out by Shibata \emph{et al.} \cite{Shibata:1993fx} for axisymmetric spacetimes with a cosmological constant and an initially large distribution of gravitational waves. While their boundary conditions (asymptotically de Sitter) and the presence of a cosmological constant meant that the end state is always de Sitter \cite{Wald:1983ky}, they showed that there can be an initial gravitational wave radiation dominated period (i.e. the GW do not always collapse to form a Schwarzschild-de Sitter spacetime.) It would therefore be interesting to revisit the work with a more general inflationary model and/or boundary setup. Another early work in 1D planar symmetry showed that GWs cannot prevent inflation with a cosmological constant or massive field \cite{Shinkai1993,Shinkai1994}.
More general 3+1D initial conditions, including those that lead to BH formation, were considered first in East \emph{et al.} \cite{East:2015ggf} and studied later by others \cite{Clough:2016ymm, Clough:2017efm, Bloomfield:2019rbs, Aurrekoetxea:2019fhr, Joana:2020rxm, Corman:2022alv,Elley:2024alx}. These showed that high-scale inflation can occur in the presence of large inhomogeneities even where BHs form, see Fig. \ref{fig:inflation}. Some recent work by Garfinkle, Ijjas \& Steinhardt has opposed this conclusion on the basis that the initial conditions and boundary choices made may bias the results towards successful inflation in some \emph{non-generic} regions of low initial curvature \cite{Garfinkle:2023vzf,Ijjas:2024oqn}, see Joana \cite{Joana:2024ltg} for a partial response. The key issue that needs to be addressed is whether the observations of successful inflation in other studies arise only from those parts of the spacetime which were ``non generic'' - i.e. the regions for which the initial curvatures were forced to be low by the imposition of periodic boundary conditions. In the previous simulations, large curvature regions often collapsed to BHs, but for the reasons discussed in the introductory sections of this review, directly tracking the causally related parts of the spacetime is non trivial and cannot be done simply by comparing aligned coordinate regions, as was done in \cite{Garfinkle:2023vzf}. It should, however, be possible to consider only the regions causally connected to a region of high curvature (by tracking null geodesics from the initial slice, for example), and then evaluate whether any of this restricted region undergoes inflation. If not, and the whole region ends up in a black hole, this would potentially validate the concerns expressed in \cite{Garfinkle:2023vzf}. However, in at least some scenarios that were considered in the earlier works \cite{East:2015ggf,Clough:2016ymm, Clough:2017efm, Bloomfield:2019rbs, Aurrekoetxea:2019fhr, Joana:2020rxm, Corman:2022alv,Elley:2024alx} black holes did not form and the final spacetime was fully inflationary, so at least in such cases it seems likely that inflation would survive, albeit with a limit on the magnitude of initial perturbations.


The robustness of low-scale models to inhomogeneous initial conditions is an even more important issue. Since the energy scale is smaller, the size of the causally disconnected patches is larger, and the potential energy may be orders of magnitudes smaller than other components. Consequently, it is possible that large inhomogeneities could spoil inflation before it begins. This has motivated a wealth of studies on the effects of initial conditions on low-scale inflation \cite{East:2015ggf, Clough:2016ymm, Clough:2017efm, Aurrekoetxea:2019fhr, Corman:2022alv, Elley:2024alx}. In general, low-scale models do seem to be less robust to inhomogeneities, with the field dynamics often disrupting the inflationary slow-roll. An important observation by Aurrekoetxea \emph{et al.} \cite{Aurrekoetxea:2019fhr} is that the convexity of the inflationary potential plays a strong role in its robustness, with convex models naturally more robust. This can be understood from arguments about the tendency of the field gradients to homogenise the field, in competition with the tendency of the potential gradient to enhance them.

\begin{figure}[t]
\begin{minipage}{0.4\textwidth}
\includegraphics[width=\textwidth]{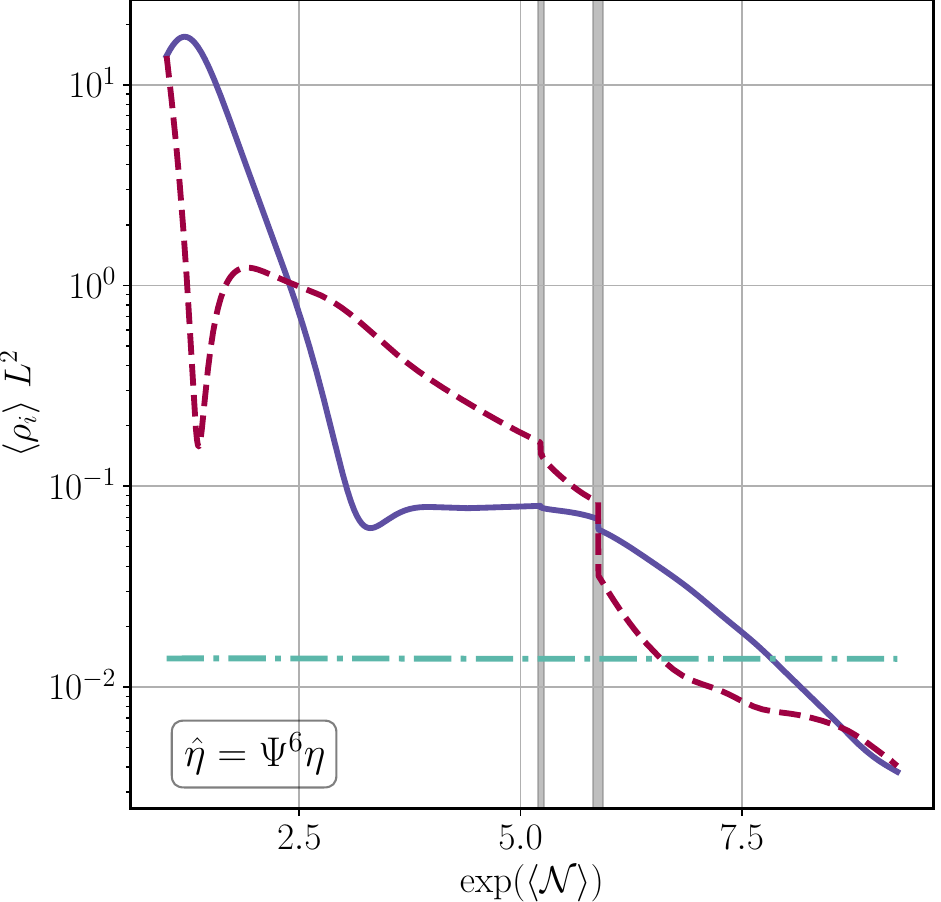}
\end{minipage}%
\hfill
\begin{minipage}{0.56\textwidth}
\vspace{6pt}
\includegraphics[width=\textwidth]{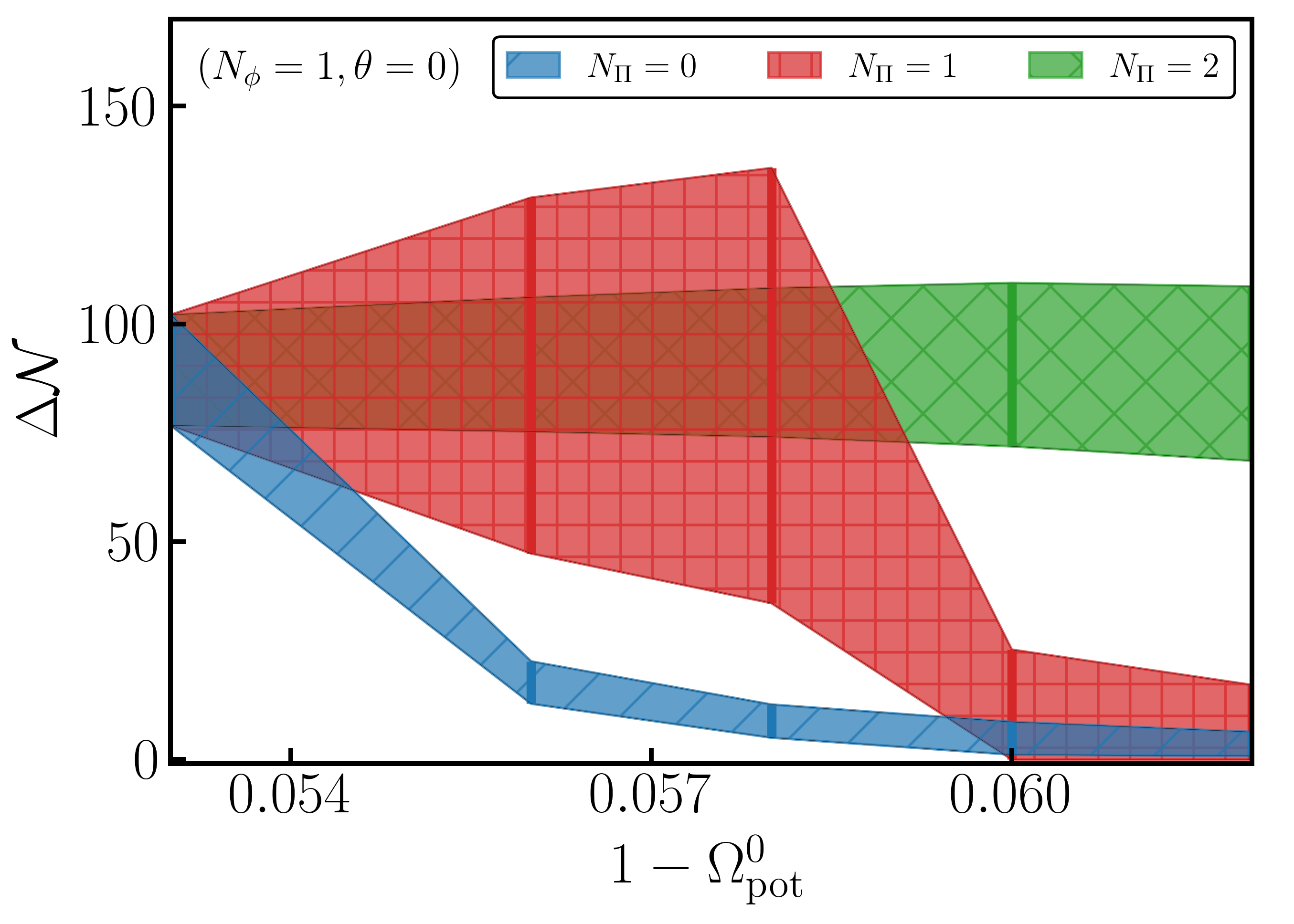}
\end{minipage}%
\caption{(Left:) Figure from \cite{Corman:2022alv} showing gradient (blue solid), kinetic (red dashed) and potential (green dot-dash) contributions to the average energy density. Despite initially high kinetic and gradient contributions, the potential dominates at late times, permitting inflation to occur. Gray regions indicate black hole formation. (Right:) Figure from  \cite{Elley:2024alx} showing the spread in the resulting number of e-folds in the case of a single mode of initial field perturbations, with homogeneous (blue), single mode (red) and double mode (green) perturbations in the kinetic term. A homogeneous kinetic term has the most significant impact, and can significantly reduce the number of e-folds of inflation. In other cases some parts of the spacetime still always experience significant inflation.} \label{fig:inflation2}
\end{figure}

The effect of large non-uniform initial inflaton velocity in addition to an inhomogeneous initial field profile was first studied in Corman \emph{et al.} \cite{Corman:2022alv} for the large-field case in both single and two-field inflationary models, and it was found that these models are generally robust, a conclusion that was later supported by the work of Elley \emph{et al.} \cite{Elley:2024alx} using different methods for generating the initial conditions. In Corman \emph{et al.} \cite{Corman:2022alv}, 
several methods based on CTS with a spatially constant mean curvature $K$ were used to generate initial data -- firstly choosing a constant time derivative for the scalar field, and secondly by rescaling the momentum such that the ratio of the gradient and kinetic energy after solving for the constraints is constant, which allowed them to look at scenarios where there is the same amount of energy in the gradient and kinetic terms. They also used a rescaled scalar field ansatz for the momentum constraint suggested by Garfinkle \& Mead \cite{Garfinkle:2020iup}. The paper by Elley \emph{et al.} \cite{Elley:2024alx} used the CTTK method of Aurrekoetxea \emph{et al.} \cite{Aurrekoetxea:2022mpw} that creates a spatially varying mean curvature $K$ and greater gradients in the extrinsic curvature $A_{ij}$. Despite the different initial data choices, and the use of GHC versus BSSN formulations, both works showed that the effect of velocity on low-scale models resulted in a larger spread of e-folds in different regions \cite{Elley:2024alx}, but did not significantly improve or harm the success of such models, in comparison to the zero velocity case -- see Fig. \ref{fig:inflation2}. This work addressed the concerns of Ijjas \cite{Ijjas:2022qsv} regarding the effect of kinetic inhomogeneities (but does not respond to the more recent criticisms in \cite{Garfinkle:2023vzf,Ijjas:2024oqn}). 

Beyond Corman \emph{et al.} \cite{Corman:2022alv}, little work so far has focussed on the multi-field case. Joana \cite{Joana:2022uwc} investigated the robustness of the Higgs inflationary model to inhomogeneous multi-field initial conditions, with a non-minimal coupling to gravity, as well as the subsequent preheating period, showing that the dynamics from gravitational shear and tensor modes can only delay, but not prevent, cosmic inflation. Given that higher energy justifications for inflationary potentials (e.g. string theory) would generically predict a multi-field landscape, more work on the robustness of such models is justified.

The impact of high-energy modifications to GR are also an interesting direction of study, given that at the start of inflation such corrections may well have been relevant. Recent advances in well-posed formulations of more general 4-derivative models like Einstein-scalar-Gauss-Bonnet \cite{Kovacs:2020pns} and higher derivative pure tensor models \cite{Figueras:2024bba} have opened up the possibility of studying the impact of higher order EFT corrections in such models. Work to develop the initial data methods, and evolutions with a single field acting both as the inflaton and the additional scalar degree of freedom have been completed by Brady \emph{et al.} \cite{Brady:2023dgu,Brady:2024}. Assuming that inflation happens at all with such models, the additional effect of inhomogeneities via the derivative terms is not found to be significant. In particular, the dynamics remain in the weakly coupled regime, rather than being driven out of the EFT regime, and the dominant effect is the additional tilt of the effective potential created by the change to the background/average cosmology. Models with a separate inflaton field and scalar degree of freedom, and pure tensor higher derivative models may give rise to more interesting dynamics and will be a focus of future work.

Finally we note that most of the work described here is quite binary in its conclusions -- inflation either works or it does not. Braden \emph{et al.} \cite{Braden:2016tjn} provides an interesting example of using NR simulations to provide actual observational constraints, assessing the CMB quadrupole constraint on the amplitude of the initial perturbations and the size of the observable universe relative to a length scale characterising ultra large-scale structure. Such studies are challenging because of the cost of NR simulations, but this study used a toy model reduced to 1D perturbations to make progress.

\subsection{Ekpyrotic and bouncing scenarios}
\label{sec:bounces}

The \emph{ekypyrosis} model of cosmology \cite{Khoury:2001wf} posits that the hot Big Bang is generated by the collisions of a brane and its bounding orbitfold in the bulk (extra-dimension) space. During the infall phase of the brane collision, the 3D universe on the brane (on which we live) is in a matter-dominated phase and undergoes a slow contraction. This phase ends at the collision, and the universe reheats during the collision and bounces, reentering the hot Big Bang phase as it falls away from the orbitfold. This phase cools, until the brane turns around and undergoes infall again, resulting in a cyclic universe with repeated bounces.

This string-inspired model has developed into a more general paradigm in which the universe undergoes cyclic periods of expansion and collapse, separated by a non-singular bounce \cite{Ijjas:2019pyf,Ijjas:2021zwv}. In such cases the slow contraction phase leading to smoothing of inhomogeneities, rather than an exponential expansion as in inflation (Ijjas \& Steinhardt provide a pedagogical review in \cite{Ijjas:2018qbo}). It should be noted that to subsequently transition from a state of collapse to a state of expansion - i.e. to achieve a ``bounce'' - one needs to violate the null convergence condition (NCC), which in GR is equivalent to violating the null energy condition (NEC). This is a stronger condition than the violation of the strong energy condition (SEC) required for an inflationary fluid, and whilst the energy conditions are known to be violated in certain cases (e.g. the Casimir effect), violating the NEC on large scales requires some new physics or exotic matter to come into play.

It was suggested that the slow contraction phase could be an efficient \emph{smoother}, i.e. it could drive the universe from an inhomogeneous phase into a homogeneous one \cite{Khoury:2001bz}. The underlying mechanism for this smoothing is \emph{ultralocality} \cite{Ijjas:2021gkf}, in which during the collapse the universe evolves into a phase where the gradients become much less important than the kinetic term, and hence the set of PDEs become a set of ODEs as in the BKL conjecture of \cite{Belinsky:1970ew}, discussed in Sec. \ref{sec:singularities}.    To generate such an evolution, one can consider scalar matter with a \emph{negative-definite} potential $V(\phi)<0$. In a homogeneous FLRW background, the equation of state parameter $w$ is then
\begin{equation}
    w = \frac{(1/2)\dot{\phi}^2-V}{(1/2)\dot{\phi}^2+V}~.\label{eqn:eos_contract}
\end{equation}
and one can see that if $V<0$, the in general $w\geq 1$ as long as $|V| < (1/2)\dot{\phi}^2$. Meanwhile, the Hubble parameter for an FLRW spacetime is given by
\begin{equation}
    |H^{-1}| \propto a^{\epsilon}~,
\end{equation}
where $\epsilon = (3/2)(1+w)$. In \emph{standard} cosmological evolutions $\epsilon = 2$ and $\epsilon=3/2$ for radiation and cold dark matter domination respectively, inhomogeneities grow.  However, in slow contraction ($\epsilon > 3$), the scale factor shrinks much more slowly than the Hubble radius, and hence the universe gets driven to an ultralocal phase which homogenises it\footnote{In an inflationary phase where $\epsilon<1$, the converse occurs,  in which the scale factor expands exponentially while the Hubble radius remains constant.}. Clearly, this argument \emph{assumes} a homogeneous universe to begin with, and hence numerical methods must be employed to test this hypothesis.

\begin{figure}[t]
\begin{minipage}{0.40\textwidth}
\includegraphics[width=\textwidth]{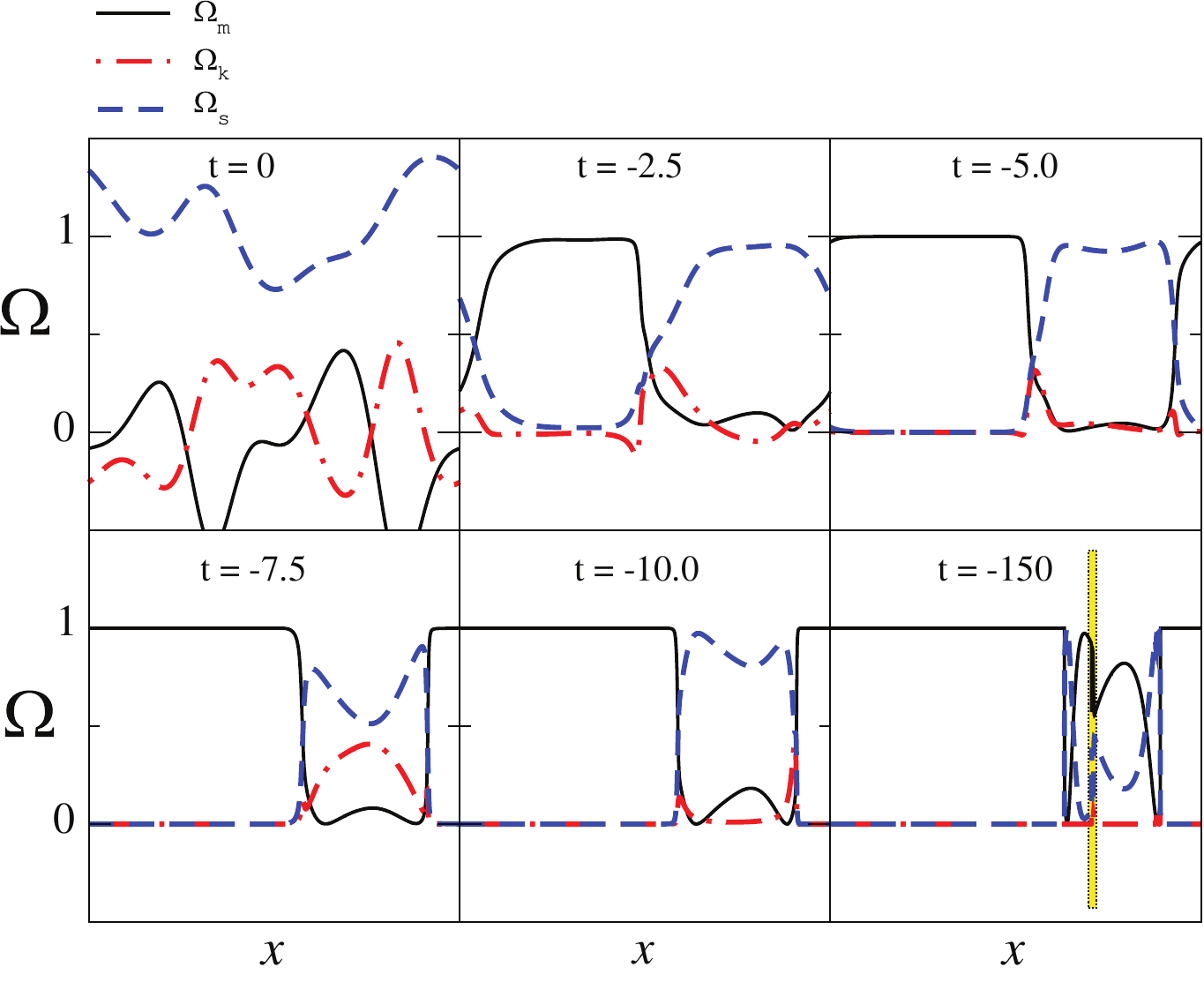}
\vspace{6pt}
\end{minipage}
\hfill
\begin{minipage}{0.49\textwidth}
\includegraphics[width=\textwidth]{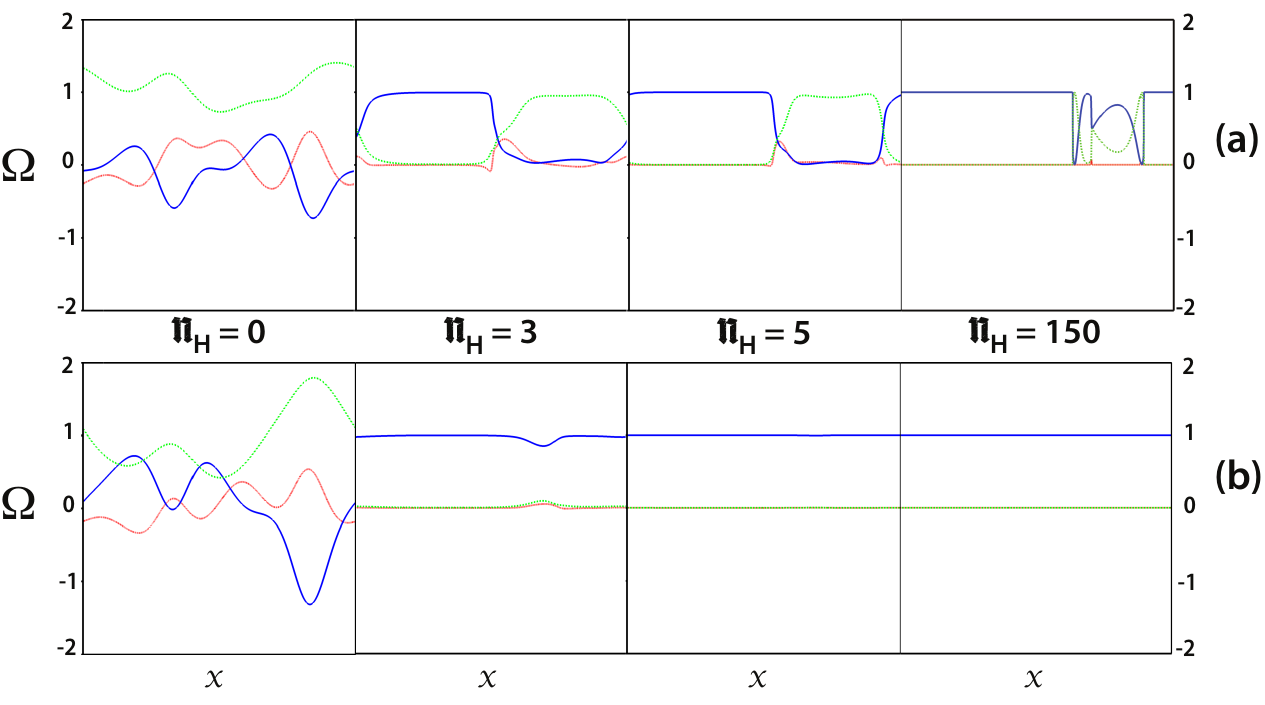}
\end{minipage}
\caption{(Left:) Figure from the first work on slow contraction by Garfinkle \emph{et al.} \cite{Garfinkle:2008ei}. The time evolution of the normalized energy densities in matter (solid), curvature (dot-dash) and shear (dashed), showing that some regions become homogeneous and isotropic. These regions eventually dominate the proper volume. (Right:) Figure from Cook \emph{et al.} \cite{Cook:2020oaj}, recreating the original simulations ((a) top row) and showing that the inhomogeneous regions disappear if one chooses the scalar to be rolling down the potential at all points initially ((b) bottom row), which the authors argue is more physically motivated.} \label{fig:bounce1}
\end{figure}

The first NR work to simulate slow contraction by Garfinkle \emph{et al.} \cite{Garfinkle:2008ei} employed the tetrad formalism of Uggla \emph{et al.} \cite{Uggla:2003fp,Lim:2003ut}. Using a scalar potential of the form
\begin{equation}
    V = -V_0\exp(-k\phi)~, \label{eqn:slow_cont_V}
\end{equation}
they simulated the evolution of slow contraction spacetimes with initial inhomogeneities in the scalar and extrinsic curvature terms in a CMC gauge (see Sec. \ref{sect:gauge_choice}) with perturbations in a single spatial direction. They showed that, for some initial inhomogeneous configurations at least, the domain bifurcate into homogeneous $w\gg 1$ slow-contraction regions and an inhomogeneous $w\approx 1$ regions, with the ratio of \emph{proper volume} of the former to the latter regions growing exponentially fast. In particular, in the slow-contraction region, the space becomes smoother, as expected. See Fig. \ref{fig:bounce1} left panel.

This work was later extended by Cook, Ijjas and others \cite{Cook:2020oaj,Ijjas:2020dws}, who argue that the initial condition for the scalar field in Garfinkle \emph{et al.} \cite{Garfinkle:2008ei} was inconsistent with starting slow contraction. Specifically, the scalar field must roll towards increasingly negative $V$ in \eqn{eqn:slow_cont_V}, and if the initial conditions are set such that this is true initially at all points in the initial hyperslice, then all the regions generically enter the slow-contraction phase that exhibits efficient smoothing, see Fig. \ref{fig:bounce1} right panel. In Kist \& Ijjas \cite{Kist:2022mew}, it was shown that the slow contraction phase is generic and does not depend strongly on the shape of the potential, as long as it is sufficiently steep $\mpl|V,{\phi}/V|\geq 5$, while in Ijjas \emph{et al.} \cite{Ijjas:2021wml} it was shown that these results are robust to perturbations in two spatial dimensions and additional perturbations modes.

In addition to smoothing, one of the key requirements for any model of cosmogenesis is the generation of a scale invariant spectrum of perturbations. Unlike inflationary models, in which the perturbations of the inflaton can provide this spectrum, slow contraction models require another scalar field $\chi$ which sources the scale-invariant perturbations, with the following non-minimal coupling to the kinetic term of the $\phi$ field \cite{Li:2013hga,Levy:2015awa}
\begin{equation}
   {\cal L}_{\mathrm{int}} = -\frac{1}{2}e^{-\phi/m}\nabla_{\mu}\chi\nabla^{\mu}\chi~,
\end{equation}
for some mass scale $m\leq M$, where $M$ is the mass of $\phi$. In Ijjas \emph{et al.} \cite{Ijjas:2021zyf}, this model was numerically studied and it was found that, unlike in the single field case, the presence of the $\chi$ field can destabilise the system as $\phi$ energy can leak into $\chi$. Nevertheless, they demonstrated that this timescale of instability is longer that the bounce cycle time as long as the mass of $\chi$ is small but non-zero. 

Finally, while the above works focus on the smoothing physics of the slow contraction phase of a bouncing cosmology scenario, Xue \emph{et al.} \cite{Xue:2013bva} focused  on the physics of the bounce itself. To execute the bounce after a slow contraction phase, an additional massless scalar ghost field $\chi$ (i.e. a scalar field with \emph{wrong sign} kinetic term) was added  
\begin{equation}
    {\cal L} = -\frac{1}{2}(\partial \phi)^2 -V(\phi) + \frac{1}{2}(\partial \chi)^2~,
\end{equation}
to the usual slow contraction field $\phi$. This ghost field violates the Null Energy Condition, but since its  energy density grows negative faster than the positive energy of $\phi$ as the scale factor $a(t)\rightarrow 0$, it will dominate at small $a(t)$ and turns the contraction phase into an expansion phase, executing the bounce \cite{Allen:2004vz}. Xue \emph{et al.} \cite{Xue:2013bva} studied the evolution of perturbations through the bounce using the harmonic gauge (see Sec. \ref{sect:gauge_choice}), and found that if the initial perturbations are large, the induced non-linearities can cause inhomogeneous regions to collapse into a singularity while relatively homogeneous regions will undergo the bounce successfully. In a slightly different direction Corman \emph{et al.} \cite{Corman:2022rqo} studied the behaviour of BHs through such a two-field bounce model with black hole horizons ranging from 1/8th the size of the minimum cosmological horizon (if there was no black hole) to order 1, and found that for sufficiently large BHs, the apparent horizons can merge and disappear for a short time during the contracting phase. Nevertheless, this state of affairs is temporary and as the bounce enters the expanding phase, the horizons separate and the BH end state is similar to its initial pre-contraction phase value.

\begin{figure}[t]
\begin{minipage}{0.6\textwidth}
\includegraphics[width=\textwidth]{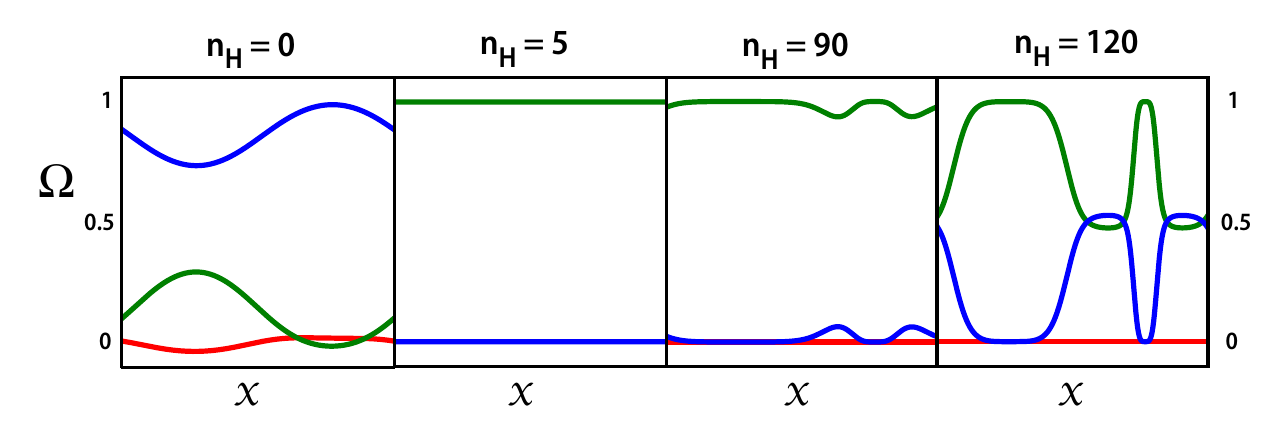}
\end{minipage}
\hfill
\begin{minipage}{0.39\textwidth}
\includegraphics[width=\textwidth]{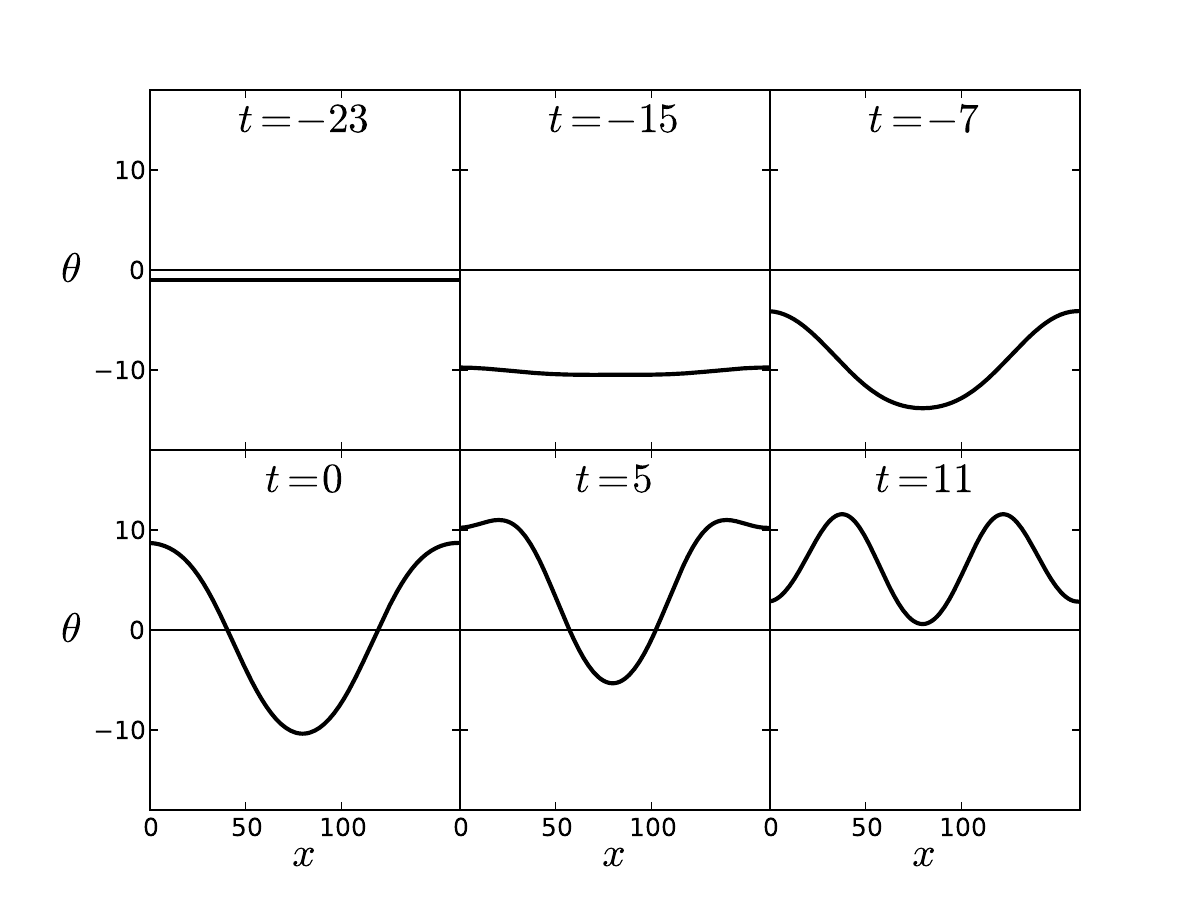}
\vspace{1pt}
\end{minipage}
\caption{(Left:) Figure from Ijjas \emph{et al.} \cite{Ijjas:2021zyf} showing slow contraction in the multi-field case needed to generate scale invariant perturbations. Time evolution of the normalized energy densities in the fields (green) the curvature (red) and the shear (blue) are shown. The model drives the universe towards RW, but a non-zero gradient in the auxilliary field destabilizes the RW state on longer timescales. (Right:) Figure from Xue \emph{et al.} \cite{Xue:2013bva} showing the local expansion as a function of space, evolving over time. Whilst the whole of the space bounces (going from negative to positive expansion), the space in the middle bounces at a much later time compared to the other regions, creating some inhomogeneity at late times.} \label{fig:bounce2}
\end{figure}

A recent paper by Ijjas \emph{et al.} \cite{Ijjas:2024oqn} compared slow contraction and inflationary smoothing, concluding that the latter is more robust and thus strongly favoured. It is hard to assess such claims without running into philosophical questions about which models and which initial conditions are most likely to occur in nature (issues that include the measure problem mentioned above). The work also only considers the smoothing phase and does not take into account the difficulties of subsequently obtaining a bounce and generating primordial perturbations, which on the face of it would seem to favour inflation. Nevertheless, it provides a useful discussion of possible criteria for directly comparing two possible early universe models.

\subsection{Signatures of inflationary bubble collisions}\label{sec:bubbles}

When a spacetime region residing in a metastable vacuum transitions quantum mechanically to a lower energy state, via the nucleation and expansion of a bubble of the new phase \cite{Coleman:1977py,Coleman:1980aw}, eternal inflation can occur \cite{1983Steinhardt,Vilenkin:1983xq,Linde:1986fc}. In such a case our observable universe is just one of many \emph{bubble universes} within an infinite inflating space. If the rate of nucleation is lower than the expansion rate of the \emph{parent} vacuum, the original phase is never consumed, leading to \emph{eternal inflation} (see Refs. \cite{Guth:2007ng,Aguirre:2007gy} for a review). Collisions between expanding bubbles could leave distinct observational signatures in the CMB \cite{Aguirre:2009ug,Kleban:2011pg}, which may allow us to confirm this hypothesis with data. Due to the non-linear amd dynamical nature of the problem, numerical simulations are needed.

The spacetime of two colliding bubbles possesses SO(2,1) symmetry, arising from the intersection of the hyperbolic SO(3,1) symmetry of individual bubbles. This allows the collision to be treated as a 1+1D problem. Early numerical simulations have studied the dynamics of isolated vacuum bubbles \cite{Carone:1989nj} and collisions between them \cite{Hawking:1982ga,Blanco-Pillado:2003fyt,Takamizu:2006gm,Takamizu:2007ks,Freivogel:2007fx,Takamizu:2017asp} (particularly in the context of branes). However, these bubbles lack an interior cosmology and hence are not adequate for making phenomenological predictions. Johnson \emph{et al.} \cite{Johnson:2011wt} performed the first extensive study of the outcome of cosmic bubble collisions using 1+1 dimensional NR simulations. They focused on the collision between \emph{thin-wall} bubbles for potentials with different features and compared their simulations to several commonly used approximations. Where the interior of the expanding bubble was vacuum, their results showed good agreement with the Israel Junction Condition formalism \cite{Israel:1966rt} that matches GR solutions across a boundary, but this did not hold
when the bubbles contained an interior cosmology. In that case, they confirmed the validity of the free-passage approximation \cite{Easther:2009ft,Giblin:2010bd,Amin:2013dqa,Amin:2014eta}, in which the field profiles of the colliding bubbles are superposed.
The authors concluded that the structure of the potential was more important for determining the outcome of a collision than the kinematics, with most of the energy going into the formation of oscillons and pockets of false vacuum, as opposed to scalar radiation. They found that bubble collisions could disrupt inflation if it was driven by a small-field model, but were less likely to do so for large-field inflationary models.

\begin{figure}[t]
\centering
\includegraphics[width=\textwidth]{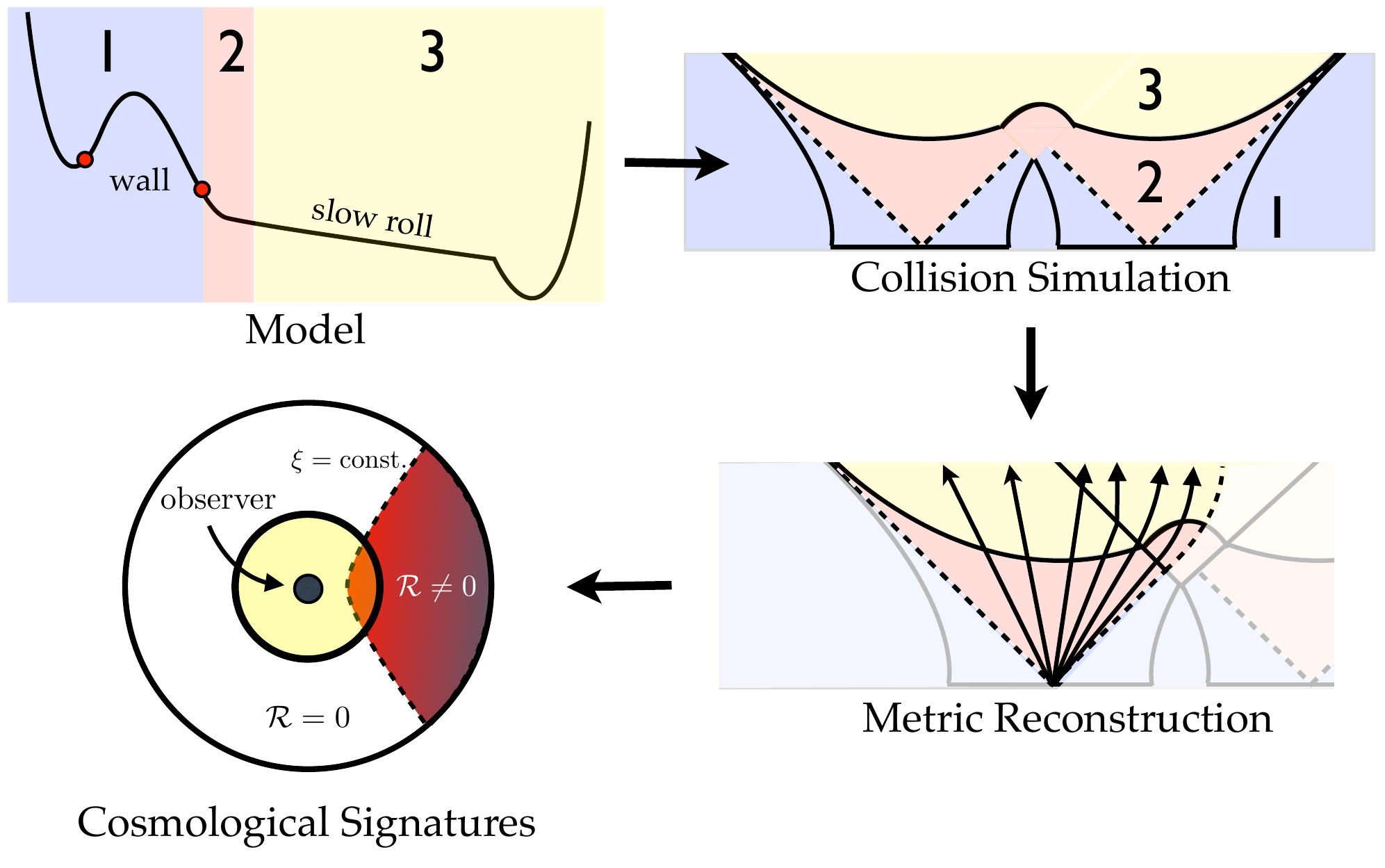}
\caption{Schematic figure depicting the procedure from a scalar inflationary potential to a prediction of the observable comoving curvature perturbation $\mathcal{R}$. 
NR simulations are performed to construct the full collision spacetime (upper right). The perturbed RW metric inside the observation bubble is reconstructed by evolving a set of geodesics through the simulation (lower right panel). A gauge transformation is used to extract the comoving curvature perturbation $\mathcal{R}$ late in the inflationary epoch. Evolving the comoving curvature perturbation, the observable universe is split into a region that is, and a region that is not, affected by the collision. The lower left panel depicts the surface of last scattering inside the observation bubble in a reference frame where the observer is at the origin of coordinates. Figure from Wainwright \emph{et al.} \cite{Wainwright:2014pta}.} \label{fig:bubble_cartoon}
\end{figure}

In a series of papers, Wainwright \emph{et al.} \cite{Wainwright:2013lea,Wainwright:2014pta,Johnson:2015gma} 
made the first quantitative connection between the scalar field inflationary model and the signature imprinted on the CMB -- an azimuthally symmetric temperature anisotropy. The anisotropy can be computed in the Sachs-Wolfe limit as
\begin{equation}
    \frac{\delta T}{T} = \frac{\mathcal{R}(\vec{X}_\mathrm{ls})}{5}\,,
\end{equation}
where $\mathcal{R}(\vec{X}_\mathrm{ls})$ is the comoving curvature perturbation on the surface of last-scattering. In the first of the papers \cite{Wainwright:2013lea}, the authors 
studied the setup depicted schematically in Fig. \ref{fig:bubble_cartoon} to obtain $\mathcal{R}$ in the observer frame.  
They found that the comoving curvature perturbation is well-described by a power law, with its shape and amplitude determined by four phenomenological parameters: the total number of predicted collisions, the shape of the power law, the radius of our past lightcone at last-scattering, and the angular size.

The signatures depend on several features that depend on the scalar potential: the post-nucleation slow-roll inflationary period, the post-inflationary cosmology, the properties of the cosmology in the colliding bubble, and the details of the potential barriers surrounding the false vacuum. In the second paper of the series \cite{Wainwright:2014pta} the authors focused on models that predict only one observable collision and studied the effect of varying the potential barrier in collisions between identical and non-identical bubbles. 
The collision between identical bubbles led to an advance of the inflaton, a positive curvature perturbation $\mathcal{R}>0$ and thus a hot spot in the CMB. The collision between non-identical bubbles, which tunnel in opposite directions, caused a delay of the inflaton, $\mathcal{R}<0$, and resulted in a cold spot instead.

The earlier works were limited to extracting the signatures in the vicinity of the collision boundary, and could not compute the observables far inside the collision region. In the final paper \cite{Johnson:2015gma}, the authors extended their techniques to explore the full bubble-collision spacetime, allowing them to make predictions for all observers. 
In the case of identical bubbles they found that observers in the instanton- and overlap-born region near the collision boundary see different signatures, and oscillatory contributions to the curvature perturbation for non-identical bubbles.

A similar numerical framework was used by Johnson \& Lin \cite{Johnson:2015mma} to compute the observational signatures of a \emph{classical transition} in which a collision between bubbles produced a region with a new vacuum, different from either of the vacua in the colliding bubbles. They found that the main observables were negative spatial curvature and an approximately quadratic comoving curvature perturbation with planar symmetry, which again could be mapped to a temperature quadrupole in the CMB. 

An alternative approach to studying bubble collisions with gravitation beyond the thin-wall approximation is the use of the double-null formalism \cite{Christodoulou1993BoundedVS,Hamade:1995ce}, which consists in decomposing the spacetime interval in the null $(u,v)$ coordinates. Hwang \emph{et al.} \cite{Hwang:2012pj} used these techniques to study the collision between vacuum bubbles with spherical, planar and hyperbolic symmetry. Kim \emph{et al.} \cite{Kim:2014ara} followed a similar approach to compute the gravitational waves emitted from the collision using the quadrupole approximation (see Sec. \ref{sec:interpretation}). They found that the field value of the false vacuum controlled both the frequency of the gravitational waves and the amplitude of the waveforms.  However, due to the highly relativistic nature of the bubbles when they collided, the results obtained using the quadrupole approximation should be considered qualitative rather than precise predictions.

\section{The transitional universe}\label{sec:mid_universe}

The era of Big Bang Nucleosynthesis (BBN) at around $T=1~\mathrm{MeV}$ marks the beginning of the period in which we can say that (broadly speaking) we understand how the universe has evolved on large scales. Before this era, it is \emph{here-be-dragons} observation-wise until the epoch in which the primordial spectrum of perturbations are seeded. The fact that the universe is expanding, and that BBN is hot, means that at some point in the far past the universe must either begin hot, or have undergone an epoch of \emph{reheating} after the perturbations were seeded.  In particular, in the inflationary model of cosmogenesis, the reheating process can feature non-perturbative dynamics that give rise to exotic compact objects and a background of GWs. Given that the early universe was opaque to light before the release of the CMB, such primordial GWs may be the only way to directly probe such early epochs, so characterizing their amplitude and spectral shape is important. 

As the universe cooled, it is expected that it underwent phase transitions that broke down a unified physical model into the forces that govern our universe today. During this process various topological defects may have arisen, such as domain walls, cosmic strings, and monopoles, depending on the symmetry of the vacuum. These objects can leave imprints on the CMB and in GW backgrounds. Primordial black holes (PBHs) may also have formed during the early phases of the universe  from the collapse of large density fluctuations generated during inflation or as the end products of phase transitions and dynamics of topological defects. In some mass bands, PBHs could constitute all (or a substantial part) of the dark matter of the universe and provide a mechanism for the origin of supermassive BHs.

Many simulations of the early universe neglect the local effects of gravity, accounting only for the overall FLRW expansion. Whilst in many cases such an approximation is justified, in cases where high density contrasts are reached, in particular in PBH formation, NR simulations will be needed. In this section, we will review the results from cosmological NR studies of reheating, topological defects and PBH formation.

\subsection{Reheating and oscillon formation}
\label{sec:reheating}

After inflation, the universe is dominated by the scalar field that drives the exponential expansion -- the inflaton. At some point, the universe must transition between the inflaton dominated era to a radiation dominated universe, converting the energy density in a process known as \emph{reheating}. This mechanism sets the scene for the hot Big Bang, converting the vacuum energy in the inflaton into radiation via its decay into other lighter fields. However, a universe filled with an oscillating inflaton condensate is governed by Mathieu-like equations, which can, depending on the inflationary potential, grow the field amplitudes exponentially and drive explosive particle production via parametric resonance. This non-perturbative process is known as \emph{preheating} (see Amin \emph{et al.} \cite{Amin:2014eta} for a review).

Preheating takes place far from thermal equilibrium and is an inherently non-linear process. 
Several levels of approximation can be used: (1) Linear perturbation theory: The field amplitudes are assumed to be small so that mode-mode couplings can be ignored. (2) Lattice field theory on a rigid FLRW background: Mode-mode couplings are allowed but the backreaction onto spacetime is neglected. (3) Lattice field theory with GR: The full dynamics of the field and the spacetime are evolved. The final case can be used to quantify the impact of the assumptions made on the conclusions reached in the first two cases.
Some of the earliest preheating studies that included GR focused on comparing these three cases. Gravitational effects from larger inhomogeneities may lead to collapse on smaller scales, and on scales much larger than the Hubble length, non-linear effects might enhance metric perturbations, with potentially observable impacts on the CMB.

\begin{figure}[t]
\begin{minipage}{0.48\textwidth}
\includegraphics[width=\textwidth]{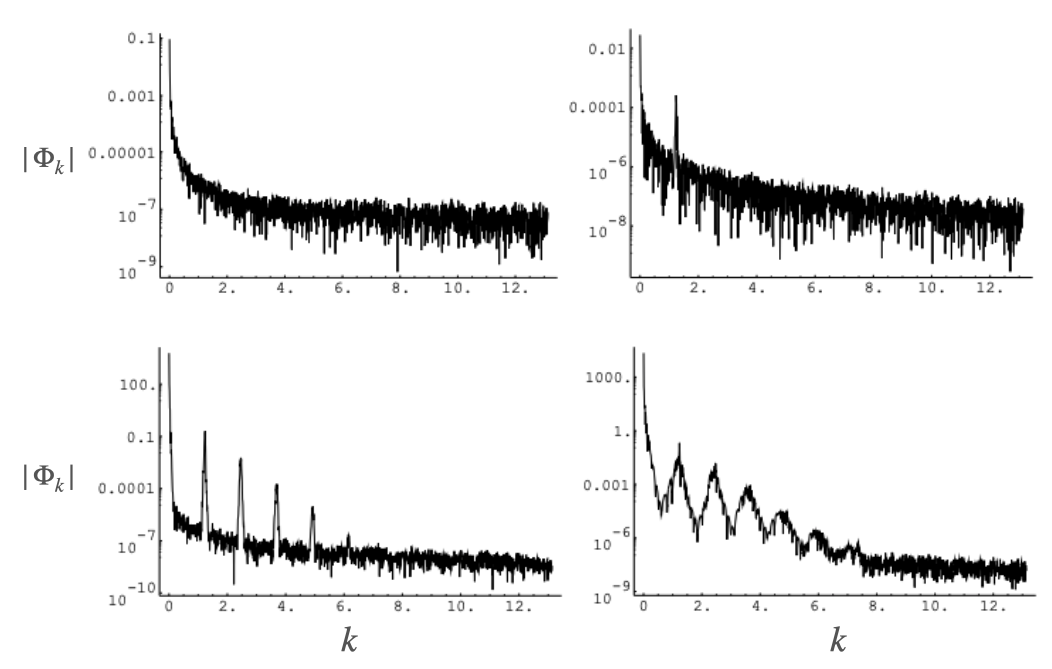}
\end{minipage}
\hfill
\begin{minipage}{0.48\textwidth}
\includegraphics[width=\textwidth]{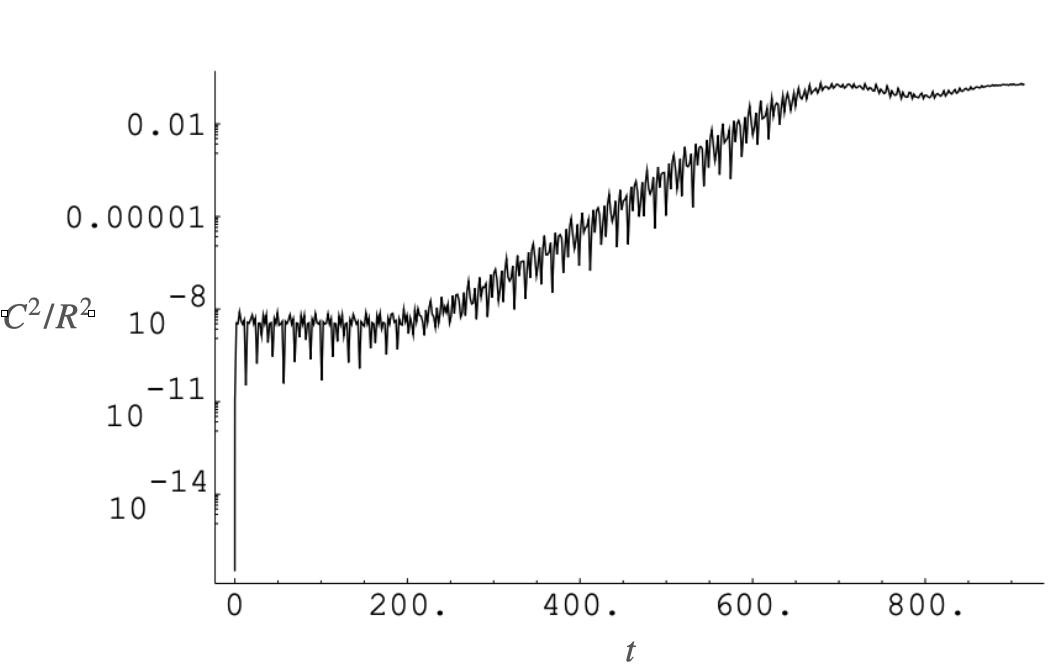}
\end{minipage}
\caption{(Left:) The power spectrum of the gauge-invariant metric perturbation $\Phi$ is plotted against $k$, with all modes initially excited. Each panel corresponds to a different time. (Right:) The growth of $C^2/R^2$ during the simulation. Figures from Easther \& Parry \cite{Easther:1999ws}.} \label{fig:easther_parry}
\end{figure}

Parry \& Easther \cite{Parry:1998pn} pioneered numerically solving Einstein field equations in 1+1D with planar symmetry and periodic boundary conditions to study preheating after the end of $m^2\phi^2$ inflation. 
The simulations showed that wavelengths considerably larger than the Hubble length $k\ll < 1/H$ would not undergo parametric amplification for the $m^2\phi^2$ model, thus verifying the perturbative treatment of the post-inflationary era in the absence of a resonance \cite{Finelli:1998bu}. However, by studying perturbations with larger $k$
they demonstrated that non-linear effects can transfer power between resonant and non-resonant modes. In over-dense regions the onset of gravitational collapse was observed and they concluded that non-linear gravitational effects would broaden any resonance band seen in perturbative analyses. This was investigated further by the same authors for the $\lambda \phi^4$ model \cite{Easther:1999ws}. 
In the case of a single excited mode, the perturbative evolution 
initially agreed with the non-linear treatment, However, while perturbation theory predicted indefinite exponential growth, non-perturbative effects in the field dynamics eventually terminated the resonance.
Comparing the rigid background and non-linear gravity simulations showed that 
when self-gravity was included, the amplitude of perturbations remained comparatively large, whereas without it they decayed.
When multiple inhomogeneous modes were initially excited, power was transferred to long wavelength modes in both the rigid FLRW background and the full GR simulations, see Fig. \ref{fig:easther_parry}. In the GR case, the overall growth was found to be larger and began sooner, although the authors could not conclude whether this was due to backreaction effects or other choices made for the simulations, such as periodic boundary conditions.
To study the impact of gravity, they suggested tracking the ratio of the domain integrals of the Weyl curvature scalar $C^2\equiv C_{\mu\nu\gamma\lambda}C^{\mu\nu\gamma\lambda}$ and the usual Ricci curvature scalar, $\mathcal{C}^2/\mathcal{R}^2$, as a measure of how far the spacetime departs from perfectly FLRW, see Fig. \ref{fig:easther_parry}. As they did not observe this quantity diverging, they concluded that no PBH formation occurred for the $\lambda \phi^4$ theory.
This is in contrast to Finelly \& Khlebnikov \cite{Finelli:2000gi}, who performed general relativistic 1+1D simulations of preheating following hybrid inflation in spherical symmetry.
They found that in this scenario, super-Hubble metric perturbations could experience significant growth, with potential evidence for BH formation. 

\begin{figure}[t]
\centering
\includegraphics[width=\textwidth]{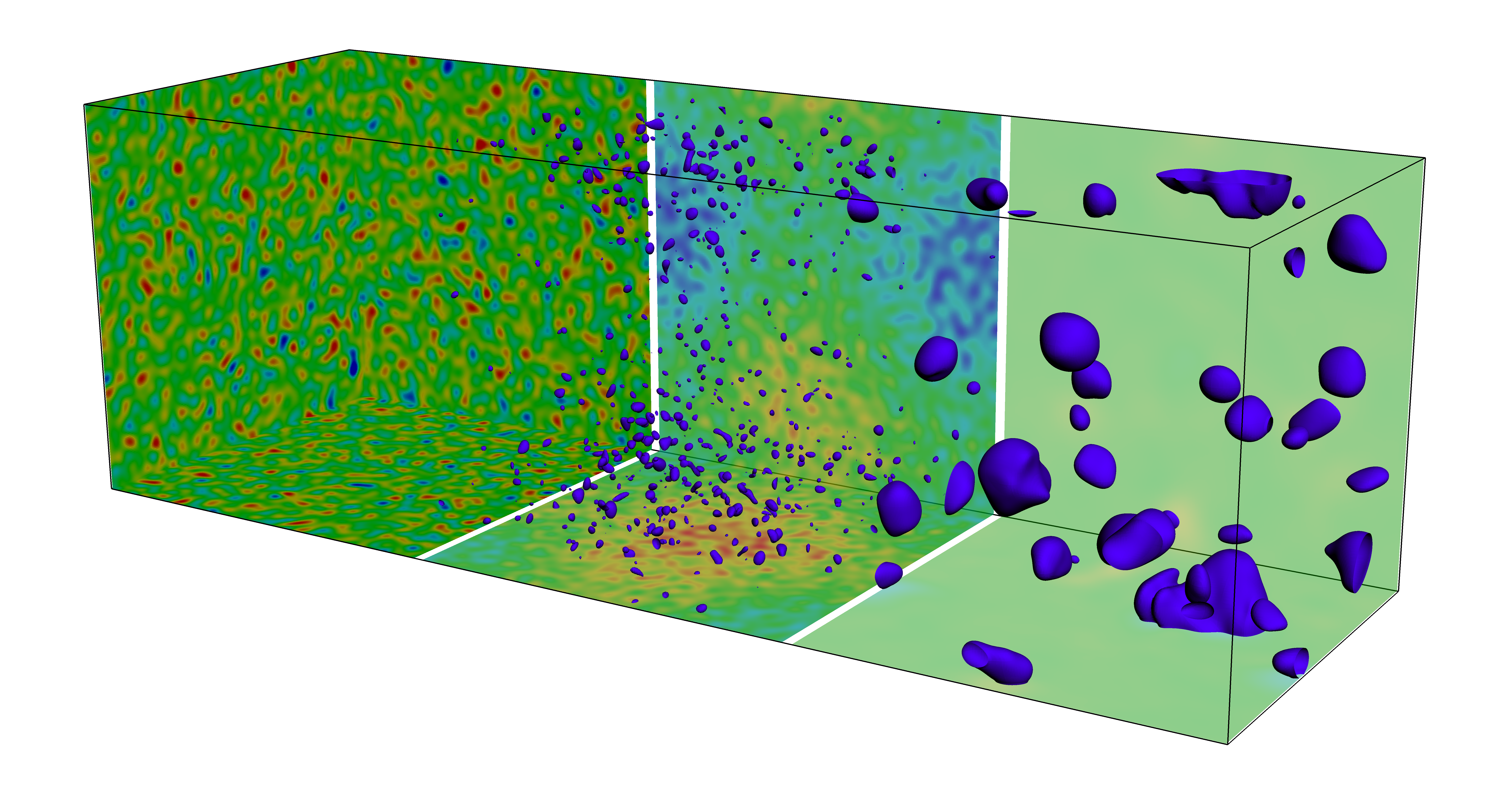}
\caption{Oscillon formation during preheating. The faces of the box show the evolution of inflationary fluctuations and the inner regions correspond to the isocontours of density, illustrating the formation of oscillons. Figure produced with simulations from Aurrekoetxea \emph{et al.} \cite{Aurrekoetxea:2023jwd}.}\label{fig:oscillons}
\end{figure}

The first 3+1D studies of preheating by Bastero-Gil  \emph{et al.} \cite{Bastero-Gil:2007lsx,Bastero-Gil:2010tpb,Bastero-Gil:2011jyw} employed the ADM formalism with the synchronous gauge to study  $m^2\phi^2+\lambda\phi^4$ and hybrid models of inflation. For the chaotic model they found that the amplification of superhorizon modes observed in 1+1D simulations  \cite{Easther:1999ws,Finelli:2000gi} persisted in 3+1D. For the hybrid model, where the resonance is more explosive, scalar metric perturbations were found to have little impact on the evolution of field fluctuations. 
They compared the obtained energy density in gravitational waves to integrating the tensor modes with a rigid FLRW scheme and found that it was approximately an order of magnitude larger\footnote{Note, however, that as discussed in Sec. \ref{sec:interpretation}, the synchronous gauge can introduce redundant gauge modes that artificially enhance the GW spectrum.}.

More recently, Giblin \& Tishue \cite{Giblin:2019nuv} performed a detailed comparison of the evolution of preheating between cosmological perturbation theory, rigid FLRW and NR.
They used the BSSN formalism with the modified version of the 1+log gauge condition in \eqn{eq:gauge_Kbar}. To translate between the results of the three schemes they prepared a useful appendix expressing their Bardeen potential diagnostics in terms of BSSN quantities. 
They used the variance of fields as a measure of the inhomogeneities,
e.g. tracking the variance of the lapse to look for signs of BH formation.
Although they found a disagreement with perturbation theory, they were not able to conclude that this led to the formation of BHs.

When the decay of the inflaton field happens very rapidly, the process is highly non-perturbative and can feature resonances that lead to the fragmentation of the homogeneous condensate into lumps, which then evolve to form oscillons: massive, localized, metastable configurations of scalar field \cite{Amin:2011hj}, see Fig. \ref{fig:oscillons}. Many inflationary potentials can be self-resonant, where resonance generates quanta of the inflaton field itself more efficiently than particles coupled to the inflaton. In a class of well-motivated self-resonant models, the universe may become dominated by oscillons, delaying thermalization. Several questions about oscillons have been explored using NR simulations: (1) Do non-relativistic (and/or non-backreacting) simulation results still apply when these are taken into account; (2) what is the spectrum of GWs emitted by these objects?  (3) can they can collapse to black holes?

\begin{figure}[t]
\begin{minipage}{0.48\textwidth}
\includegraphics[width=\textwidth]{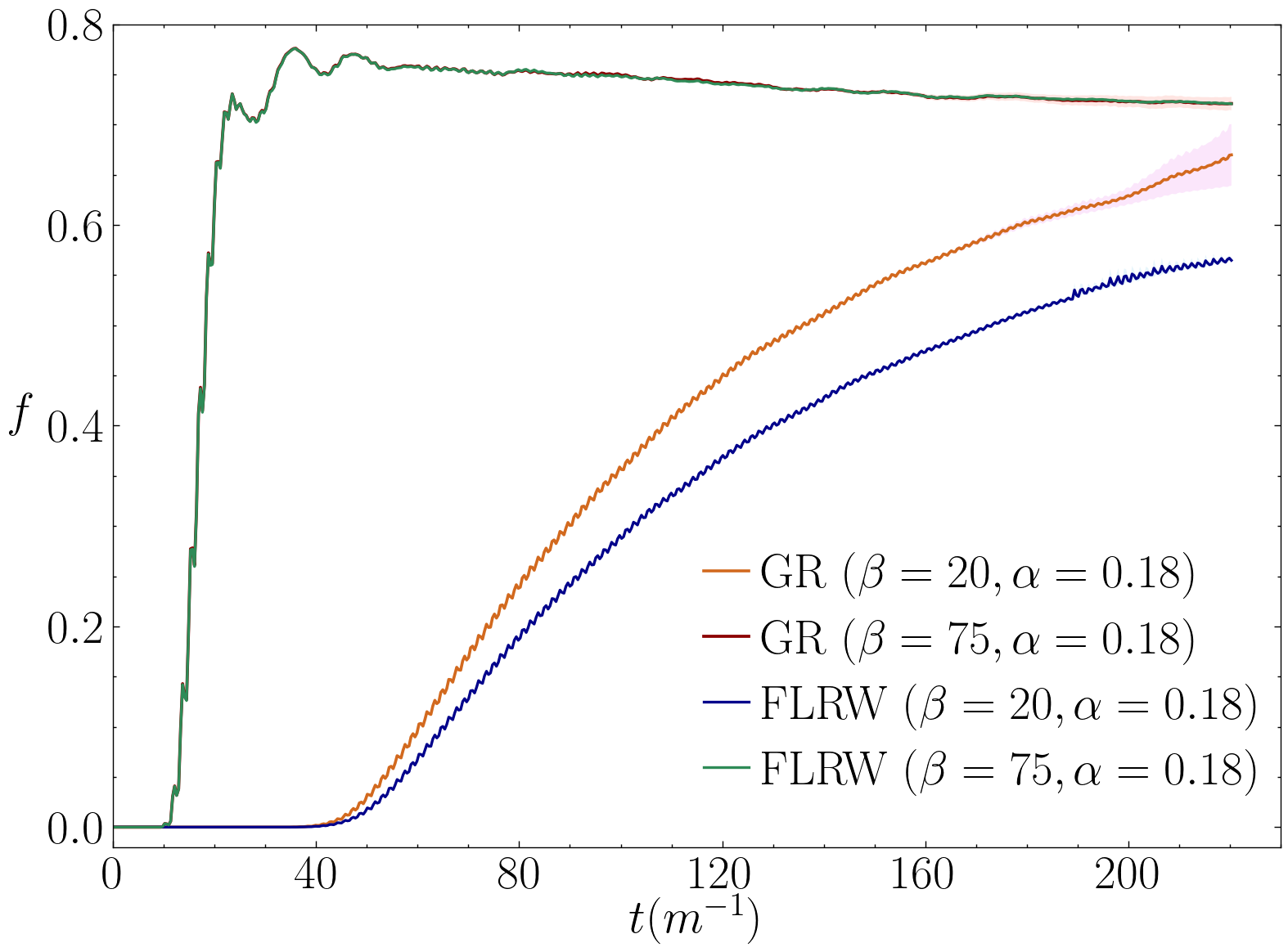}
\end{minipage}
\hfill
\begin{minipage}{0.48\textwidth}
\includegraphics[width=\textwidth]{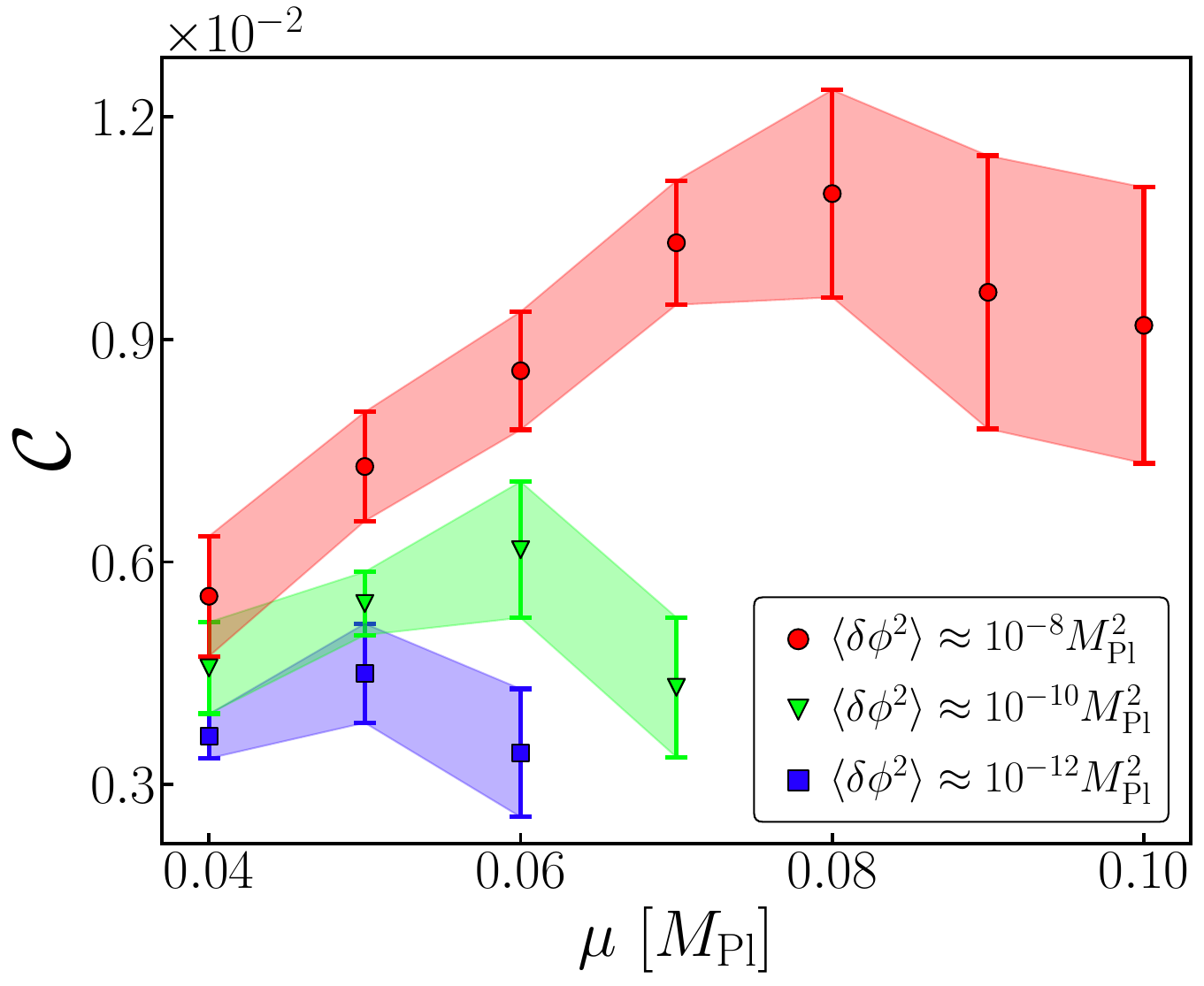}
\end{minipage}
\caption{(Left:) Comparison of the time evolution of the fractions of energy contained in oscillons between the full GR and FLRW schemes, from Kou \emph{et al.} \cite{Kou:2019bbc}. (Right:) Compactness $\mathcal{C} = GM/R$ of oscillons formed during inflationary preheating for different scales of the potential $\mu$ and initial amplitude of the fluctuations $\langle\delta\phi^2\rangle$, from Aurrekoetxea \emph{et al.} \cite{Aurrekoetxea:2023jwd}.} \label{fig:osc_frac}
\end{figure}

Kou \emph{et al.}  \cite{Kou:2019bbc}, performed NR simulations of oscillon preheating starting from a standard spectrum of the initial vacuum fluctuations $\phi$ and $\dot{\phi}=0$, with an ultraviolet cutoff to ensure convergence. They compared their results using the BSSN formulation with the adapted 1+log gauge conditions in \eqn{eq:gauge_Kbar} implemented within \texttt{CosmoGRaPH} \cite{Mertens:2015ttp} to traditional schemes that use a rigid FLRW background.
To quantify whether oscillons dominate the universe, the authors computed the energy fraction contained in oscillons as the fraction of energy in regions where the energy density is greater than twice the average
\begin{equation}
    f = \frac{\bigintssss_{\rho>2\langle\rho\rangle} dV\rho}{\int dV\rho}
\end{equation}
They found that for some parameters of the models studied, significant deviations from FLRW were observed (see the left panel of Fig. \ref{fig:osc_frac}). These were interpreted as GR corrections producing more numerous and denser oscillons, although they could also be a consequence of the gauge and slicing dependency of the quantity $f$.
The authors also suggested that these oscillons could collapse into BHs. However, this conclusion was drawn from studying the evolution of a single box-size mode with periodic boundary conditions, a setup that could favour the formation of BHs as it does not permit the oscillons to radiate.

Aurrekoetxea \emph{et al.} \cite{Aurrekoetxea:2023jwd} studied preheating with the simplest and most Planck-consistent initial conditions in both the field and velocity at the end of inflation ($\dot{\phi}\neq 0$). It was found that the assumption $\dot{\phi}=0$ (mainly the lack of a homogeneous mode, not inhomogeneities) can impact the strength of the resonance and the growth of overdensities. They focused on studying the compactness of the oscillons 
\begin{equation}
    \mathcal{C} = \frac{GM}{R}\,.
\end{equation}
Their main result is shown in Fig. \ref{fig:osc_frac} (right panel), where the compactness of the oscillons is plotted against $\mu$ for different initial amplitude of the fluctuations $\langle\delta\phi^2\rangle$. Here $\mu$ is the characteristic scale of the inflationary  potential and parameterises whether the model is small field ($\mu\ll\mpl$) or large field ($\mu\approx\mpl$). Their simulations showed that the compactness in bounded by a maximum, which is not sufficient for the oscillons to collapse into BHs. The authors also observed differences in the evolution of overdensities when comparing their results to those obtained using a rigid FLRW background scheme.

Kou \emph{et al.} \cite{Kou:2021bij} extended their simulations to extract the gravitational waves emitted during and after the formation of these oscillons. In their work, they compared the extracted density $\rho_\mathrm{GW}$ using a rigid FLRW scheme to NR simulations in different gauges and concluded that the extracted GWs are highly sensitive to the choice of gauge. In particular, the synchronous gauge was found to be a very unreliable gauge condition to be used to extract GW spectra, as it contained redundant gauge modes that artificially enhanced the GW power spectrum. The authors argued that both 1+log and the radiation gauges can provide accurate GW predictions if their parameters are suitably chosen, but further work would be needed to connect these gauge dependent results to observational quantities. 

\begin{figure}[t]
\centering
\includegraphics[width=0.65\textwidth]{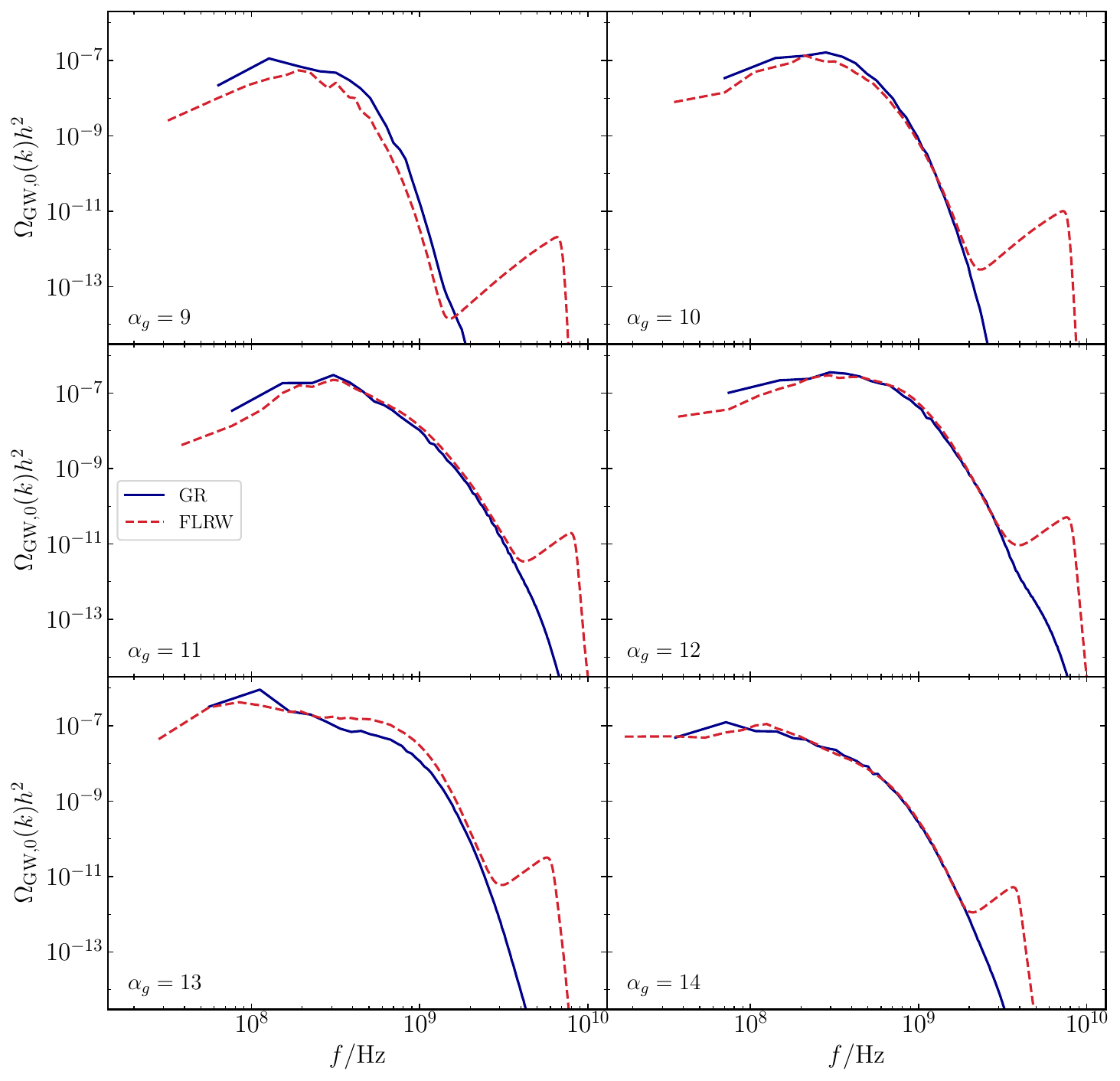}
\caption{Present-day gravitational-wave spectra extracted from simulations with different couplings $\alpha_g$. The figure, from Adshead \emph{et al.} \cite{Adshead:2023mvt}, compares the results obtained from FLRW and full GR simulations.} \label{fig:gw_bssn}
\end{figure}

Another example of a preheating process is gauge preheating, when the energy of the inflaton is transferred into gauge fields. This can be sourced via a coupling $\alpha_g\phi F_{\mu\nu} \tilde{F}^{\mu\nu}$ in the Lagrangian,
where $\alpha_g$ parameterizes the strength of the coupling and the range of momenta for which the process features a tachyonic instability.
This instability allows the full vacuum energy to be transferred within a single oscillation of the inflaton field, and the strongest effects occur when the inflaton rolls the fastest ($\dot{\phi}$ is the largest) -- typically at the end of inflation. 
In Adshead \emph{et al.} \cite{Adshead:2023mvt}, the authors showed that linearized gravity is quickly violated during this process. However, despite reaching large density contrasts $\delta\rho/\rho\approx 30$, PBHs did not form, indicating that the internal pressure of the gauge fields was playing an important role. Using a similar approach to Aurrekoetxea \emph{et al.} \cite{Aurrekoetxea:2023jwd} but based on the over-mass $\delta M$ enclosed in a region of radius $R$, they computed the compactness of the overdensities formed and found a maximum value of $\mathcal{C}\approx 0.1$. The authors compared the extracted GW spectrum from FLRW and BSSN lattice simulations finding remarkable agreement and suggesting that non-linear gravity at most had a $\mathcal{O}(1)$ effect, see Fig. \ref{fig:gw_bssn}. Finally, we note that although motivated by dark matter, the work of Alcubierre \emph{et al.} \cite{Alcubierre:2015ipa} showing the growth of scalar fluctuations on an expanding background is also relevant to this section, as is the early work on nucleosynthesis in 1D planar symmetry by Centrella \emph{et al.} \cite{CentrellaMatzner1986}, which showed that inhomogeneities affect predicted nucleosynthesis abundances.

\subsection{Phase transitions and topological defects}

As the universe expands and cools down, it can undergo phase transitions that lead to the formation of exotic objects. Among the most studied relics are topological defects \cite{Kibble:1976sj,Vilenkin:1984ib,Vilenkin:2000jqa} -- structures formed when the universe experienced symmetry breaking. The specific details of these structures depend on the topology of the vacuum. For example, if the vacuum breaks a parity symmetry, two-dimensional defects known as domain walls can form. We previously encountered such relics in Sec. \ref{sec:bubbles}, where we discussed the observational signatures in the CMB from inflationary bubble collisions. NR has also been used to study the evolution of individual bubbles during vacuum decay, such as the 1+1D work by Blanco-Pillado \emph{et al.} \cite{Blanco-Pillado:2019xny}, and Giombi \emph{et al.} \cite{Giombi:2023jqq}. In the context of the Standard Model, East \emph{et al.} \cite{East:2016anr} studied  the evolution of unstable fluctuations in the Higgs field during inflation using numerical relativity simulations in axisymmetry. Recently,  Lin \emph{et al.} \cite{Lin:2021ubu,Lin:2022ygd} have performed 3+1D NR simulations to investigate how the evolution of inhomogeneities in the early universe can populate the landscape of multiple vacua. The dynamics of the formed bubbles may serve as seeds for supermassive black holes, as studied by Huang \emph{et al.} \cite{Huang:2023mwy}, and enhance the formation of black hole binaries, as suggested by Yuwen \emph{et al.} \cite{Yuwen:2024gcf}.

If the transition breaks a U(1) symmetry, line-like defects known as cosmic strings are formed \cite{Hindmarsh:1994re,Copeland:2009ga}. The vast separation of scales involved in these phenomena already poses a challenge to their study in a rigid expanding background, as the objects quickly fall below the grid resolution. In absence of AMR techniques, it is common to use the \emph{fat-string} approximation, which introduces an artificial scale that increases the width of the string over time. While this approximation can capture the large-scale dynamics of cosmic strings, it may not accurately represent the small-scale behaviour. There is a need for more sophisticated simulation techniques to resolve these objects across different scales.

The dynamics of topological defects sources gravitational waves. If these are weak, we might observe the incoherent sum of all the events that are too weak to be directly detected -- the stochastic GW background. However, local dynamics might also source strong gravitational-wave signatures that can be directly detected. In both cases, an NR treatment may be necessary: either to validate the predictions of a weak-field treatment of the source, or to predict the detailed signatures to be expected. However, much of the current research focuses on the linearised treatment of these gravitational-wave signatures, and to the best of our knowledge, there is no simulation that fully incorporates NR in a cosmological context -- hence the brevity of this section. However, asymptotically flat NR simulations have been performed to understand the GR dynamics of these objects and construct gravitational-wave templates \cite{Laguna-Castillo:1987eab,Helfer:2018qgv,Aurrekoetxea:2020tuw,Aurrekoetxea:2023vtp,Aurrekoetxea:2022ika}.

The dynamics of bubbles after a cosmological phase transition has also recently been studied using holography. In this approach, one solves Einstein's equations in asymptotically AdS$_5$ to study the out-of-equilibrium dynamics of strongly-coupled gauge theories on the boundary  \cite{Casalderrey-Solana:2020vls,Bea:2021zol,Bea:2021zsu,Bea:2022mfb}.

\subsection{Primordial black holes}

Gravitational collapse of matter has long been a target of NR simulations, in particular the phenomenon of critical collapse uncovered by Choptuik \cite{Choptuik:1992jv} (see \cite{Gundlach:2007gc} for a review). The formation of primordial black holes (PBHs) in the early universe is a case of gravitational collapse that is of particular cosmological interest. PBHs can form through various processes in the early universe, but the primary mechanism considered is the collapse of large primordial fluctuations generated during or at the end of inflation. This is in contrast to ordinary BHs that form as the endpoint of stellar evolution. If sufficiently large fluctuations re-enter the cosmological horizon during the radiation era, they can collapse to form BHs with a wide range of masses. PBHs could be a candidate for dark matter or provide seeds for supermassive BHs. Important questions for NR include quantifying the size and scale of density perturbations that lead to the formation of a BH, and characterising their subsequent evolution, e.g. whether they accrete further matter.

An overdense region is parameterised with respect to the background density by $\delta\rho(R)\equiv \rho/\bar{\rho} - 1$. The amplitude of overdensities is commonly defined by integrating this quantity
\begin{equation}
    \delta \equiv \frac{1}{V}\int_0^{R_0} 4\pi \delta\rho R^2 dR
\end{equation}
where $V=4\pi R_0^3/3$ and $R_0$ is the radius of the overdensity. Some early analytical estimates by Carr \& Hawking \cite{Carr:1974nx,Carr:1975qj} suggested a threshold amplitude for collapse as $\delta_c\approx 1/3$, which is significantly above the amplitude of the scale-invariant density fluctuations generated during inflation that we measure on larger (CMB) scales and so requires an enhancement mechanism. Even in such scenarios, PBHs are produced from the high-amplitude tail of the probability distribution of the curvature perturbations. The threshold amplitude $\delta_c$ therefore needs to be accurately determined, since even a small error in its value can lead to a significant change in the predicted number of PBHs. NR simulations have been used to refine this estimate, and set bounds on the possible production of PBHs in different models.

One of the most common approaches to simulate the collapse of a spherically symmetric overdensity is using the approaches developed by Misner \& Sharp \cite{Misner:1964je} and May \& White \cite{PhysRev.141.1232}. The spherically symmetric line element is parameterised as 
\begin{equation}
    ds^2 = A^2(t,r) dt^2 + B^2(t,r) dr^2 + R(t,r)^2 d\Omega^2\,,
\end{equation}
where $d\Omega^2 = d\theta^2 + \sin^2\theta d\phi^2$. The operators $D_t \equiv A^{-1}\partial_t$ and $D_r \equiv B^{-1}\partial_r$, which represent derivatives with respect to proper time and radial proper distance, are introduced and applied to the areal radius function $R(t,r)$ to define the quantities
\begin{equation}
    U \equiv D_t R = \frac{1}{A}\partial_t R\,, \qquad \Gamma \equiv D_r R = \frac{1}{B}\partial_r R\,.
\end{equation}
The quantity $U$ measures the velocity of the fluid with respect to the centre of the coordinates and $\Gamma$ gives a measure of the spatial curvature. These are related to the Misner-Sharp mass $M$ as
\begin{equation}
    \Gamma^2 = 1 + U^2 - \frac{2M}{r}\,,
\end{equation}
where $M = 4\pi \int \rho R^2 dR$. The system of equations for relativistic hydrodynamics is closed after specifying an equation of state that relates the pressure and energy density of the fluid. However, one known problem of the Misner-Sharp formalism is the appearance of coordinate singularities soon after the formation of BHs, which prevents any further evolution.

The first numerical study of PBH formation was conducted by Nadezhin \emph{et al.} \cite{1978SNadezhin}. Using a similar code to that of May \& White \cite{PhysRev.141.1232} and incorporating excision to avoid the singularity, the authors explored how the formation process depended on the amplitude and profile of a deviation from FLRW. They established a local criterion for BH formation, determining the proportion of matter collapsing into the PBH and defining its boundary in terms of the \emph{visibility horizon}. Their results indicated that pressure is a significant impediment to PBH formation, so that when a PBH forms, it is much smaller than the cosmological horizon, and the accretion afterwards is negligible. Novikov \& Polnarev \cite{1980Novikov} investigated the impact of varying the equation of state of the fluid. Softer equations of state resulted in similar behaviours but with more massive PBHs that formed earlier, whereas for hard equations of state with $P\approx \rho$, they observed a discontinuity in the flow near the PBH that halted the accretion of matter. Pressure gradients caused the outer layer of matter to be ejected to infinity, and so they concluded that \emph{catastrophic accretion} (where the PBH grows at the same rate as the cosmological horizon) was impossible.  Bicknell \& Henriksen \cite{1979Bicknell} used a method of hydrodynamics characteristics to avoid the BH singularity. They showed that regions of overdensity considerably smaller than $\delta\rho/\rho\approx 1$ could collapse to PBHs with mass greater than or of the order of the horizon mass.

\begin{figure}[t]
\begin{minipage}{0.48\textwidth}
\includegraphics[width=0.96\textwidth]{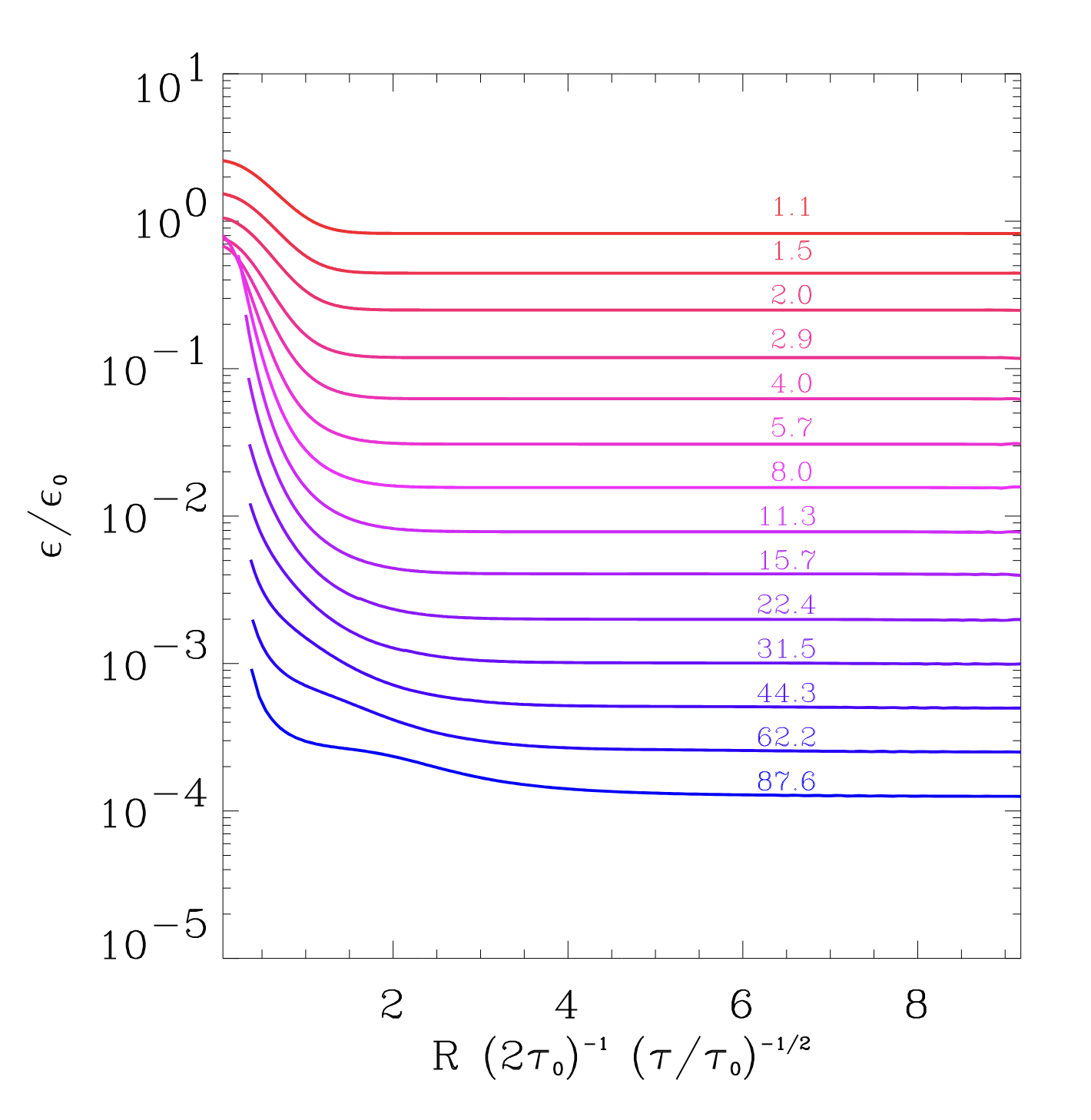}
\end{minipage}
\hfill
\begin{minipage}{0.48\textwidth}
\includegraphics[width=\textwidth]{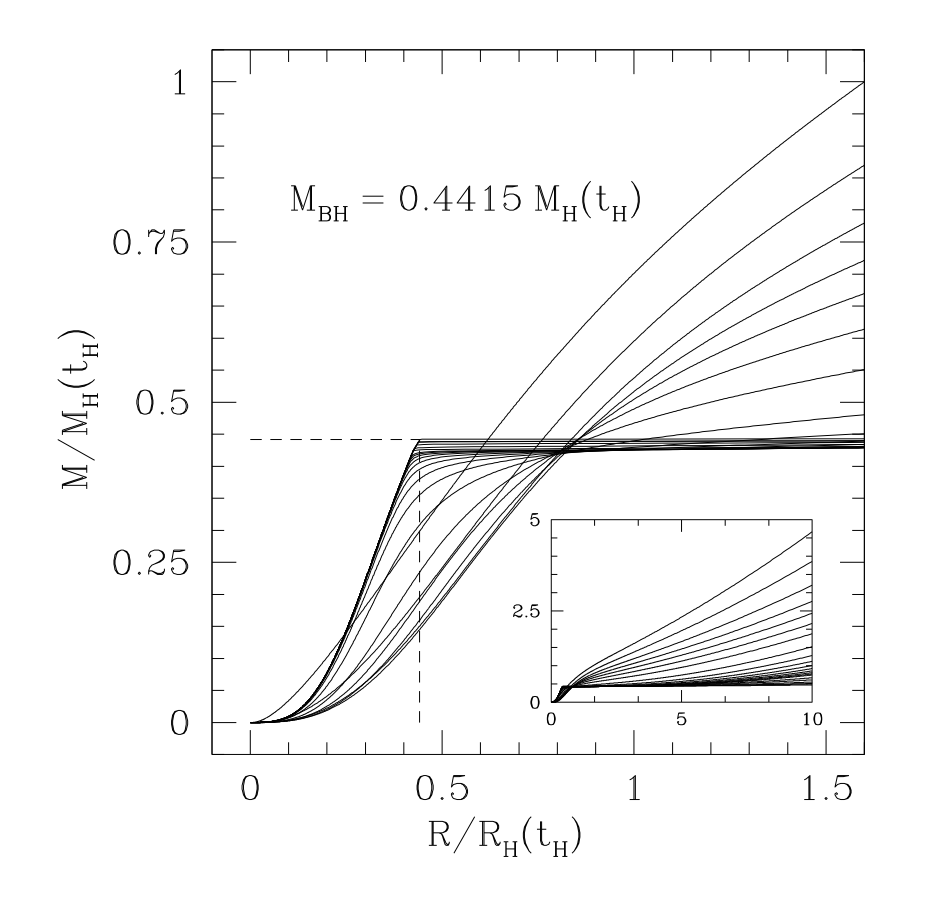}
\end{minipage}
\caption{(Left:) Time evolution of an overcritical Gaussian perturbation that forms a PBH, simulated using the the Hernandez-Misner formulation. Figure from Niemeyer \& Jedamzik \cite{Niemeyer:1999ak}.  (Right:) Figure from Musco \emph{et al.} \cite{Musco:2004ak} illustrating the evolution of the mass-energy during PBH formation using an explicit Lagrangian hydrodynamics code, in which they found $\delta_c\approx 0.4$. \label{fig:pbh}}
\end{figure}

To avoid the singularity, the Hernandez-Misner \cite{Hernandez:1966zia} version of the Misner-Sharp equations is often used, in which the time coordinate $t$ is replaced by the outgoing null coordinate $u$.
A useful comparison between the Misner-Sharp and the Hernandez-Misner system of equations can be found in Table 1 of Baumgarte \emph{et al.} \cite{1995Baumgarte} or section 2 of Musco \emph{et al.} \cite{Musco:2004ak}. A drawback of this approach is that specifying initial data on a null hypersurface is not trivial, but methods to do this have been developed \cite{1989Miller,1995Baumgarte}.
Niemeyer \& Jedamzik \cite{Niemeyer:1999ak} used this approach to investigate the collapse of horizon-sized density fluctuations to PBHs (see Fig. \ref{fig:pbh}). They found that the growth of the PBH mass due to accretion was insignificant, and identified the threshold for BH formation to be $\delta_c\approx 0.7$, higher than the commonly used value $\delta_c\approx 1/3$. Shibata \& Sasaki \cite{Shibata:1999zs} studied the problem using constant mean curvature time-slicing conditions in 1+1D, and suggested that these differences were due to the presence of an unphysical decaying mode in the initial data. 
Further work by Green \emph{et al.} \cite{Green:2004wb}, Hawke \& Stewart \cite{Hawke:2002rf} and Musco, Miller \& Rezzola \cite{Musco:2004ak,1989Miller,Miller:1994xy} confirmed this, finding that the effective $\delta$ should be in the range $0.3 < \delta_c < 0.5$ (see Fig. \ref{fig:pbh}). 

These works also considered whether the critical scaling found by Choptuik \cite{Choptuik:1992jv} in asymptotically flat spacetimes and massless scalars
\begin{equation}\label{eq:crit_coll}
    M_\mathrm{bh} = K (\delta-\delta_c)^\gamma
\end{equation}
holds for fluids and in an asymptotically FLRW spacetime, which is the relevant case for PBH formation \cite{Niemeyer:1997mt}. Niemeyer \& Jedamzik \cite{Niemeyer:1999ak} demonstrated that near-critical behaviour applies to PBH formation, extending the results of Evans \& Coleman \cite{Evans:1994pj} to an asymptotic FLRW background. Their simulations yielded the critical exponent $\gamma\approx 0.36$ and $K$ ranging from $2.7$ to $12$ depending on the shape of the initial overdensity. Hawke \& Stewart found consistent results, but noted that the scaling relation did not hold for sufficiently small $(\delta-\delta_c)$, reaching a minimum PBH mass $M_\mathrm{bh}\approx 10^{-4}M_\mathrm{h}$. Musco \emph{et al.} \cite{Musco:2004ak} also confirmed the scaling law relation and found that a positive cosmological constant leads to lower values of $\gamma$ and higher critical $\delta_c$. 

Work has continued using the Misner-Sharp and Hernandez-Misner equations to study PBH formation in 1+1D in a variety of contexts \cite{Polnarev:2006aa,Musco:2008hv,Musco:2012au,Musco:2018rwt,Musco:2021sva,Bloomfield:2015ila,Deng:2016vzb,Deng:2017uwc,Atal:2019erb,Escriva:2019nsa,Escriva:2019phb,Escriva:2021pmf,Escriva:2022bwe,Escriva:2022pnz,Escriva:2022yaf,Escriva:2023qnq,Harada:2023ffo}. 
The 3+1D NR techniques that are now standard in the simulation of BBH mergers have not yet been extensively used by the PBH community, with the exception of the early work by Shibata \& Sasaki \cite{Shibata:1999wi}, and simulations performed with \texttt{COSMOS} \cite{Yoo:2013yea,Okawa:2014nda,Uehara:2024yyp} (based on the NR code \texttt{SACRA} \cite{Yamamoto:2008js}) and \texttt{GRChombo} \cite{Andrade:2021rbd} NR codes. These techniques have been used to go beyond 1+1D simulations and study the effect that non-spherical perturbations may have on the results.

Using spheroidal super-horizon perturbations, Yoo \emph{et al.} \cite{Yoo:2020lmg} studied the effect of ellipticity on the threshold of PBH formation, finding that it is negligible in standard situations during the radiation-dominated era. Yoo later extended the work \cite{Yoo:2024lhp} by adding a tidal torque to the non-spherical profile of the initial curvature perturbation as a mechanism to generate angular momentum and form a spinning remnant, but the final spin remained consistent with zero, in agreement with expectations for the radiation-dominated era. More recently, Escriv\'a \& Yoo \cite{Escriva:2024lmm} have used 3+1 relativistic simulations to study the collapse of ellipsoidal perturbations and found that while the BH formation threshold is modified, the modification is small.  However, in the case of a softer equation of state, small deviations from spherical symmetry may evolve differently due to the absence of pressure gradient forces. De Jong \emph{et al.}  \cite{deJong:2021bbo,deJong:2023gsx} used a simple model of two scalar fields to study the mass and spin of PBHs from the collapse of perturbations with angular momentum during an early matter-dominated era $P\approx 0$. They found that although the initial mass and spin at formation can be of order $\mathcal{O}(10\%)$, the accretion of background matter post-collapse dilutes the dimensionless spin of the PBH. Hence, unless the matter era is short, they concluded that the final spin of the PBHs formed will be negligible. 

There appears to be significant scope for these studies to be extended to more general cases, using relativistic hydrodynamics and modern 3+1D NR methods.

\section{The late universe}\label{sec:late_universe}

Observations of large-scale structure in the cosmology of the late universe -- by which we broadly mean cosmological evolution past the epoch of matter-radiation equality -- provide a key probe of cosmological parameters and fundamental physics. Observations of the CMB \cite{Planck:2018vyg,ACT:2020gnv}, galaxy clustering and matter distribution \cite{DESI:2024mwx} and galaxy lensing \cite{DES:2022qpf} constrain the parameters of the $\Lambda$CDM model of cosmology, which is presently considered the standard description of our universe. 

Much of the interpretative power of these observations come from their \emph{statistics} , e.g. their 2-point and higher correlation functions. While some predictions can be made analytically, the gold standard results are derived from numerical simulations of large-scale structure using N-body/fluid codes (for a review see \cite{Angulo:2021kes}). These simulations encompass scales on which the large-scale (Hubble-Lema$\hat{i}$tre) space-time curvature plays a role, but usually make several simplifying assumptions compared to a full NR evolution. 
First, it is assumed that the matter component is non-relativistic and (for the dark matter) collisionless. In addition, since on the largest scales the universe is approximately homogeneous it is assumed that the gravitational fields of perturbations are weak. This means that, locally at least, gravitational forces are captured by Newtonian gravity, specifically the Poisson equation, 
\begin{equation}
    \nabla^2 \Phi = 4\pi G \rho~,
    \end{equation}
where $\Phi$ is the local gravitational field sourced by the matter density $\rho$. Dark matter is represented by massive point particles that are evolved using Newton's 2nd Law of motion ${\bf a} = \nabla \Phi$ where ${\bf a}$ is the acceleration. This so-called \emph{Poisson-Vlasov limit}, is employed by most of the main N-body simulation codes e.g.  \texttt{GADGET-4} \cite{Springel:2020plp} and \texttt{RAMSES} \cite{Teyssier:2001cp}. 
Since the effects of expansion are important at scales comparable to the Hubble scale $H^{-1}$ even in the Poisson-Vlasov limit, most N-body codes capture this effect by allowing the background space to homogeneously expand, following the Friedman equations of some pre-specified $\Lambda$CDM cosmology.

There are two related motivations for going beyond these assumptions in late-time cosmology with NR. 
First, to quantify the effects of the backreaction of the (possibly $\mathcal{O}(1)$) small-scale inhomogeneities on the large-scale evolution. This was initially related to the idea that small-scale inhomogeneities might somehow manifest on the largest scales as a smooth negative pressure fluid (hence mimicking a cosmological constant) \cite{Buchert:1999er,Kolb:2004am}. Part of this question involves understanding how such inhomogeneities will impact on actual observations by comoving (or in the more general case geodesic) observers. Second, higher order correlation functions beyond the power spectrum can be sourced by non-linear evolution, thus in principle any search for non-Gaussianities \cite{Desjacques:2010jw,Desjacques:2016bnm,Leistedt:2014zqa,Yoo:2012se} in the large-scale structure data must account for non-linear effects.  Even if non-linear effects are too small to be detected, the increasing size and precision of large-scale structure datasets, especially in the upcoming Stage 4 class surveys, means that \emph{all} possible \emph{systematics} must be accounted for \cite{Euclid:2020tff}. This makes an investigation of large-scale structure evolution with NR highly timely. 

There exist several codes that are half-way houses between the Poisson-Vlasov approach and full NR, which are worth highlighting since they capture many of the potential relativistic effects with fewer overheads than a full NR simulation. For example, in \texttt{gevolution} \cite{Adamek:2015eda}, the Poisson equation (which arises from the $t-t$ component of the Einstein equations in the non-relativistic weak-field limit), is extended by the full $t-t$ and $i-j$ Einstein equations up to quadratic order, to better capture the dynamics of the gravitational degrees of freedom. The first equation is an elliptic equation that captures the dynamics of the Newtonian potential, while the second captures the tensor degrees of freedom as a hyperbolic equation.  Another code is \texttt{GRAMSES} \cite{Barrera-Hinojosa:2019mzo,Barrera-Hinojosa:2020arz}, in which the metric degrees of freedom are obtained by solving the momentum and Hamiltonian constraints in-between time steps (instead of using the Poisson equation). This stratagem captures the gravitational degrees of freedom in the non-relativistic limit.  In both codes, the matter sector is modelled by collisionless particles that move according to their corresponding geodesic equations. Much of the NR work described below has used pressureless (or very low pressure) fluids to model the dark matter evolution, building on existing NR capabilities. This is not ideal since the description and imposition of a fixed equation of state breaks down once collapsed structures begin to form. For precision work, hybrid NR-N-body codes that account for kinetic pressure and permit virialisation are required. However, such studies have provided a useful first step in looking at the impact of general relativistic effects on large cosmological scales.

\subsection{Backreaction and averaging in non FLRW spacetimes}

The first systematic investigations using NR to study late-time cosmological evolution were done using the \texttt{CosmoGRaPH} code of Giblin, Mertens \& Starkman \cite{Mertens:2015ttp,Giblin:2015vwq}, and in parallel by Bentivegna \& Bruni \cite{Bentivegna:2015flc} using the Einstein Toolkit \cite{Brown:2008sb,Loffler:2011ay},  closely (and independently) followed by Macpherson, Lasky \& Price \cite{Macpherson:2016ict} also using the Einstein Toolkit.

In these works, matter was modelled as a pressureless fluid with density parameter $\rho$, and evolved via the continuity equation -- that is, matter was assumed to flow along geodesics, with zero initial coordinate velocity, such that the momentum flux $S_i=0$ identically. Both works successfully used the synchronous gauge ($\alpha=1, \beta^i = 0$), which does not completely remove issues of gauge dependence in inhomogeneous spacetimes but means that the coordinate observers track geodesics. Therefore in the limit of small perturbations one should recover the FLRW slicing in the comoving gauge.  In a later paper Giblin \emph{et al.} \cite{Giblin:2016mjp} used ray-tracing of null geodesics in the same simulations to properly quantify the effects of inhomogeneities on observables and to construct an \emph{average Hubble diagram}, and found consistent results. 

\begin{figure}[t]
\begin{minipage}{0.27\textwidth}
\includegraphics[width=\textwidth]{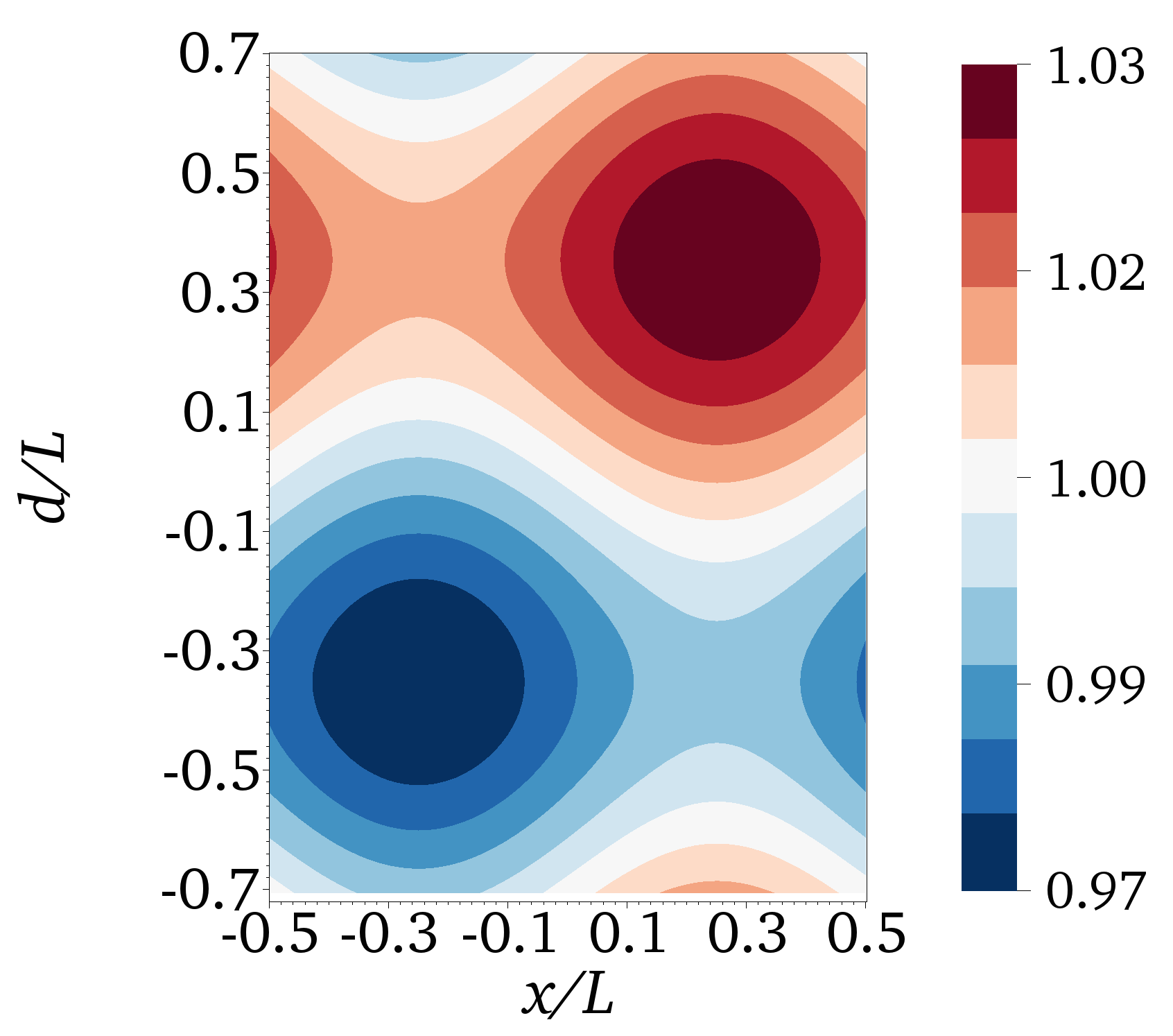}
\end{minipage}
\begin{minipage}{0.27\textwidth}
\includegraphics[width=\textwidth]{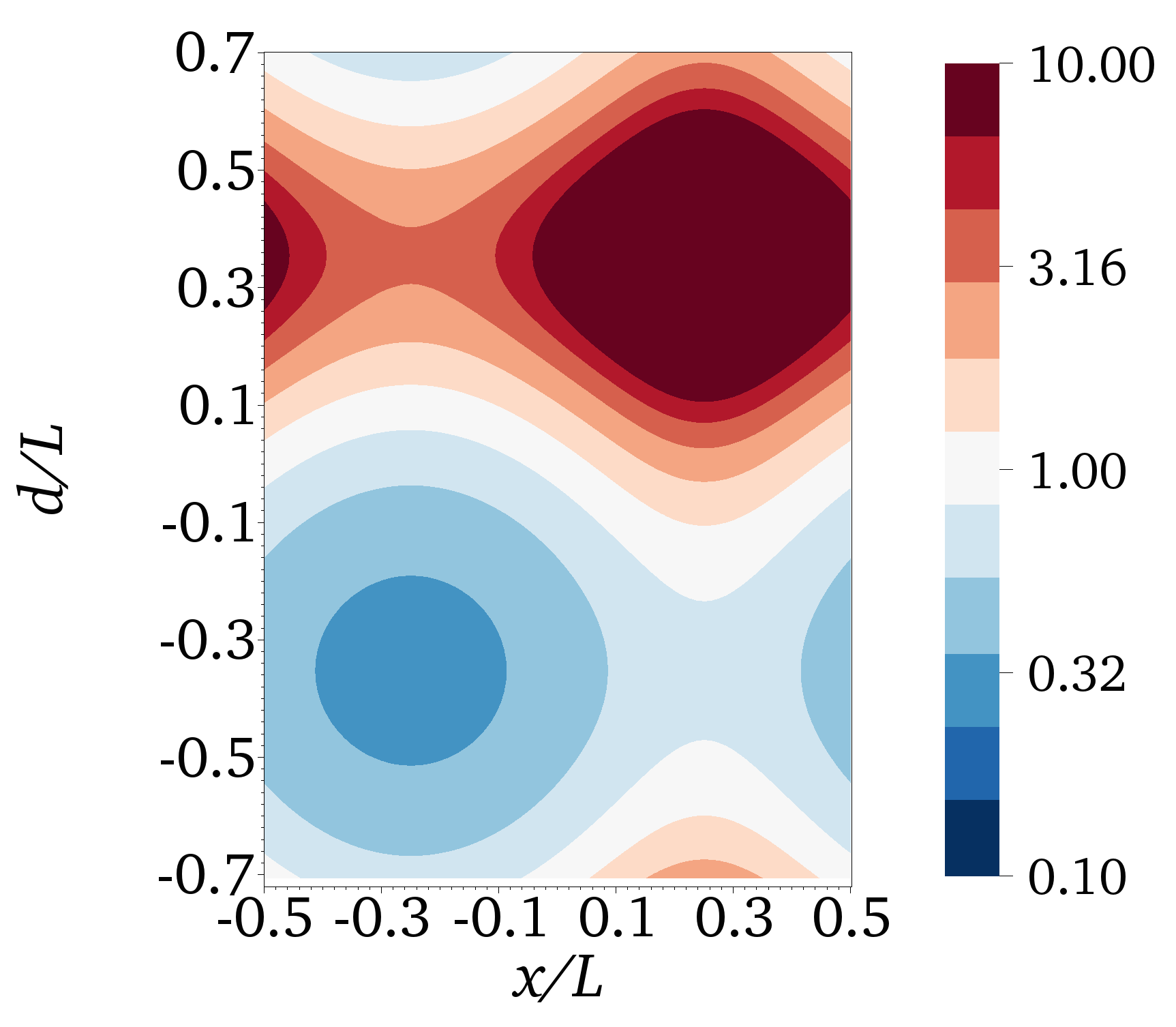}
\end{minipage}
\hfill
\begin{minipage}{0.38\textwidth}
\includegraphics[width=\textwidth]{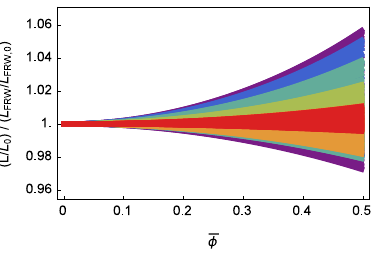}
\end{minipage}
\caption{(Left:) Profile of the inhomogeneous matter density ratio $\rho / \bar\rho$ on the $y=z$ plane at the start of the simulation and at a later time, from  Bentivegna \& Bruni \cite{Bentivegna:2015flc}.  (Right:) The ratio of the proper length to the initial length of six sets of paths in the simulation, compared with the FLRW estimate, from shorter paths (purple) to longer (red) in spectral order. Shorter paths deviate more from the FLRW expectation, up to a maximum of $\sim 5\%$ over one e-fold. From Giblin \emph{et al.} \cite{Giblin:2015vwq}. \label{fig:latetime}}
\end{figure}

In order to match numerical simulations to the FLRW quantities, the authors defined \emph{volume averaged} quantities $\langle X \rangle_S$ over the spatial hyperslices $S$ \cite{Buchert:1999er}
\begin{equation}
\langle X \rangle_S  \equiv \frac{\int_S X \sqrt{\gamma}d^3x}{\int_S  \sqrt{\gamma}d^3x}~,
\end{equation}
where the normalisation factor in the denominator is simply the spatial volume, from which we can obtain the approximate scale factor-like average
\begin{equation}
    a^3_{\cal{S}} \equiv \int_{\cal{S}} \sqrt{\gamma}d^3x 
    ~\label{eqn:average_aD}.
\end{equation}
In Giblin \emph{et al.} \cite{Giblin:2015vwq} it was shown that the averaged quantities closely tracked those of an equivalent FLRW spacetime. When perturbations (in the form of a peaked power spectrum) with variance  $\langle \delta \rho \rangle/ \langle \rho \rangle \approx 0.3 $ were introduced, they found that the deviation from FLRW, captured in the variance of the local expansion parameter $\langle \delta K \rangle / \langle K \rangle$ was of order $0.3\%$ suggesting that even in the presence of large perturbations, spacetime tracks FLRW to a good approximation.

In Bentivegna \& Bruni \cite{Bentivegna:2015flc,Bentivegna:2016stg}, the authors quantified the deviation from FLRW by computing the correction terms to the Friedman equation. Taking the 2nd time derivative of \eqn{eqn:average_aD}, yields an equation of motion that is analogous to the 2nd Friedman equation
\begin{equation}
\frac{\ddot{a}_{\cal{S}}}{a_{\cal{S}}} = -\frac{4\pi}{3}\langle \rho\rangle a_{\cal{S}}^{-3} + \frac{\cal{Q}_{\cal{S}}}{3}~,
\end{equation}
where the correction term 
\begin{equation}
{\cal{Q}_{\cal{S}}} = \frac{2}{3}\left(\langle \Theta^2 \rangle - \langle \Theta\rangle^2 - 2\langle \sigma^2 \rangle\right)~,
\end{equation}
measures the \emph{back-reaction} from non-linear evolution, with $\Theta$ the expansion of the fluid and $\sigma^2 = \sigma_{ij}\sigma^{ij}$ its shear squared.  They found that for small values of initial perturbations, ${\cal{Q}}_D \propto a_{\cal{D}}^{-1}$, which agrees with 2nd order perturbation theory calculations \cite{Li:2007ny}.
Using a simplified model (a sine wave), they checked the validity of the standard model of collapse, the \emph{spherical top-hat} model, in which a top-hat shaped overdensity is allowed to evolve under global expansion in a matter-dominated universe. The competition between infall and expansion leads to a moment of \emph{turn around}  when the gravitational collapse overcomes the local expansion. The critical density at the turn-around time $t_{TA}$ is $\delta_{OD} \equiv \delta \rho/\rho$. They found that $\delta_{OD}$ differed from the estimate of $\delta_{OD}\approx 1.06$, due to the presence of a growing and decaying mode in the initial data (because of the zero velocity assumption, as pointed out by East \emph{et al.} \cite{East:2017qmk}). A later work by Munoz \& Bruni \cite{Munoz:2023rwh} resolved this difference by fixing the initial data, finding $\delta_{OD} = 1.055$ in good agreement with the semi-analytic results, thus validating the standard collapse model. 

The main results of these two works were subsequently confirmed by Macpherson \emph{et al.} \cite{Macpherson:2016ict,Macpherson:2018btl}, who also used the Einstein Toolkit but incorporated the \texttt{GRHydro} thorn for more general matter (allowing the addition of a small pressure to the dust model) and developed a new \texttt{FLRWSolver} thorn to generate the initial data, along with separate diagnostics tools. In this and later works  \cite{Macpherson:2018btl,Macpherson:2018akp} Macpherson \emph{et al.} simulated late-time structure formation in a matter-dominated universe, using a realistic matter power spectrum of perturbations derived from observations of the CMB as initial conditions. Due to limited resolution, they chose to simulate cosmological scales ranging from $k\sim 10^{-2}~\mathrm{Mpc}^{-1}$ to $k\sim 1~\mathrm{Mpc}^{-1}$. By implementing an averaging scheme which depends on the scale $r_D$, they found that while large-scale spatial homogeneity and isotropy is maintained, at scales $r_D\leq 180h^{-1}~\mathrm{Mpc}$, which correspond to local galaxy survey scales, deviation from homogeneity could be up to $6\%$. Nonetheless, this deviation only affects the measured value of the Hubble constant at the sub-percent level, and so is not sufficient on its own to resolve the current tension in its value, although in Macpherson \cite{Macpherson:2024zwu} the value is found to be more significant using ray tracing instead of spatial averages. A study that added tensor perturbations was carried out by Wang \cite{Wang:2018qfr}.

The codes used in these works undertook a comparison exercise \cite{Adamek:2020jmr} together with \texttt{gevolution} and \texttt{GRAMSES}. All codes ran a simulation with identical initial conditions (a plane wave) and similar resolutions and compared the paths of two null, spatially separated trajectories, launched from the initial hyperslice.
As the null trajectory is sensitive the entire spacetime metric, this serves as a good test of equivalence in relativistic cases. All codes showed excellent agreement (within $0.03\%$).

\begin{figure}[t]
\begin{minipage}{0.48\textwidth}
\includegraphics[width=\textwidth]{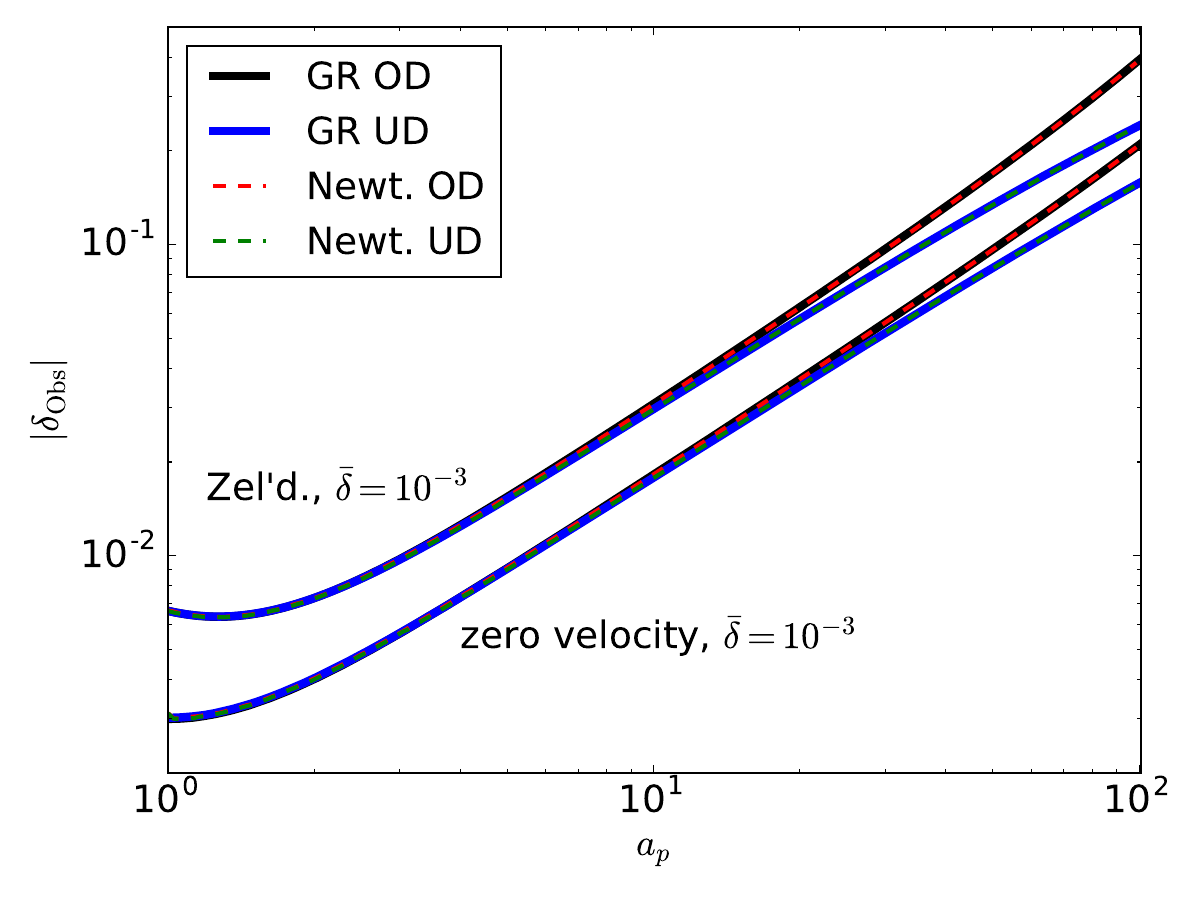}
\end{minipage}
\hfill
\begin{minipage}{0.48\textwidth}
\includegraphics[width=\textwidth]{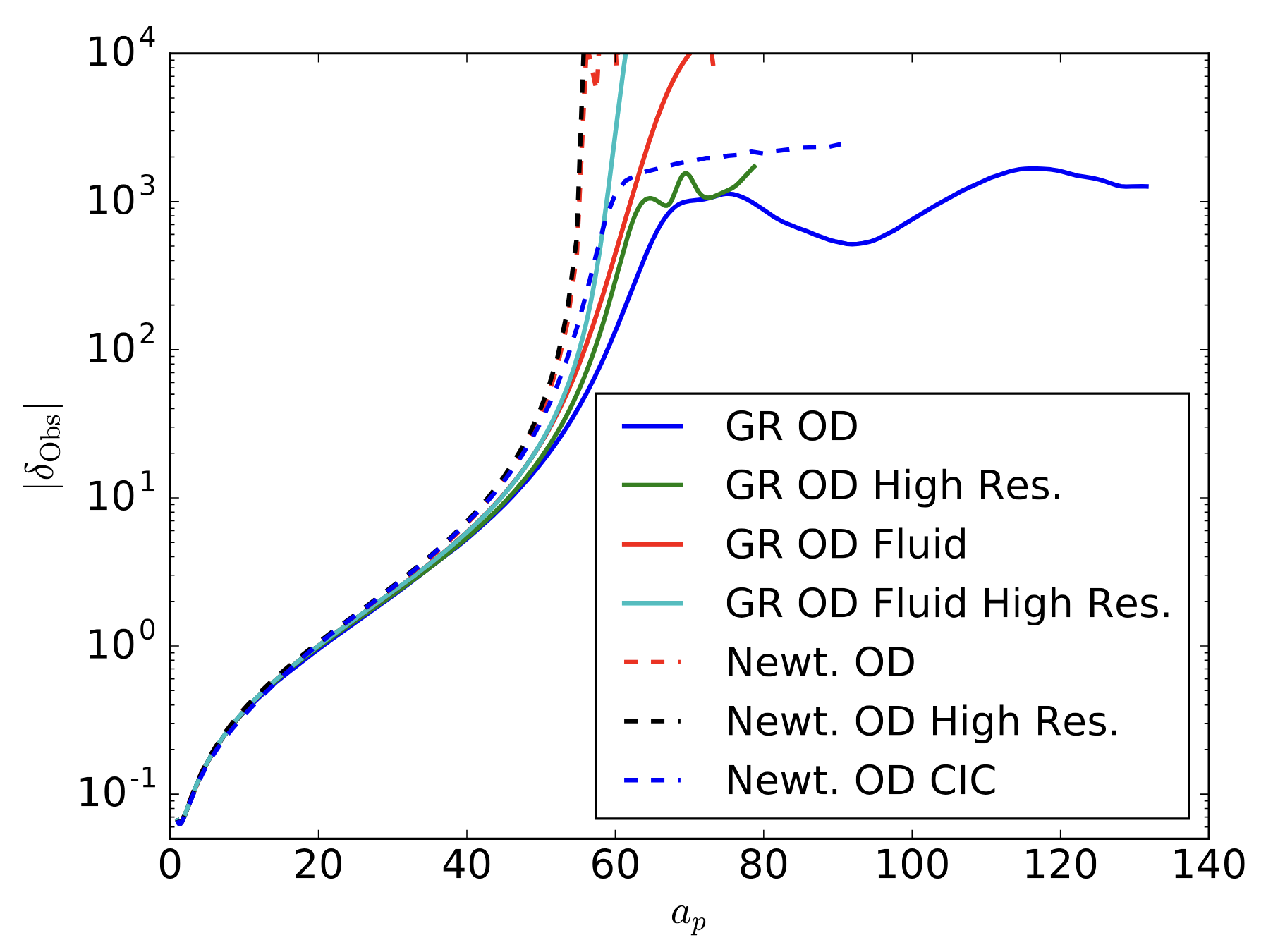}
\end{minipage}
\caption{(Left:) The density contrast as measured by an observer comoving with matter at the point of maximum overdensity (OD) or underdensity (UD) for two cases and different initial velocity profiles. The GR (solid lines) and Newtonian (dotted lines) calculations closely track each other. Using fluid simulations in GR, from East \emph{et al.} \cite{East:2017qmk}.  (Right:) The same results but for all different treatments including the Einstein-Vlassov system (blue and green lines), again showing good agreement between the Newtonian and GR cases. From East \emph{et al.} \cite{East:2019chx}. \label{fig:latetime-East}}
\end{figure}

A direct comparison between NR and N-body simulations was carried out by East \emph{et al.} \cite{East:2017qmk}. In this work, a full hydrodynamical treatment was used to model the collisionless dark matter \cite{East:2011aa}, while the metric variables were evolved using the GHC formulation \cite{East:2015ggf}.  Meanwhile, \texttt{GADGET-2} \cite{Springel:2005mi} was used to generate results for a non-relativistic N-body simulation for comparison to the NR results.
Instead of computing deviations from linear theory using averaged quantities, the authors chose a set of \emph{fiducial observers} that were co-moving with the matter, and tracked the evolution of $\delta \rho$ over their worldlines.  They compared their results to both the Poisson-Vlasov results and expectations from linear perturbation theory. Furthermore, they solved for the geodesics of null rays, and therefore allowing them to compute the cosmological distance as a function of redshift for free-falling observers located along the photon paths. They found that, as long as the Newtonian potential remains in the weak-field limit, the difference between the NR simulations to both the N-body simulations and linear theory expectation is at the sub-percent level, see Fig. \ref{fig:latetime-East}. Their results were also found to be consistent with the spherical top-hat model of collapse.

Where conservative hydrodynamical equations are used to model the dark matter, as in East \emph{et al.} \cite{East:2017qmk}, direct comparison between the fluid picture employed in the NR simulations and the particle picture in the N-body simulations become unreliable once shell-crossing occurs, as the virialisation process cannot be captured in the fluid description. 
To overcome this, one can introduce the \emph{Einstein-Vlasov} formalism, in which matter is modelled as a distribution of $N$ collisionless point-like particles. 
In this formalism, the stress-tensor is given by the sum of all the energy of all particles labelled by $i$ 
\begin{equation}
    T^{\mu\nu} = \sum_{i=1}^{i=N}\epsilon_{i} l^{\mu}_il^{\nu}_i~, \label{eqn:einsteinvlasov_ST}
\end{equation}
where $\epsilon_i$ is the energy density of the particle, and $l^{\mu}_i$ is the tangent to geodesic curve of the particle $x_i^{\mu}(\tau)$. For particles of mass $m$, $l_i^{\mu}l_i{}_{\mu} = -m^2$. The particles follow the geodesic equation as usual,
\begin{equation}
    l_i^{\mu}\nabla_{\mu}l_i^{\nu} = 0~.
\end{equation}
An averaging prescription is then implemented to coarse-grain \eqn{eqn:einsteinvlasov_ST} into an average value per grid point. 
This was first implemented by Pretorius \& East \cite{Pretorius:2018lfb} to study the collapse of collisionless massless particles and extended to massive particles in a later work \cite{East:2019chx}. In parallel, Giblin \emph{et al.} \cite{Giblin:2018ndw} implemented this formalism into the \texttt{CosmoGRaPH} framework for collisionless massive particles, and more recently by Magnall \emph{et al.} \cite{Magnall:2023tzm} using the \texttt{phantom} SPH code. 

The Einstein-Vlasov framework allows one to directly compare along the trajectories of particles beyond shell crossing.
As in the earlier work with fluids \cite{East:2017qmk}, these works found that the Einstein-Vlasov formalism agrees with the Poisson-Vlasov formalism up to sub-percent level, as long as the Newtonian potential does not exceed the weak-field approximation $|\Psi_N| \leq 0.1$. In East \emph{et al.} \cite{East:2019chx}, setting up identical initial conditions in both formalisms allowed them to explicitly compare the trajectories, and they found that as long as the evolution remained non-relativistic (in the Einstein-Vlasov picture), they were in excellent agreement with each other. Interestingly, they found that in the non-linear limit, the collapse of overdensities occurred faster in the Newtonian picture when compared to the GR one, although even then differences remained small, see Fig. \ref{fig:latetime-East}.

The use of NR to probe the dynamics of expanding universes and the impact of backreaction in inhomogeneous cases has also been tackled from the perspective of a lattice of BHs. Roughly speaking, if the scale of the BHs is much shorter than the curvature scale of the universe $M_{BH}\ll H^{-1}$, then a lattice of static BHs mimics (at the largest scales) that of cold dark matter (see e.g. \cite{Green:2014aga}), which can drive cosmic expansion. Thus such a spacetime is an interesting \emph{toy model}, with which one can study the physics of dark matter and the role of averaging in FLRW, at least on the largest scales. This was first proposed by Linquist and Wheeler \cite{RevModPhys.29.432}, and then numerically simulated in several works \cite{Yoo:2012jz,Bentivegna:2012ei,Bentivegna:2013jta,Bentivegna:2016fls,Yoo:2014boa,Durk:2016yja,Yoo:2018pda}, see Bentivegna \emph{et al.} \cite{Bentivegna:2018koh} for a review. These simulations have studied effects such as the \emph{dressing} (differences from the naive superposition of constituents) of mass and other physical parameters, and the biases in optical measurements that inhomogeneous gravitational fields can produce. They find that as the number of BHs increase, backreaction decreases.

Other notable work on averaging focusing on a confirmation of Isaacson’s formula (that gravitational waves behave as a radiation fluid on average) was performed by Ikeda \emph{et al.} \cite{Ikeda:2015hqa}.

\subsection{Cosmological observables in the non-linear regime}

The results  of \cite{East:2019chx} described in the previous section showed that the Poisson-Vlasov formalism is remarkably accurate in capturing most of the relevant gravitational effects in late-time cosmological evolution, \emph{even in the non-linear regime}.  Significant deviations from this approximate formalism as compared to a fully relativistic treatment do not occur until deep in the strong gravity regime -- expected at very short wavelength scales when local regions \emph{decouple} fully from the Hubble flow. Indeed, these results indicate that scale-separation is robust in cosmological evolution -- non-linearity of GR does not induce UV-IR couplings where UV modes significantly affect large-scale cosmological evolution, at least not sufficiently to explain the observed acceleration of the expansion of the universe. However, they cannot entirely rule out percent level effects that may be be relevant for future precision cosmological observations, for example galaxy surveys.

NR still has a role to play both in terms of understanding the systematics of $N$-body predictions and as a useful cross-check. The first such cross-check was undertaken by Giblin \emph{et al.} \cite{Giblin:2017ezj} where they computed the weak lensing power spectrum using NR. Weak lensing power computations often rely on approximations \cite{Lewis:2006fu}. In particular, line-of-sight integrals often rely on the so-called \emph{Born approximation}, where the light ray is assumed to propagate along the background geodesic and is not perturbed by higher order gravitational effects. Attempts to incorporate \emph{post-Born} effects has led to disagreement\footnote{Although \cite{Su:2014mga,Denton-Turner:2021zce} argued that the disagreement is not related to the Born approximation (which still applies) but a disagreement in the labeling of what constitutes lens-lens couplings.} \cite{Pratten:2016dsm,Marozzi:2016uob}. Giblin \emph{et al.} \cite{Giblin:2017ezj} attempted to resolve this issue by  computing the weak lensing power spectrum by ray-tracing a universe evolved using NR. They found that the approximation may lead to percent level error at angular scales of $l=20\sim 30$ when comparing full NR to perturbation theory computations, potentially signalling important systematics not captured in higher order perturbation theory. More recent efforts to develop tools to probe cosmological distances using ray tracing have been carried out \cite{Macpherson:2022eve,Grasso:2021iwq}, and support the conclusion that effects may be at percent level for scales above 100 Mpc.

\begin{figure}[t]
\includegraphics[width=\textwidth]{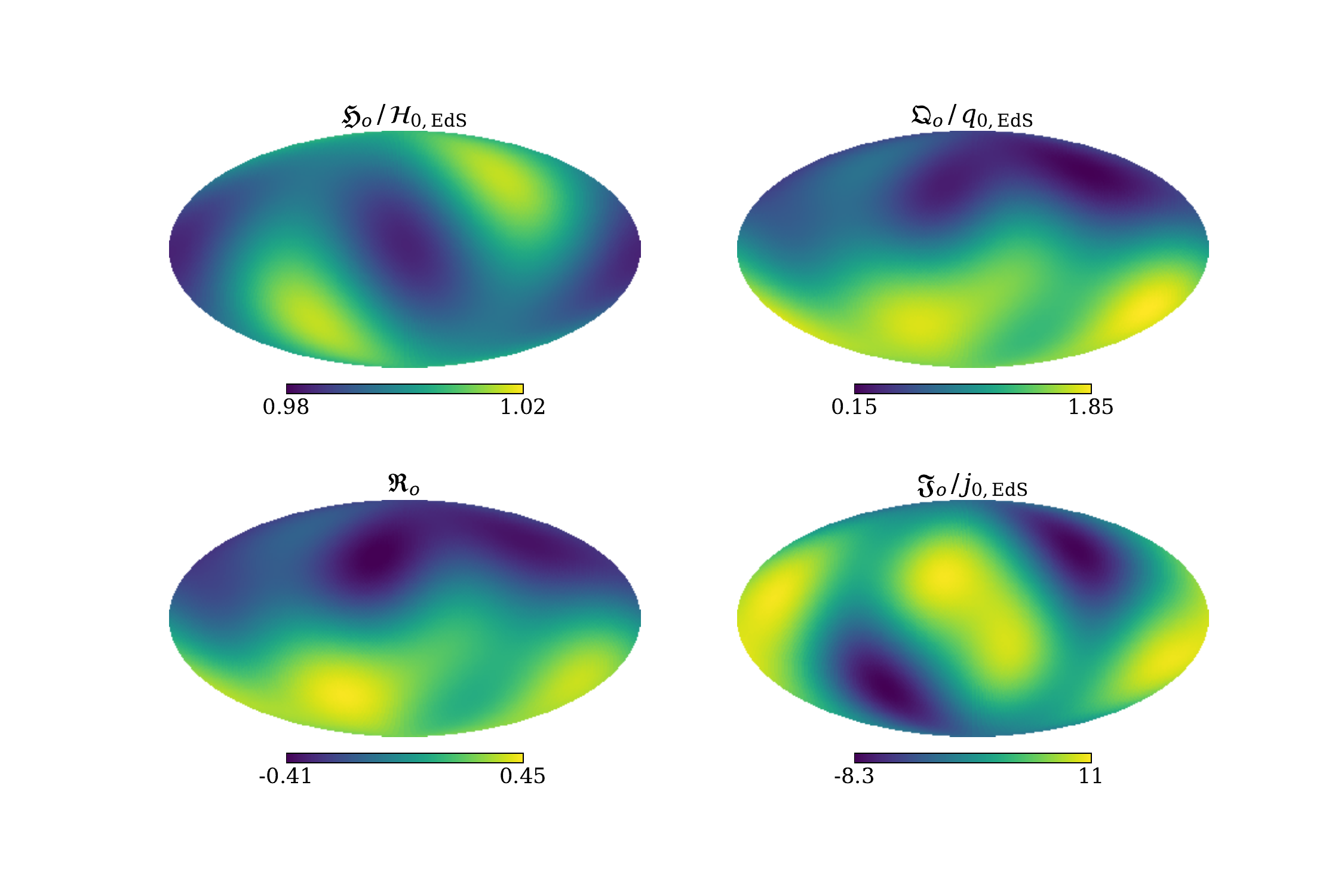}
\caption{Sky-maps of the effective Hubble, deceleration, curvature, and jerk parameters (top-left to bottom-right, respectively) for an observer. Each parameter (with the exception of the curvature) is normalised by its respective EdS value. From Macpherson \& Heinesen \cite{Macpherson:2021gbh}. \label{fig:latetime-observable}}
\end{figure}

Macpherson \& Heinesen \cite{Macpherson:2021gbh} studied the anisotropic effects due to small-scale structure, from initial data that were statistically homogeneous and isotropic, and computed observables such as the Hubble parameter and its higher moments with different levels of sky coverage. They found up to percent level cosmic variance in the measurements. For example, there is a $7\%$ probability that an observer can measure up to a $3\%$ difference in the Hubble parameter of a typical Einstein-de-Sitter universe, see Fig. \ref{fig:latetime-observable}. In related works, Macpherson \cite{Macpherson:2024zwu} and Williams \emph{et al.} \cite{Williams:2024vlv} have studied the impact of nearby inhomogeneities and voids on observers' inferences of the Hubble constant, where they found that the voids are in general negatively curved and void statistics such as size distribution roughly follows that of $N$-body simulations.  Koksbang \emph{et al.} \cite{Koksbang:2024xfr} extended this by computing the \emph{red-shift drift} -- where the redshift of an object evolves through time due to cosmic expansion. They showed that NR predicts that at redshift $z=0.1$, fluctuations in the redshift drift due to inhomogeneities are at the $\sim 10\%$ level, and so potentially observable in the next generation spectroscopic sky surveys. Another observable of interest is frame dragging, arising from cosmological vector modes caused in the late universe by non-linearities. Frame dragging is the leading-order correction in a post-Newtonian formalism and is expected to be generated after the first shell-crossing in nonlinear structure formation, even in the weak-field regime \cite{Thomas:2015kua,Bruni:2013mua,Jelic-Cizmek:2018gdp,Zheng:2022azi}. Some initial work using the relativistic codes \texttt{GRAMSES} and \texttt{gevolution} has been done in this direction \cite{Adamek:2015eda,Barrera-Hinojosa:2021msx,Barrera-Hinojosa:2020gnx} and the first steps using NR are taken in \cite{Munoz:2022duf,Munoz:2023rwh}.

Another use of NR has been to probe the question of cosmological spatial curvature $\Omega_k$ -- how spatially flat is the universe -- in an inhomogeneous universe, as in Tian \emph{et al.} \cite{Tian:2020qnm}. Since curvature perturbations violate FLRW symmetry, averaged quantities over a spatial hyperslice do not necessarily correspond to the true local spatial curvature. The authors found that these inhomogeneities introduce an effective $\Omega_k$ of the order of the perturbations, in general agreement with analytic arguments. This has recently been extended by Oestreicher \& Koksbang \cite{Oestreicher:2024mtw} who found more significant curvature perturbations of order 10\%.

\section{Closing remarks}\label{sec:discussion}

Cosmologists seem to have struck lucky.  They profit from the simplicity of the FLRW model, due to the approximate homogeneity and isotropy of our observable universe.
Beyond this realm, where the light of the CMB peters out and gravity may be the only messenger, there is no guarantee that the FLRW model still holds. To explore this regime we are forced to rely on new methods, and NR provides a particularly useful tool where dynamical strong gravity effects are at play. The use of NR introduces conceptual complications, challenges to implementation, and a potentially high cost per simulation. But used efficiently it can (and already has) shed light on parts of the Universe that are dark to us.

In this review, our aim is twofold. First, we provide a starting point for both cosmologists and numerical relativists who are interested in using NR in the context of cosmology. While much of the specialised NR knowledge developed for BH spacetimes carries over to cosmology, the mapping is not perfect and there is an opportunity for new techniques to be developed. Finding physically well-motivated constraint-obeying initial conditions and developing gauge conditions adapted to cosmological spacetimes far from the FLRW universe are two good examples of areas that deserve further study.

Second, we give a comprehensive overview of cosmological simulations that have used NR, covering a range of work: from tentative steps to explore the unknown darkness beyond the observable universe, to important work checking GR backreaction effects in numerical simulations of the late universe. There is much more work to be done, and many areas where we have barely scratched the surface -- exploring cosmology with modified gravity, understanding string cosmology, extending gravitating cosmic strings simulations to networks, to name a few. 

We hope that this Living Review will encourage cosmologists and numerical relativists alike to break new ground, and we look forward to expanding this Review in the future with the fruits of their labour.

\section{Summary table and open questions}\label{sec:summary_tables}

A summary table of the work that has been completed in the various areas covered by the review is provided in Tab. \ref{tab-summary} for reference.  We also provide here an (incomplete) list of open questions and outstanding directions that have been highlighted by the review:\\

\begin{table}[t]
\resizebox{\columnwidth}{!}{%
\begin{tabular}{c|>{\centering\arraybackslash}p{1.5cm}>{\centering\arraybackslash}p{1.5cm}|>{\centering\arraybackslash}p{1.5cm}>{\centering\arraybackslash}p{1.5cm}|>{\centering\arraybackslash}p{1.5cm}>{\centering\arraybackslash}p{1.5cm}|>{\centering\arraybackslash}p{1cm}>{\centering\arraybackslash}p{1cm}>{\centering\arraybackslash}p{1cm}|}
\hhline{~---------}
 &
  \multicolumn{2}{c|}{Dimension} &
  \multicolumn{2}{c|}{Boundaries} &
  \multicolumn{2}{c|}{Matter type} &
  \multicolumn{3}{c|}{GR formulation} \\ \hhline{~---------}
 &
  \multicolumn{1}{c|}{No 3+1D} &
  \multicolumn{1}{c|}{3+1D} &
  \multicolumn{1}{c|}{Periodic} &
  \multicolumn{1}{c|}{FLRW} &
  \multicolumn{1}{c|}{Fluids} &
  \multicolumn{1}{c|}{Fund. fields} &
  \multicolumn{1}{c|}{BSSN/CCZ4} &
  \multicolumn{1}{c|}{GHC} &
  \multicolumn{1}{c|}{Other} \\ \hhline{-=========}
\multicolumn{1}{|c|}{Singularities} &
  \sinA & \multicolumn{1}{|>{\centering\arraybackslash}p{1.5cm}|}{\sinB} &
  \sinC & \multicolumn{1}{|>{\centering\arraybackslash}p{1.5cm}|}{\sinD} &
  \sinE & \multicolumn{1}{|>{\centering\arraybackslash}p{1.5cm}|}{\sinF} &
  \multicolumn{1}{>{\centering\arraybackslash}p{1.5cm}}{\sinG} & \multicolumn{1}{|>{\centering\arraybackslash}p{1.5cm}|}{\sinH} & \multicolumn{1}{>{\centering\arraybackslash}p{1.5cm}|}{\sinI} \\ \hline
\multicolumn{1}{|c|}{Inflation} &
  \infA & \multicolumn{1}{|>{\centering\arraybackslash}p{1.5cm}|}{\infB} &
  \infC & \multicolumn{1}{|>{\centering\arraybackslash}p{1.5cm}|}{\infD} &
  \infE & \multicolumn{1}{|>{\centering\arraybackslash}p{1.5cm}|}{\infF} &
  \multicolumn{1}{>{\centering\arraybackslash}p{1.5cm}}{\infG} & \multicolumn{1}{|>{\centering\arraybackslash}p{1.5cm}|}{\infH} & \multicolumn{1}{>{\centering\arraybackslash}p{1.5cm}|}{\infI} \\ \hline
\multicolumn{1}{|c|}{Slow contraction/Bounces} &
  \bouA & \multicolumn{1}{|>{\centering\arraybackslash}p{1.5cm}|}{\bouB} &
  \bouC & \multicolumn{1}{|>{\centering\arraybackslash}p{1.5cm}|}{\bouD} &
  \bouE & \multicolumn{1}{|>{\centering\arraybackslash}p{1.5cm}|}{\bouF} &
  \multicolumn{1}{>{\centering\arraybackslash}p{1.5cm}}{\bouG} & \multicolumn{1}{|>{\centering\arraybackslash}p{1.5cm}|}{\bouH} & \multicolumn{1}{>{\centering\arraybackslash}p{1.5cm}|}{\bouI} \\ \hline
\multicolumn{1}{|c|}{Bubble collisions} &
  \bubA & \multicolumn{1}{|>{\centering\arraybackslash}p{1.5cm}|}{\bubB} &
  \bubC & \multicolumn{1}{|>{\centering\arraybackslash}p{1.5cm}|}{\bubD} &
  \bubE & \multicolumn{1}{|>{\centering\arraybackslash}p{1.5cm}|}{\bubF} &
  \multicolumn{1}{>{\centering\arraybackslash}p{1.5cm}}{\bubG} & \multicolumn{1}{|>{\centering\arraybackslash}p{1.5cm}|}{\bubH} & \multicolumn{1}{>{\centering\arraybackslash}p{1.5cm}|}{\bubI} \\ 
\hhline{==========}
\multicolumn{1}{|c|}{Reheating} &
  \rehA & \multicolumn{1}{|>{\centering\arraybackslash}p{1.5cm}|}{\rehB} &
  \rehC & \multicolumn{1}{|>{\centering\arraybackslash}p{1.5cm}|}{\rehD} &
  \rehE & \multicolumn{1}{|>{\centering\arraybackslash}p{1.5cm}|}{\rehF} &
  \multicolumn{1}{>{\centering\arraybackslash}p{1.5cm}}{\rehG} & \multicolumn{1}{|>{\centering\arraybackslash}p{1.5cm}|}{\rehH} & \multicolumn{1}{>{\centering\arraybackslash}p{1.5cm}|}{\rehI} \\ \hline
\multicolumn{1}{|c|}{Phase transitions} &
  \phaA & \multicolumn{1}{|>{\centering\arraybackslash}p{1.5cm}|}{\phaB} &
  \phaC & \multicolumn{1}{|>{\centering\arraybackslash}p{1.5cm}|}{\phaD} &
  \phaE & \multicolumn{1}{|>{\centering\arraybackslash}p{1.5cm}|}{\phaF} &
  \multicolumn{1}{>{\centering\arraybackslash}p{1.5cm}}{\phaG} & \multicolumn{1}{|>{\centering\arraybackslash}p{1.5cm}|}{\phaH} & \multicolumn{1}{>{\centering\arraybackslash}p{1.5cm}|}{\phaI} \\ \hline
\multicolumn{1}{|c|}{Primordial BH} &
  \pbhA & \multicolumn{1}{|>{\centering\arraybackslash}p{1.5cm}|}{\pbhB} &
  \pbhC & \multicolumn{1}{|>{\centering\arraybackslash}p{1.5cm}|}{\pbhD} &
  \pbhE & \multicolumn{1}{|>{\centering\arraybackslash}p{1.5cm}|}{\pbhF} &
  \multicolumn{1}{>{\centering\arraybackslash}p{1.5cm}}{\pbhG} & \multicolumn{1}{|>{\centering\arraybackslash}p{1.5cm}|}{\pbhH} & \multicolumn{1}{>{\centering\arraybackslash}p{1.5cm}|}{\pbhI} \\
  \hhline{==========}
\multicolumn{1}{|c|}{Backreaction} &
  \bacA & \multicolumn{1}{|>{\centering\arraybackslash}p{1.5cm}|}{\bacB} &
  \bacC & \multicolumn{1}{|>{\centering\arraybackslash}p{1.5cm}|}{\bacD} &
  \bacE & \multicolumn{1}{|>{\centering\arraybackslash}p{1.5cm}|}{\bacF} &
  \multicolumn{1}{>{\centering\arraybackslash}p{1.5cm}}{\bacG} & \multicolumn{1}{|>{\centering\arraybackslash}p{1.5cm}|}{\bacH} & \multicolumn{1}{>{\centering\arraybackslash}p{1.5cm}|}{\bacI} \\ \hline
\multicolumn{1}{|c|}{Observables} &
  \obsA & \multicolumn{1}{|>{\centering\arraybackslash}p{1.5cm}|}{\obsB} &
  \obsC & \multicolumn{1}{|>{\centering\arraybackslash}p{1.5cm}|}{\obsD} &
  \obsE & \multicolumn{1}{|>{\centering\arraybackslash}p{1.5cm}|}{\obsF} &
  \multicolumn{1}{>{\centering\arraybackslash}p{1.5cm}}{\obsG} & \multicolumn{1}{|>{\centering\arraybackslash}p{1.5cm}|}{\obsH} & \multicolumn{1}{>{\centering\arraybackslash}p{1.5cm}|}{\obsI} \\ \hline
\end{tabular}%
}
\vspace{10pt}
\caption{Summary of the work in this review.}
\label{tab-summary}
\end{table}

\noindent \textbf{NR techniques for cosmology}
\begin{enumerate}
    \item What are the optimum boundary and gauge conditions for cosmological simulations, and how do they depend on the physics being studied?
    \item How can we access a wider range of physical scales or make approximations to allow us to study GR effects across scales, e.g. in string networks?
    \item What are the best diagnostics for inhomogeneous spacetimes? Can spatial averages be trusted or do we always need to use geodesic observers? If so, how should they be positioned to sample the spacetime?
    \item How can we accurately quantify GWs or other stochastic background observables in cosmological spacetimes?
    \item How can we construct initial data that is physically ``correct'', as well as constraint satsfying? To what extent do arbitrary choices like conformal flatness limit the physics we can study? Can we go beyond these?
    \item Can we develop open source tools that are general purpose for different cosmological spacetimes, e.g. a general initial condition solver for 3+1D periodic spacetimes?
\end{enumerate}

\noindent \textbf{The early universe}
\begin{enumerate}
    \item Does the BKL conjecture continue to hold for asymptotically flat spacetimes?
    \item What is the role of periodic boundary conditions in findings about the robustness of inflation simulations? Do they provide a bias towards inflation?
    \item What is the impact of considering multi-field models on the robustness of inflation?
    \item How can the smooth contraction phase connect to the bounce in Ekpyrosis scenarios?
    \item What are the observable consequences of either inflation or bouncing scenarios in the CMB or other observations like GW backgrounds?
    \item{Do modifications to gravity at higher energies change these behaviours? Can we connect to quantum gravity/string theory predictions?}
\end{enumerate}

\noindent \textbf{The transitional universe}
\begin{enumerate}
    \item How does gravity and accurately resolving the small-scale dynamics of early universe relics (e.g. the cosmic string core) affect the evolution of the network?
    \item Are there any (p)reheating scenarios that generate significant amounts of PBHs with small amplitude fluctuations?
    \item How can we best quantify the stochastic GW background produced from early universe phenomena like reheating?
    \item What are the effects of non-spherical perturbations (i.e. going to 3+1D simulations) in the formation of PBHs?
    \item Can we make quantitative predictions that connect observational data to parameters of the fundamental theory?
    \item How does the GW spectrum of a cosmological phase transition change when studying the dynamics of the gauge theory using holography?
\end{enumerate}

\noindent \textbf{The late universe}
\begin{enumerate}
    \item What is the impact of including GR effects on systematics and percent level effects that may arise in next generation cosmological surveys? e.g. frame dragging, spatial curvature, voids and overdensities, red-shift drift.
    
    \item How does one disentangle which effects are ``backreaction'' effects, and which are numerical systematics such as differences in methodologies?
    \item Does the assumption of periodicity that is used in most simulations impact on the (lack of) development of non-FLRW features?
    \item Are there other GR effects that could explain cosmological tensions?
    \item What are the signatures of modified gravity effects in the large scale structure statistics, e.g. non-linear screening?
\end{enumerate}

\bmhead{Acknowledgements}

We acknowledge helpful input from many of the authors of the papers presented here, in particular Thomas Baumgarte, Marco Bruni, Maxence Corman, Will East, David Garfinkle, Tom Giblin, Sofie Marie Koksbang, Hayley Macpherson and Robyn Munoz. We also thank the anonymous referees for their comments, which helped us improve this manuscript. JCA acknowledges funding from the Beecroft Trust and The Queen’s College via an extraordinary Junior Research Fellowship (eJRF). KC acknowledges funding from from the European Research Council (ERC) under the European Union’s Horizon 2020 research and innovation programme (grant agreement No 693024), an STFC Ernest Rutherford Fellowship project reference ST/V003240/1 and STFC Research Grant ST/X000931/1 (Astronomy at Queen Mary 2023-2026). EAL acknowledges support from a Leverhulme Trust Research Project Grant.  

\phantomsection
\addcontentsline{toc}{section}{References}
\bibliography{LRCosmo}

\end{document}